\documentclass[aps,twocolumn,superscriptaddress,raggedbottom,showpacs,floatfix]{revtex4-1}
\usepackage[pdftex]{graphicx}
\usepackage{mmap}
\usepackage{amssymb}
\usepackage{amsmath}
\usepackage{color}
\usepackage{hyperref}
\usepackage{textcomp}
\hypersetup{colorlinks=true,linkcolor=blue,urlcolor=blue,citecolor=blue}

\begin{document}
\title{Plasmonic modes at inclined edges of anisotropic 2D materials}
\author{Alexey A. Sokolik}%
\affiliation{Institute for Spectroscopy RAS, 142190 Troitsk, Moscow, Russia}%
\affiliation{National Research University Higher School of Economics, 109028 Moscow, Russia}%
\affiliation{Institute of Microelectronics Technology and High Purity Materials RAS, 142432 Chernogolovka, Russia}%
\author{Oleg~V. Kotov}%
\email{oleg.v.kotov@yandex.ru}%
\affiliation{Institute of Microelectronics Technology and High Purity Materials RAS, 142432 Chernogolovka, Russia}%
\affiliation{N.~L. Dukhov Research Institute of Automatics (VNIIA), 127055 Moscow, Russia}%
\author{Yurii~E. Lozovik}%
\email{lozovik@isan.troitsk.ru}%
\affiliation{Institute for Spectroscopy RAS, 142190 Troitsk, Moscow, Russia}%
\affiliation{Institute of Microelectronics Technology and High Purity Materials RAS, 142432 Chernogolovka, Russia}%
\affiliation{National Research University Higher School of Economics, 109028 Moscow, Russia}%
\affiliation{N.~L. Dukhov Research Institute of Automatics (VNIIA), 127055 Moscow, Russia}%

\begin{abstract}
Confined modes at the edge arbitrarily inclined with respect to optical axes of nonmagnetic anisotropic 2D materials are considered. By developing the exact Wiener-Hopf and approximated Fetter methods we studied edge modes dispersions, field and charge density distributions. The 2D layer is described by the Lorentz-type conductivities in one or both directions, which is realistic for natural anisotropic 2D materials and resonant hyperbolic metasurfaces. We demonstrate that, due to anisotropy, the edge mode exists only at wave vectors exceeding the nonzero threshold value if the edge is tilted with respect to the direction of the resonant conductivity. The dominating contribution to field and charge density spatial profiles is provided by evanescent 2D waves, which are confined both in space near the 2D layer and along the layer near its edge. The degree of field confinement along the layer is determined by wave vector or frequency mismatch between the edge mode and continuum of freely propagating 2D modes. Our analysis is suitable for various types of polaritons (plasmon-, phonon-, exciton-polaritons, etc.) at large enough wave vectors. Thanks to superior field confinement in all directions perpendicular to the edge these modes look promising for modern plasmonics and sensorics.
\end{abstract}

\maketitle
\section{Introduction} \label{Sec1}
Light trapping and manipulation at the nanoscale, below the diffraction limit, are the key techniques of nanophotonics opening the way to local amplification of the electric field. The resulting enhanced light-matter interaction gives rise to enhanced emission and absorption probabilities, high optical sensitivity, large photonic forces, and strong nonlinearities \cite{schuller2010}. Numerous applications of strong light confinement include biosensing, super-resolution imaging, nanofocusing, nanoscale heat transfer, surface-enhanced Raman scattering, modification of the spontaneous emission rate of quantum emitters, and optical manipulation and trapping of nanoparticles. Realization of these phenomena and techniques is feasible by using excitation of confined light-matter hybrid modes (plasmon-, phonon-, exciton-, magnon-polaritons, etc.) at interfaces separating media with permittivities or/and permeabilities of opposite signs \cite{basov2016,low2016}. The negative permittivity is provided by the charge redistribution and corresponding local currents caused by various physical mechanisms, e.g., coherent oscillations of free carriers for plasmon-polaritons (PPs), or ionic charge oscillations for phonon-polaritons (PhPs). Polaritonic modes consisting of collective oscillations of polarization charges in matter, coupled with electromagnetic (EM) waves, in the non-retarded limit can be called in a broad sense ``plasmonic'' modes. 

Traditionally, PPs are excited in noble-metal structures, where local or surface PPs are intrinsically bound to a 2D metal-dielectric interface \cite{Maier,Bozhevolnyi}. The recent rise of atomically thin 2D materials \cite{Roadmap2D} hosting different types of polaritons opens up new opportunities in design of polaritonic systems with stronger light confinement, higher transparency, and dynamic tunability compared to conventional plasmonic films. By stacking 2D layers in van der Waals (vdW) heterostructures \cite{Geim2013,novoselov2016}, one can engineer artificial thin films with an unusual combination of polaritons. Besides, the emergence of layered 2D materials led to great progress in phonon-polaritonics \cite{Rev2015_PhP,Rev2019_PhP} realized in thin films of polar materials such as hexagonal boron nitride (hBN) \cite{dai2014,li2015,Rev2019_hBN,nikitin2018} or molybdenum trioxide ($\rm MoO_3$) \cite{ma2018,zheng2019,nikitin2020}. These polaritonic platforms have at least two advantages over conventional metal films or even graphene. First, while in graphene due to the linear electron spectra the field confinement is ultimately limited by Landau damping, in such materials with quadratic spectrum the Landau damping region is shifted to much larger wave vectors, and polaritons confinement is limited only by the intrinsic material loss \cite{dubrovkin2018,lee2020}. Second, as optical phonons exhibit much longer lifetimes than free carriers, especially in isotopically pure hBN \cite{giles2017}, the optical losses of PhPs are significantly lower than for PPs. Thus isotopically enriched polar vdW films may become the best ``plasmonic'' materials which provide both ultrastrong field confinement and large propagation length of surface waves, thus overcoming the traditional trade-off between these properties. However, PhPs allow us to work in the mid-IR or lower frequencies, whereas exciton-polaritons in thin semiconductors (e.g., $\rm MoSe_2$ \cite{hu2017}), while retaining strong confinement and large propagation lengths, can be excited in the visible range. 

Thanks to the inherent anisotropy, many vdW materials exhibit hyperbolic regime, when they behave as a dielectric along one direction and as a metal along the orthogonal one \cite{Smith_2003,Jacob_2012}. Hyperbolic behavior results in the unique optical phenomena, such as negative refraction \cite{hoffman2007,fang2009,lin2017}, strong enhancement of spontaneous emission \cite{noginov2010,jacob2012}, super-Planckian thermal emission \cite{guo2012}, enhanced superlensing effects \cite{smith2004,liu2007,dai2015}, polaritons self-collimation \cite{stein2012,forati2014,alu2017}, and spin control \cite{ITMO_spin}. A broad class of natural hyperbolic materials covers very wide spectral
range \cite{narimanov2015,korzeb2015,gjerding2017,sun2014}. For instance, polar crystals (hBN, $\rm MoO_3$, etc.) possess hyperbolicity in the mid-IR and THz owing to strong phonon resonances, while black phosphorus (BP) \cite{Alu_BP, Katsnelson_BP} and tetradymites $\rm Bi_2Se_3$, $\rm Bi_2Te_3$ \cite{esslinger2014} depending on thickness can show hyperbolic regime in UV, visible, near-IR, or mid-IR, which originates from highly anisotropic interband electronic transitions. The majority of vdW materials exhibit only the out-of-plane hyperbolicity, but still the in-plane one was predicted in thin BP \cite{Alu_BP, Katsnelson_BP} and experimentally measured in $\rm MoO_3$ \cite{ma2018,zheng2019}. However, so far to obtain the in-plane hyperbolicity in the visible or microwave ranges it is better to use artificial thin hyperbolic materials (metasurfaces) composed of planar anisotropic arrays of metallic subwavelength scatters \cite{RevMeta_Holloway,RevMeta_ITMO,Alu_Rev,Kotov2017,RevHMS_2019,Kotov2019}. Although hyperbolic behaviour may occur in BP monolayer in the visible range, it has a too low figure of merit \cite{Katsnelson_BP}, while silver-grating \cite{HMS_Liu2013,Visible_HMS} or gold-nanodisks metasurfaces \cite{ITMO_PRB,ITMO_expOpt} provide pronounced hyperbolic PPs in the visible range. 

The research of polaritonic modes in 2D materials or metasurfaces has given a new impetus to the topic of edge polaritonic modes \cite{nikitin2011,mason2014,nikitin2016,angelis2016,mortensen2017,aken2016,lu2018,aken2020,bisharat2017}. These 1D modes are interesting due to their superior field confinement in all directions perpendicular to the edge, even stronger than for 2D modes propagating along the layer \cite{nikitin2016,mortensen2017}. Historically, the most-studied edge modes are edge magnetoplasmons (EMPs) \cite{Fetter,Volkov,wassermeier1990,ashoori1992,talyanskii1992,kukushkin2008,wang2012,yan2012,lin2013,kumada2014,jin2016,cohen2018,sokolik2019}, first theoretically described in the nonretarded limit at the boundary of a conducting isotropic 2D layer half-plane placed in perpendicular magnetic field \cite{Fetter,Volkov}. In Ref.~\cite{Volkov} Volkov and Mikhailov obtained the exact solution of integro-differential equation for the electrostatic potential of EMPs by means of Wiener-Hopf method, which allows us to write EMPs dispersion equation in explicit form. The simplified approach allowing us to find the mode frequencies in an algebraic way was proposed by Fetter in Ref.~\cite{Fetter}. This approximation provides qualitatively correct EMPs dispersions, however with incorrect behavior of the field and charge density in close vicinity to the edge. In recent papers on edge EM modes both the Fetter approximation \cite{wang2011,cohen2018,stauber2019,zabolotnykh2016} and the Wiener-Hopf exact method \cite{mason2014,cohen2018,sokolik2019,stauber2020,margetis2020} have been applied. The effect of EM field retardation in the edge plasmon-polaritons problem was taken into account generalizing the Fetter approach \cite{zabolotnykh2016}, as well as the Wiener-Hopf method \cite{margetis2020}. Recent interest to 2D anisotropic (hyperbolic) materials gave rise to the development of both techniques for the anisotropic case \cite{stauber2019,stauber2020}. However, in these works the results are given only for the edge parallel to the anisotropy axis and for the Drude conductivities in both directions. 

Drude-type conductivities are not suitable for most realistic implementations of 2D anisotropic materials. Anisotropic metasurfaces composed of 2D array of scatters (e.g., metal nanodisk array) or natural polar materials with in-plane anisotropy (e.g., $\rm MoO_3$) possess Lorentz-type optical responses in both in-plane directions. Metasurfaces based on 1D metallic array (e.g., graphene ribbon arrays) or natural materials with in-plane anisotropy of interband electronic transitions (e.g., phosphorene) have Drude-type response in one direction and resonant-like in the orthogonal one.

In this article we consider plasmonic modes localized at the edge which is arbitrarily inclined to the anisotropy axis of a nonmagnetic 2D layer with anisotropic conductivity. We assume the resonant (Lorentz-type) conductivities in one or both directions of the layer. To analyze the edge modes in a time-reversal symmetric anisotropic material, we apply both exact Wiener-Hopf and approximated Fetter methods, which were initially developed for isotropic systems in magnetic field. We obtain and compare the corresponding exact and approximated solutions for the edge modes dispersions, as well as for the field and density distributions. We show that the edge modes exist only in the inductive elliptic region of the spectrum, where the conductivities in both directions have positive imaginary parts. 

We demonstrate that a resonant behavior of the conductivity in one of the directions and a nonzero tilt of the edge with respect to this direction result in existence of the edge modes only at wave vectors exceeding a nonzero threshold. By comparing with the exact Wiener-Hopf solution we confirm the validity of the Fetter approximation but also reveal its limitations. The Fetter approximation in this case provides red-shifted edge mode dispersions and incorrect values of the threshold wave vector. Yet we find it very important and remarkable that the simple Fetter approximation remains quite good in the anisotropic case giving qualitatively correct results. We demonstrate that the degree of edge mode field confinement along the 2D layer to the edge is determined by wave vector or frequency mismatch between the edge and 2D modes. Our results can be applied to a wide class of anisotropic 2D nonmagnetic materials and to various types of polaritons (plasmon-, phonon-, exciton-polaritons, etc.) at large enough wave vectors.

The article is organized as follows. In Sec.~\ref{sec_theory} we describe the theoretical approach to calculate dispersions of edge as well as 2D waves; the Fetter approximated approach is also described. In Sec.~\ref{sec_results} we demonstrate and analyze the results of numerical calculations of mode dispersions, field and density profiles, and field localization lengths. Sec.~\ref{sec_discussion} is devoted to discussion and conclusions, and Appendices~\ref{Appendix_A}--\ref{Appendix_C} provide details of calculations.

\section{Theory}\label{sec_theory}
\subsection{Edge modes}\label{sec_edge_modes}
We consider the 2D layer with optical response described by the Lorentz-type conductivities \cite{ITMO_PRB}
\begin{equation}
\sigma_{\perp,\parallel}(\omega)=\frac{cA_{\perp,\parallel}}{4\pi}
\frac{i\omega}{\omega^2-\Omega_{\perp,\parallel}^2+i\omega\gamma}
\label{sigma_perp_para}
\end{equation}
along mutually perpendicular $\perp$ and $\parallel$ axes. Here $c$ is the light velocity, $A_{\perp,\parallel}$ count for the spectral weights of two resonances with frequencies $\Omega_{\perp,\parallel}$, where $\Omega_\parallel>\Omega_\perp$. For our approach, it is essential that the conductivity $\sigma_\parallel(\omega)$ with the higher resonant frequency has the resonant Lorentz-type form. The other conductivity $\sigma_\perp(\omega)$ can be of the Lorentz or Drude ($\Omega_\perp=0$) type.

For simplicity of the following theoretical analysis, we assume equal decay rates $\gamma$ for both resonances. As will be shown below, when EM field retardation is neglected, the equations for the modes include only the ratios $\sigma_{\alpha\beta}(\omega)/i\omega$, so due to (\ref{sigma_perp_para}) the complex frequency $\omega$ enters these equations only in combinations like $\omega^2-\Omega_{\perp,\parallel}^2+i\omega\gamma=\tilde\omega^2-\Omega_{\perp,\parallel}^2+\gamma^2/4$, where $\omega=\tilde\omega-i\gamma/2$ is the complex frequency of decaying modes having the real part $\tilde\omega$. Therefore the damping $\gamma$ simply adds $-i\gamma/2$ to resulting $\omega$ of the modes and renormalizes resonance frequencies $\Omega_{\perp,\parallel}\rightarrow(\Omega_{\perp,\parallel}^2-\gamma^2/4)^{1/2}$. In the following calculations we will assume $\omega$ to be real, bearing in mind that the total complex frequency is obtained after the substitution $\omega\rightarrow\omega-i\gamma/2$ and that $\Omega_{\perp,\parallel}$ are already renormalized by the damping.

With this convention the conductivities (\ref{sigma_perp_para}) become purely imaginary. At $\omega<\Omega_\perp$ there is a capacitive elliptic regime ($\mathrm{Im}\,\sigma_\perp<0$, $\mathrm{Im}\,\sigma_\parallel<0$), at $\Omega_\perp<\omega<\Omega_\parallel$ the behavior is hyperbolic  ($\mathrm{Im}\,\sigma_\perp>0$, $\mathrm{Im}\,\sigma_\parallel<0$), and for $\omega>\Omega_\parallel$ the 2D layer demonstrates the inductive elliptic regime ($\mathrm{Im}\,\sigma_\perp>0$, $\mathrm{Im}\,\sigma_\parallel>0$).

\begin{figure}
\begin{center}
\includegraphics[width=0.6\columnwidth]{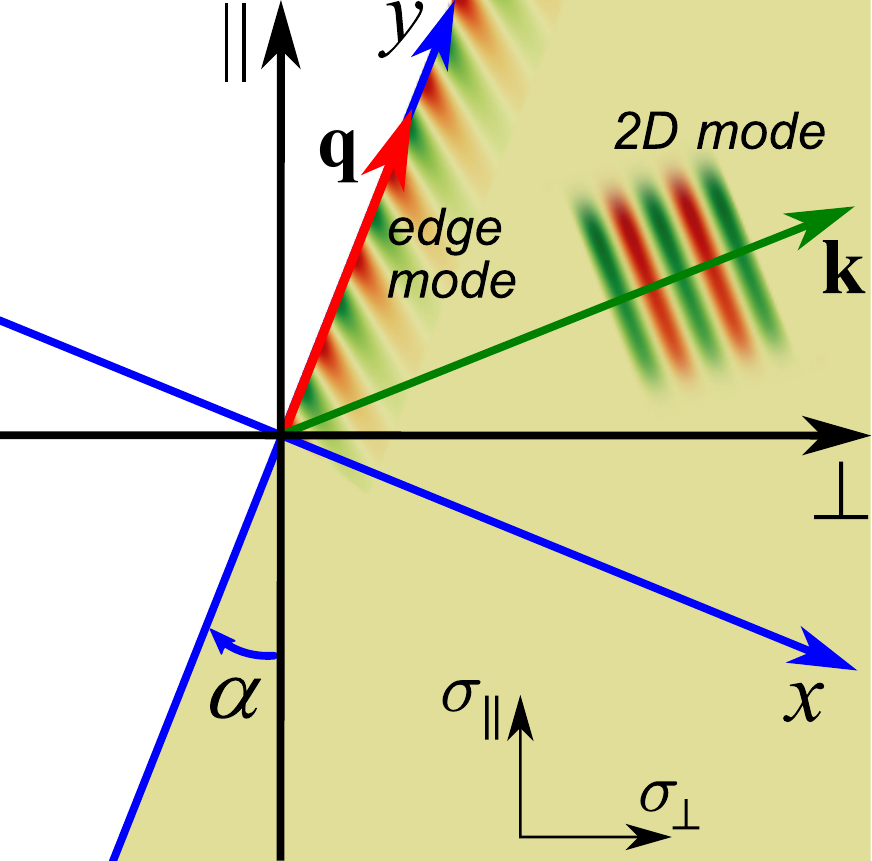}
\end{center}
\caption{\label{Fig1}The $(\perp,\parallel)$ optical axes of the anisotropic 2D layer and the $(x,y)$ axes, where the 2D layer half-plane is located in the region $x\geqslant0$, and edge modes propagate along the $y$ axis; $\alpha$ is the angle between the $y$ and $\parallel$ axes. 2D modes are not confined to the edge and can freely propagate in different directions $\mathbf{k}$ along the half-plane.}
\end{figure}

We consider the half-plane of the 2D layer occupying the $x\geqslant0$, $z=0$ region, where the axes $(x,y)$ shown in Fig.~\ref{Fig1} are rotated on the angle $\alpha$ with respect to the layer optical axes $(\perp,\parallel)$. In the rotated coordinate system, the components of the conductivity tensor are
\begin{align}
\sigma_{xx}&=\sigma_\perp\cos^2\alpha+\sigma_\parallel\sin^2\alpha,\nonumber
\\[0.2em]
\sigma_{yy}&=\sigma_\perp\sin^2\alpha+\sigma_\parallel\cos^2\alpha, \label{sigma_tensor}
\\[0.2em]
\sigma_{xy}&=\sigma_{yx}=(\sigma_\perp-\sigma_\parallel)\sin\alpha\cos\alpha.\nonumber
\end{align}

Assuming that the edge mode propagates along the $y$ axis with the wave vector $q>0$ and frequency $\omega$, we can combine the continuity equation $\partial\rho/\partial t+\mathrm{div}\mathbf\,\mathbf{j}=0$ and the Ohm's law $j_\alpha=\sigma_{\alpha\beta}E_\beta=-\sigma_{\alpha\beta}\nabla_\beta\varphi$ for the 2D charge density $\rho$ at the 2D layer and for electric field potential $\varphi(x)\equiv\varphi(x,z=0)$ at the layer, both being proportional to $e^{i(qy-\omega t)}$. Calculating $\mathrm{div}\,\mathbf{j}$, we need to take into account that, due to the abrupt edge, $\sigma_{\alpha\beta}\propto\Theta(x)$, where $\Theta(x)$ is the unit step function. The resulting equation, which is valid at $x\geqslant0$, $z=0$, is
\vspace{-0.5em}
\begin{align}
&i\omega\rho(x)=-\delta(x)(\sigma_{xx}\partial_x+iq\sigma_{xy})\varphi(x)
\nonumber
\\[0.2em]
 &-iq(\sigma_{xy}+\sigma_{yx})\varphi'(x)-
(\sigma_{xx}\partial_x^2-q^2\sigma_{yy})\varphi(x).\label{cont}
\end{align}

Another basic relation is the Poisson equation $\varepsilon_\mathrm{b}\nabla^2\varphi=-4\pi\rho\delta(z)$, which can be applied when the EM field retardation is neglected. After transforming it into the integral form, we get \cite{Fetter,Volkov}
\begin{equation}
\varphi(x)=\frac{4\pi}{\varepsilon_\mathrm{b}}\int\limits_0^\infty
dx'\:L(x-x')\rho(x'),\label{Poisson}
\end{equation}
where $\varepsilon_\mathrm{b}$ is the background dielectric constant of a three-dimensional medium surrounding the 2D layer. The kernel
\begin{equation}
L(x)=\int\frac{dk}{2\pi}\frac{e^{ikx}}{2\sqrt{k^2+q^2}}=\frac1{2\pi}K_0(q|x|),
\label{kernel}
\end{equation}
where $K_0$ is the modified Bessel function of the second king, describes the field created by a string with the charge density proportional to $e^{iqy}$.

Eqs.~(\ref{cont})--(\ref{kernel}) are the basic equations which should be solved in order to obtain dispersion and other characteristics (field and charge density distributions) of the edge mode. The Wiener-Hopf method was applied in Ref. \cite{Volkov} to the problem of such a kind in the case of edge magnetoplasmons, when $\sigma_{xy}=-\sigma_{yx}$. We apply the same method to anisotropic 2D layer with time reversal symmetry, when $\sigma_{xy}=\sigma_{yx}$, as described in Appendix~\ref{Appendix_A}. The resulting dispersion equation for edge modes is
\begin{align}
&\int\frac{dk}{k^2\sigma_{xx}-kq(\sigma_{xy}+\sigma_{yx})+q^2\sigma_{yy}}
\ln\left\{-\varepsilon(k)\right\}=0,\label{disp}
\\[0.2em]
&\varepsilon(k)=1-\frac{2\pi}{i\varepsilon_\mathrm{b}\omega}
\frac{k^2\sigma_{xx}-kq(\sigma_{xy}+\sigma_{yx})+q^2\sigma_{yy}}{\sqrt{k^2+q^2}}.
\label{epsilon}
\end{align}
In terms of $\sigma_{\perp,\parallel}$ it has the form
\begin{align}
&\int\frac{dk}{\sigma_\perp k_\perp^2+\sigma_\parallel k_\parallel^2}
\ln\left\{-\varepsilon(k)\right\}=0,
\\[0.2em]
&\varepsilon(k)=1-\frac{2\pi}{i\varepsilon_\mathrm{b}\omega}
\frac{\sigma_\perp k_\perp^2+\sigma_\parallel k_\parallel^2}{\sqrt{k^2+q^2}},
\label{epsilon_alpha}
\end{align}
where 
\begin{align}
k_\perp=-k\cos\alpha+q\sin\alpha,\quad k_\parallel=k\sin\alpha+q\cos\alpha.
\label{k_perp_para}
\end{align}
For a proper choice of the logarithm branch in $\ln\{-\varepsilon(k)\}$ an infinitesimal positive real part can be added to $\sigma$'s in these equations.

The solution $\omega_\mathrm{e}(q)$ of the edge mode dispersion equation is monotonously increasing and exists only in the inductive elliptic region, when $\mathrm{Im}\,\sigma_\perp>0$ and $\mathrm{Im}\,\sigma_\parallel>0$ [for Eq.~(\ref{sigma_perp_para}) at $\omega>\Omega_\parallel$], and at wave vectors exceeding the threshold one (see Appendix~\ref{Appendix_A})
\begin{align}
q_0=\frac{i\varepsilon_\mathrm{b}\Omega_\parallel}{2\pi\sigma_\perp(\Omega_\parallel)}|\sin\alpha|.\label{q0gen}
\end{align}
Generally, this formula is applicable at the points $\omega$ of the complex frequency plane where one of the conductivities ($\sigma_\parallel$) tends to infinity, while the other one ($\sigma_\perp$) remains finite. With the Lorentz-type expressions (\ref{sigma_perp_para}) for both conductivities we obtain
\begin{align}
q_0=\frac{2\varepsilon_\mathrm{b}}{cA_\perp}(\Omega_\parallel^2-\Omega_\perp^2)
|\sin\alpha|.\label{q0}
\end{align}

Note that we assume $q>0$ in our calculations. For the backpropagating edge mode with $q<0$, the solutions $\varphi(x)$, $\rho(x)$ of Eqs.~(\ref{cont})--(\ref{kernel}) become complex conjugated with respect to those for $q>0$.

\subsection{2D modes}
The dielectric function $\varepsilon(k)$ characterizes the degree of weakening of the external field plane wave $e^{i(-kx+qy-\omega t)}$ due to the 2D layer response at $z=0$ [the sign at $kx$ is negative due to the Fourier transform convention in (\ref{Fourier1})]. Therefore the equation
\begin{align}
\varepsilon(k)=0\label{2D_disp}
\end{align}
provides the dispersion of TM-polarized 2D modes, i.e., the modes which propagate along the uniform 2D layer with the wave vector $\mathbf{k}=-k\mathbf{e}_x+q\mathbf{e}_y$ (see Fig.~\ref{Fig1}) and have the field confined to the 2D layer in space as $e^{-\sqrt{k^2+q^2}|z|}$. These modes were studied in detail in Ref.~\cite{ITMO_PRB} with taking into account the retardation of EM field. In the $(\perp,\parallel)$ coordinate system [see Eqs.~(\ref{epsilon_alpha})--(\ref{k_perp_para})], the dispersion equation for these modes at $q\gg\omega/c$ is
\begin{align}
\frac{2\pi}{i\varepsilon_\mathrm{b}\omega}
\frac{\sigma_\perp k_\perp^2+\sigma_\parallel k_\parallel^2}
{\sqrt{k_\perp^2+k_\parallel^2}}=1.\label{2D_disp1}
\end{align}
It can be solved analytically as a biquadratic equation resulting in two positive solutions $\omega^\mathrm{2D}_\pm$, which correspond to hyperbolic modes at $\Omega_\perp<\omega^\mathrm{2D}_-<\Omega_\parallel$ and elliptic modes at $\omega^\mathrm{2D}_+>\Omega_\parallel$. We expect that the retardation \cite{ITMO_PRB} is significant only at $q\rightarrow0$ near the light cone.
\vspace{-0.5em}
\subsection{Fetter approximation}
\vspace{-0.2em}
Another method frequently used to solve the problem (\ref{cont})--(\ref{kernel}) of edge modes is the Fetter approach \cite{Fetter} (see Appendix~\ref{Appendix_B}) where the nonlocal integral equation (\ref{Poisson}) is approximated by the differential local one. In this case, the general dispersion equation for arbitrary $\sigma_{\alpha\beta}$ tensor takes the form
\begin{align}
&\left(\frac{4\pi q\sigma_{xx}}{i\varepsilon_\mathrm{b}\omega}-1\right)
\left(\frac{4\pi q\sigma_{yy}}{i\varepsilon_\mathrm{b}\omega}-2\right)\nonumber
\\[0.2em]
&-\left(\frac{4\pi q\sigma_{xy}}{i\varepsilon_\mathrm{b}\omega}-i\sqrt2\right)
\left(\frac{4\pi q\sigma_{yx}}{i\varepsilon_\mathrm{b}\omega}+i\sqrt2\right)=0.
\label{Fetter_disp_equation}
\end{align}
Specifically for time-reversal symmetric anisotropic 2D layer, the dispersion equation in terms of $\sigma_{\perp,\parallel}$ [see (\ref{sigma_tensor})] is
\begin{align}
\frac{4\pi q}{i\varepsilon_\mathrm{b}\omega}\sigma_\perp\sigma_\parallel-
\sigma_\perp(1+\cos^2\alpha)-\sigma_\parallel(1+\sin^2\alpha)=0.
\label{Fetter_disp_alpha}
\end{align}
Eqs.~(\ref{disp})--(\ref{epsilon_alpha}) and Eqs.~(\ref{Fetter_disp_equation})--(\ref{Fetter_disp_alpha}) being, respectively, the exact and approximate relations for the edge modes dispersions are the main analytical result of the paper, which can be used for any 2D material described by conductivity tensor with $\sigma_{xy}=\sigma_{yx}$. Using (\ref{sigma_perp_para}), we obtain the explicit formula for edge mode dispersion in the Fetter approximation:
\begin{align}
\omega_\mathrm{e}^\mathrm{F}(q)=\sqrt{\frac{\displaystyle\frac{cq}{\varepsilon_\mathrm{b}}+\Omega_\perp^2\frac{1+\sin^2\alpha}{A_\perp}+		\Omega_\parallel^2\frac{1+\cos^2\alpha}{A_\parallel}}{\displaystyle\frac{1+\sin^2\alpha}{A_\perp}+\frac{1+\cos^2\alpha}{A_\parallel}}}.
\label{disp_Fetter}
\end{align}
However, as shown in Appendix~\ref{Appendix_B}, a physically meaningful solution for the mode confined to the edge exists only in the frequency region $\omega>\Omega_\parallel$, which corresponds to wave vectors $q$ exceeding the threshold one
\begin{align}
q_0^\mathrm{F}=\frac{\varepsilon_\mathrm{b}(\Omega_\parallel^2-\Omega_\perp^2)}
{2cA_\perp}(3-\cos2\alpha).\label{q0F}
\end{align}
The conclusion about existence of the edge modes only in the inductive elliptic region ($\mathrm{Im}\,\sigma_\perp>0$, $\mathrm{Im}\,\sigma_\parallel>0$) is the same for exact Wiener-Hopf and approximated Fetter solution, however the respective values (\ref{q0}) and (\ref{q0F}) of the threshold wave vector differ, especially near $\alpha=0,\pi$
(see the inset in Fig.~\ref{Fig4} below).
\vspace{-0.5em}
\subsection{Isotropic surface}\label{sec_isotropic}
\vspace{-0.2em}
In the particular case of isotropic surface with the Lorentz-type response, $\Omega_\perp=\Omega_\parallel\equiv\Omega$, $A_\perp=A_\parallel\equiv A$, the exact Wiener-Hopf solution of the dispersion equation (\ref{disp}) is greatly simplified:
\begin{equation}
\omega_\mathrm{e,i}=\sqrt{\frac{cAq}{\varepsilon_\mathrm{b}\eta_0}+\Omega^2},
\label{edge_isotropic}
\end{equation}
where $\eta_0\approx2.4344$ is the solution of equation
\begin{equation}
\int\frac{d\xi}{\xi^2+1}\ln\left\{\frac{\eta_0\sqrt{\xi^2+1}}2-1\right\}=0.
\end{equation}
Dispersion in the Fetter approximation (\ref{disp_Fetter}) in the isotropic case,
\begin{equation}
\omega_\mathrm{e,i}^\mathrm{F}=
\sqrt{\frac{cAq}{3\varepsilon_\mathrm{b}}+\Omega^2},\label{Fetter_isotropic}
\end{equation}
differs from (\ref{edge_isotropic}) quantitatively by replacing $\eta_0$ with $3$.

For 2D modes, only the elliptic solution of (\ref{2D_disp1}) exists in the isotropic case:
\begin{equation}
\omega_\mathrm{+,i}^\mathrm{2D}=
\sqrt{\frac{cAq}{2\varepsilon_\mathrm{b}}+\Omega^2}.\label{2D_isotropic}
\end{equation}

Note that these results are valid for the Drude-type conductivity as well, if we take $\Omega=0$. In this case the edge mode becomes gapless with a square-root dispersion, as shown in Ref.~\cite{Volkov}.

\begin{figure}
\begin{center}
\includegraphics[width=0.9\columnwidth]{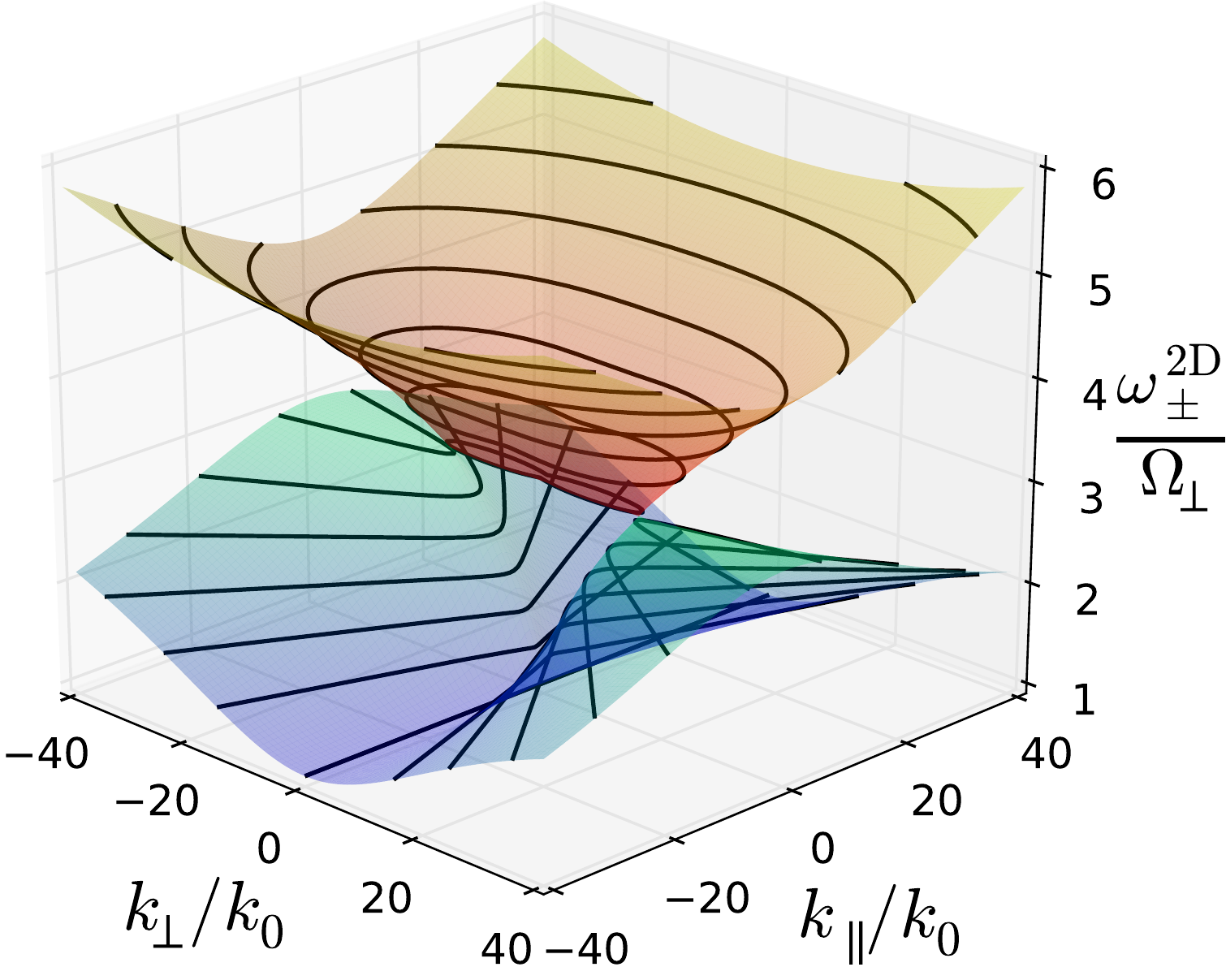}
\end{center}
\caption{\label{Fig2} Dispersions $\omega^\mathrm{2D}_\pm$ and isofrequency contours of 2D hyperbolic ($\omega^\mathrm{2D}_-$, lower surface) and elliptic ($\omega^\mathrm{2D}_+$, upper surface) modes propagating along a uniform 2D layer. Wave vectors $k_{\perp,\parallel}$ are in the units of $k_0=\varepsilon_\mathrm{b}\Omega_\perp^2/cA$, frequencies are in the units of $\Omega_\perp$. Hybridization of hyperbolic and elliptic modes at two points $\{k_\perp,k_\parallel\}=\{\pm16k_0,0\}$ is seen.}
\end{figure}

\section{Calculation results}\label{sec_results}
\subsection{Dispersions of edge and 2D modes}
Hereafter, similarly to Ref.~\cite{ITMO_PRB}, we assume equal spectral weights $A_\perp=A_\parallel\equiv A$ of both resonances $\Omega_{\perp,\parallel}$ and the anisotropy parameter $\Omega_\parallel/\Omega_\perp=3$. All wave vectors will be shown in the units of $k_0=\varepsilon_\mathrm{b}\Omega_\perp^2/cA$.

Dispersion of 2D modes, which propagate along uniform surface, are found from Eq.~(\ref{2D_disp1}) and shown in Fig.~\ref{Fig2}. The hyperbolic $\omega^\mathrm{2D}_-$ and elliptic $\omega^\mathrm{2D}_+$ waves hybridize \cite{ITMO_PRB} at the points $k_\perp=\pm2\varepsilon_\mathrm{b}(\Omega_\parallel^2-\Omega_\perp^2)/cA_\perp$, $k_\parallel=0$. The hyperbolic modes exist in the range $\Omega_\perp<\omega^\mathrm{2D}_-<\Omega_\parallel$ and are canalized at $|\mathbf{k}|\rightarrow\infty$ in the directions $\{k_\perp,k_\parallel\}\propto\{\pm[(\omega^\mathrm{2D}_-)^2-\Omega_\perp^2]^{1/2},\pm[\Omega_\parallel^2-(\omega^\mathrm{2D}_-)^2]^{1/2}\}$. The dispersion of the elliptic mode $\omega^\mathrm{2D}_+$ starts from the straight line connecting the hybridization points at $\omega^\mathrm{2D}_+=\Omega_\parallel$ and then isofrequency contours, going through $\infty$-like non-convex shape, eventually become elliptic in the high-frequency limit.

\begin{figure}
\begin{center}
\includegraphics[width=0.9\columnwidth]{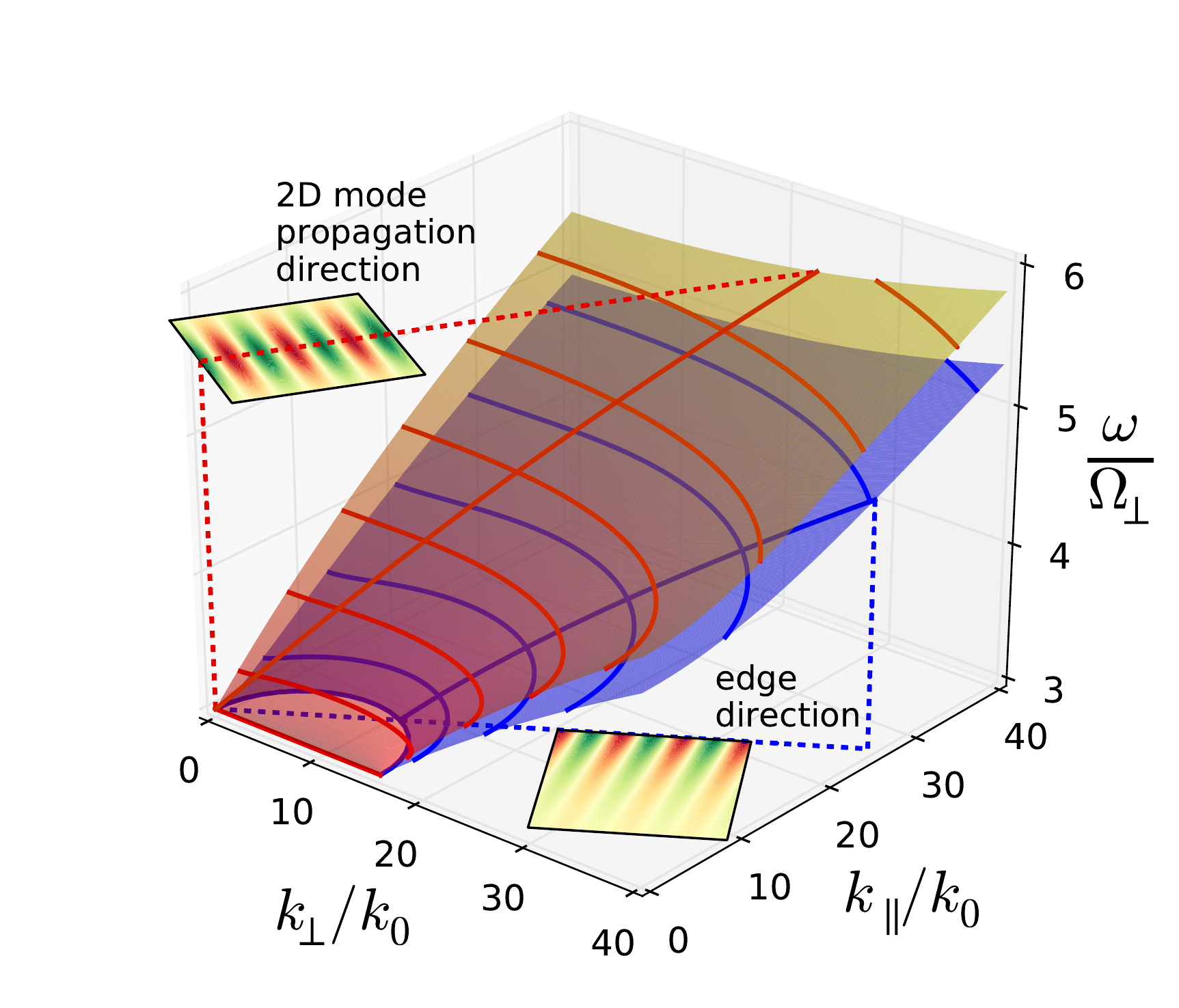}
\end{center}
\caption{\label{Fig3}Upper surface: dispersion and isofrequency contours of elliptic 2D mode propagating along the uniform 2D layer. Bottom surface: the set of 1D dispersions of edge modes $\omega_\mathrm{e}(q,\alpha)$ with the wave vectors
$q=[k_\perp^2+k_\parallel^2]^{1/2}$ along the $y$ axis and different angles $\alpha$ between the wave vector and the $\parallel$ axis. Thus at each $(k_\perp,k_\parallel)$ two surfaces show frequencies of 2D elliptic and 1D edge modes propagating with the same wave vectors in the same direction. Insets show examples of 2D and edge mode dispersions taken at some directions of the $\mathbf{k}$ wave vector (for 2D mode) or edge (for edge mode).}
\end{figure}

Edge mode dispersions found using the Wiener-Hopf method from Eq.~(\ref{disp}) at different angles of the half-infinite 2D layer cutting $0\leqslant\alpha\leqslant\pi/2$ are shown in Fig.~\ref{Fig3}. The edge mode wave vector along the $y$ axis in the $(\perp,\parallel)$ coordinates is $\{k_\perp,k_\parallel\}=\{q\sin\alpha,q\cos\alpha\}$. This picture is mirror-symmetric for negative values of $k_\perp$ and $k_\parallel$. As we see, the edge modes indeed start from, generally, nonzero wave vector (\ref{q0}). This is a signature of 2D layer anisotropy, because in the isotropic case the edge mode (\ref{edge_isotropic}) always starts from $q=0$.

\begin{figure}
\begin{center}
\includegraphics[width=\columnwidth]{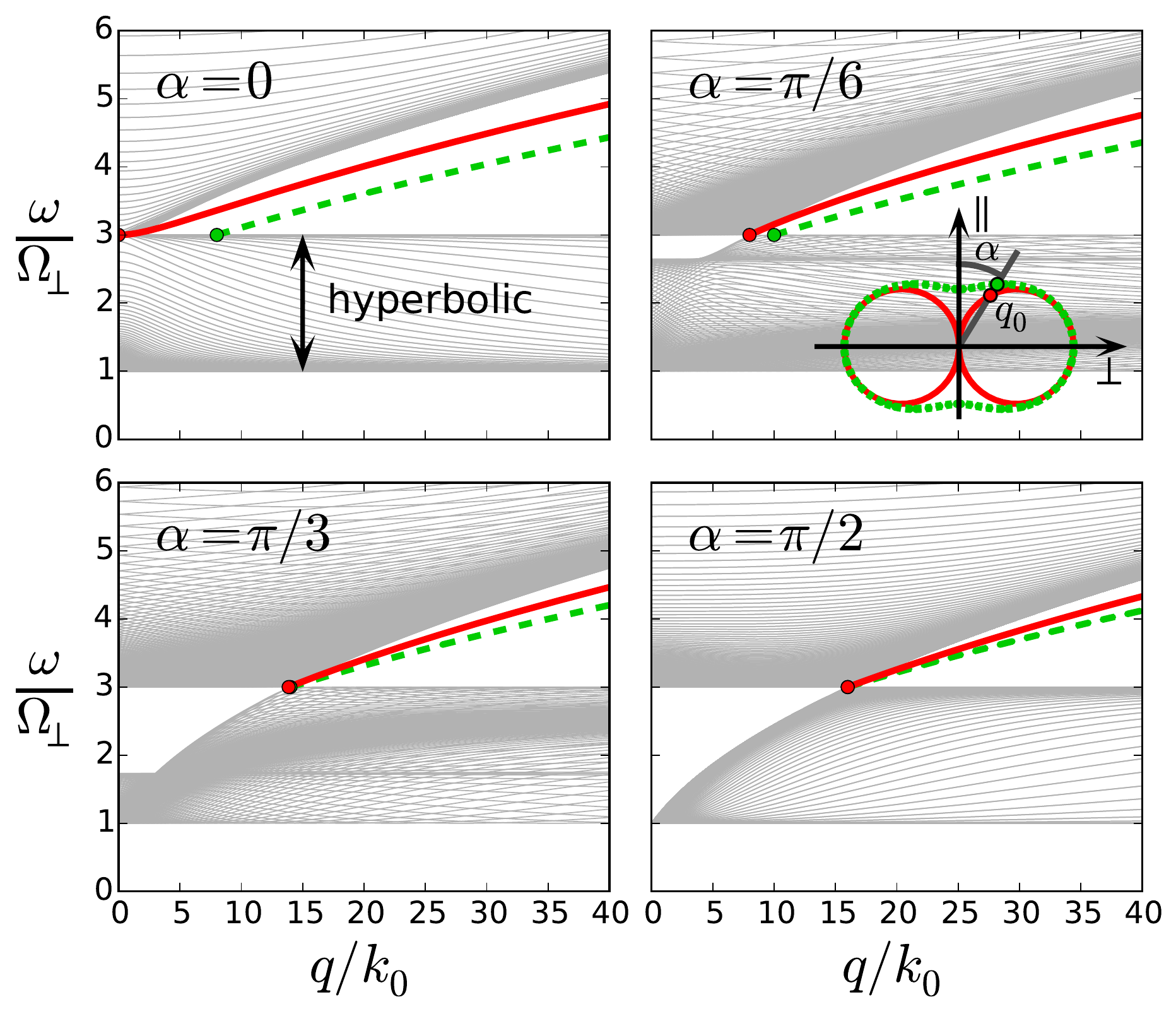}
\end{center}
\caption{\label{Fig4}Dispersions of edge mode (solid lines) starting at the threshold wave vector $q_0$ at four different directions $\alpha$ of the edge. Thin lines show continua of dispersions of elliptic ($\omega/\Omega_\perp\geqslant3$) and hyperbolic ($1\leqslant\omega/\Omega_\perp\leqslant3$) 2D modes with different $x$-projections of the wave vector $\mathbf{k}$, but with the same $y$-projection $q$ of $\mathbf{k}$ as that of the edge mode. Dashed lines show the edge mode dispersions in the Fetter approximation (\ref{disp_Fetter}). Inset shows the $\alpha$ angle dependencies of the exact threshold wave vector $q_0$ (solid line) and its counterpart $q_0^\mathrm{F}$ (dotted line) obtained in the Fetter approximation.}
\end{figure}

It is useful to compare the dispersion of the edge mode $\omega_\mathrm{e}$ vs. its wave vector $q$ along the edge (or $y$ axis) and continuum of 2D mode dispersions $\omega^\mathrm{2D}_\pm$ with the same wave vector $q$ along the edge ($y$ axis) and different wave vectors components $-\infty<k<\infty$ perpendicularly to the edge (along the $x$ axis). Examples for several 2D layer cutting angles $\alpha$ are shown in Fig.~\ref{Fig4}. We see that, first, the edge mode dispersion starts from the threshold wave vector (\ref{q0}), which coincides with the $y$-axis projection of the hyperbolic and elliptic 2D modes hybridization point. Second, the edge mode at any $q$ and $\alpha$ lies below the elliptic 2D mode continuum. It means the edge mode is stable and cannot decay into the elliptic 2D ones with conservation of $q$ and $\omega$. The Fetter approximation (\ref{disp_Fetter}) provides slightly lower edge mode frequency than the exact one and the threshold wave vector (\ref{q0F}) with the incorrect angle dependence (see the comparison in the inset of Fig.~\ref{Fig4}).

\begin{figure}
\begin{center}
\includegraphics[width=\columnwidth]{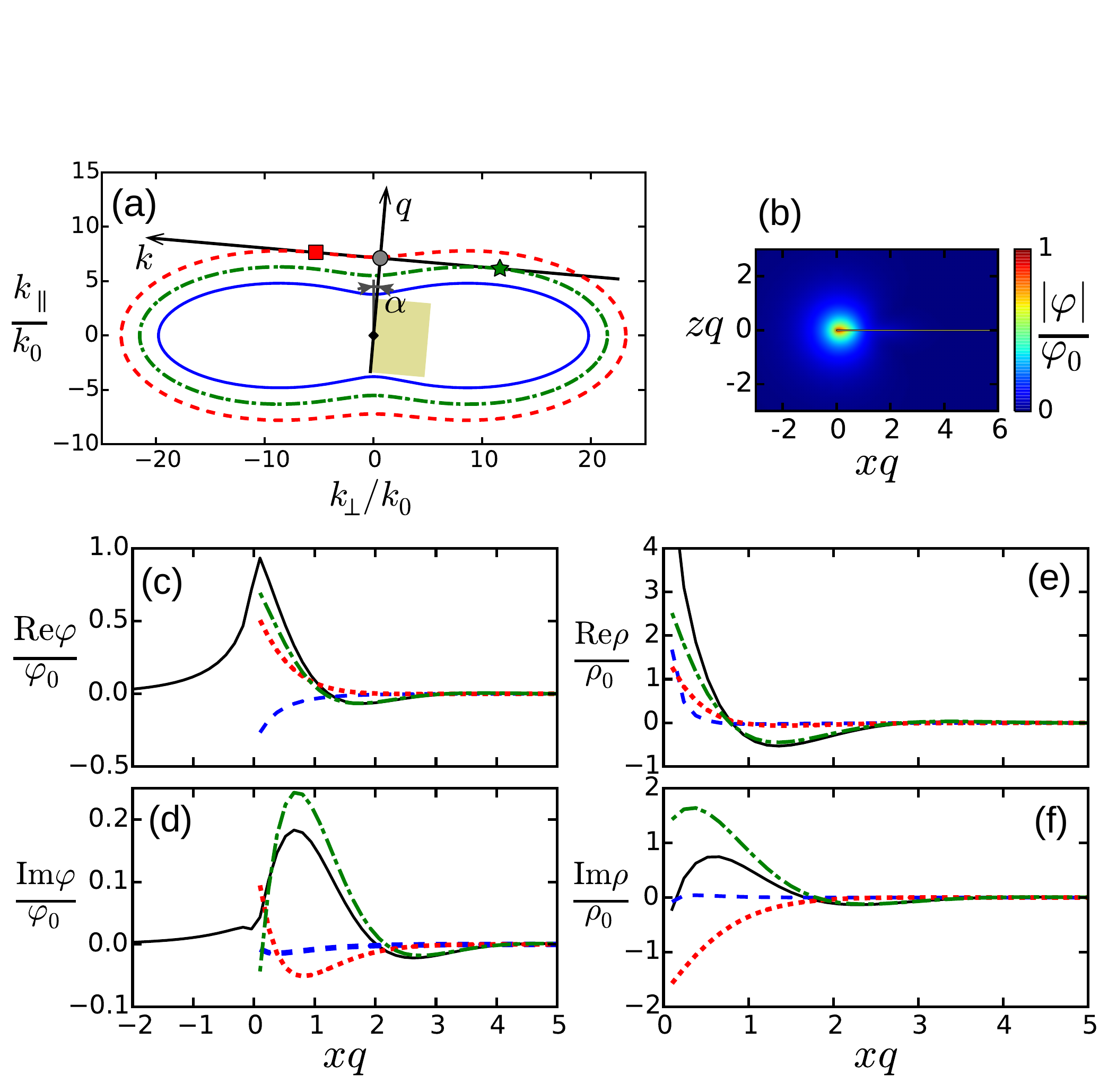}
\end{center}
\caption{\label{Fig5}Analysis of edge mode field and density distributions at $\alpha=\pi/36$, $q=7.14k_0$, 
$\omega_\mathrm{e}=3.3\Omega_\perp$. (a) 
Wave vector mismatch between the edge mode having the frequency $\omega_\mathrm{e}$ and wave vector $q$ along the edge (circle), and elliptic 2D modes with the same frequency at different wave vector directions [solid line shows isofrequency contour $\omega^\mathrm{2D}_+(\mathbf{k})=\omega_\mathrm{e}$]. Square and star show the 
real parts of the complex wave vectors 
$\{k_x,k_y\}=\{-k_i,q\}$ of two evanescent 2D modes reaching the edge mode frequency owing to nonzero $\mathrm{Im}\,k_i$. Dashed and dash-dotted curves show isofrequency contours $\omega^\mathrm{2D}_+(\mathbf{k})=\mathrm{const}\neq\omega_\mathrm{e}$ of non-evanescent 2D modes passing through these wave vectors and having not wave vector, but frequency mismatches with the edge mode. (b) Spatial profile of the 
amplitude $|\varphi|$ of the edge mode field in the $(x,z)$ plane in the units of the peak value $\varphi_0\equiv\varphi(x=0,z=0)$. (c), (d) Real and 
imaginary parts of the potential $\varphi(x)$ (solid lines) and their 
decomposition into the rapidly decaying part (dashed lines) and two oscillating parts (dotted and dash-dotted lines); the latter ones correspond to evanescent 2D waves. (e), (f) The same as (c), (d) but for the charge density $\rho(x)$ in the units of $\rho_0=\varphi_0q\varepsilon_\mathrm{b}/4\pi$.}
\end{figure}

\subsection{Field and density distributions}\label{sec_field_dens}
The Wiener-Hopf method allows us to calculate spatial profiles of electric potential $\varphi(x,z)$ and charge density $\rho(x)$ explicitly in terms of inverse Fourier transform integrals over complex plane of wave vectors, as shown in Appendix~\ref{Appendix_C}. These integrals are naturally divided into two contributions [see (\ref{phi+}) and (\ref{rho})]. The first one comes from the integration along the cut at the imaginary axis and decays on moving away from the edge at $z=0$ faster than the exponent $e^{-q|x|}$. The second one, present only at $x\geqslant0$, comes from residues in the poles $k_i$ in the lower complex half-plane $k$ and has the form $\varphi(x,z)\propto 
e^{-ik_ix-\sqrt{k_i^2+q^2}|z|}$, $\rho(x)\propto e^{-ik_ix}$ of decaying oscillations for each $k_i$. 
Here $k_i$ are the roots of the equation
\begin{equation}
\varepsilon(k_i)|_{\omega=\omega_\mathrm{e}(q)}=0\label{evanescent}
\end{equation}
with $\mathrm{Im}\,k_i<0$, where $\varepsilon(k)$ is given by (\ref{epsilon}). It means that $-k_i$ are the complex wave vectors along the $x$ axis which are needed for a 2D mode to equate both its frequency $\omega=\omega_\mathrm{e}(q)$ and the $y$-component of wave vector $q$ to those of the edge mode. Nonzero $\mathrm{Im}\,k_i$ are required to achieve this equality because at any real $k_i$ the frequencies $\omega_\mathrm{e}$ and $\omega_+^\mathrm{2D}$ do not match at the same $q$, as seen in Fig.~\ref{Fig4}.

The 2D mode with the wave vector $\{k_x,k_y\}=\{-k_i,q\}$ is evanescent in the 
positive $x$ direction and can be considered as a lower-dimensional counterpart 
of conventional evanescent modes $\propto e^{-\varkappa|z|}$ confined to the 2D plane in 3D
space. These modes can be called doubly evanescent, because their field is 
confined both to the $z=0$ plane and to the $x=0$ edge of the surface. 
Eq.~(\ref{evanescent}) for these modes can have, depending on $q$ and $\alpha$, 2 or 4 solutions $k_i$ coming in complex conjugated pairs, so we expect the presence of 1 or 2 evanescent 2D modes with $\mathrm{Im}\,k_i<0$. Both edge mode frequency and spatial distribution of field and density change smoothly on transition between the regimes of 2 and 4 solutions of Eq.~(\ref{evanescent}).

The peculiar feature of the anisotropic 2D layer is that $k_i$ have 
generally nonzero real parts (except the cases of $\alpha=0,\pi/2$) which makes 
the field and charge density of the evanescent 2D mode $\propto e^{i(-k_ix+qy)}$ to be both decaying and oscillating along the surface, i.e. having the wave front $-\mathrm{Re}(k_i)x+qy=\mathrm{const}$ inclined with respect to the $x$ axis in the $(x,y)$ plane, see the example in Fig.~\ref{Fig1}. As noted also in Ref.~\cite{stauber2019}, the qualitatively similar feature arises in the Fetter approximation, where $\varphi,\rho\propto e^{-k_\mathrm{F}x+iqy}$ at $x>0$, and $-ik_\mathrm{F}$ is the complex wave vector with $\mathrm{Im}\,(-ik_\mathrm{F})<0$ (see Appendix~\ref{Appendix_B}), which, however, does not correspond exactly to evanescent 2D wave.

\begin{figure}
\begin{center}
\includegraphics[width=\columnwidth]{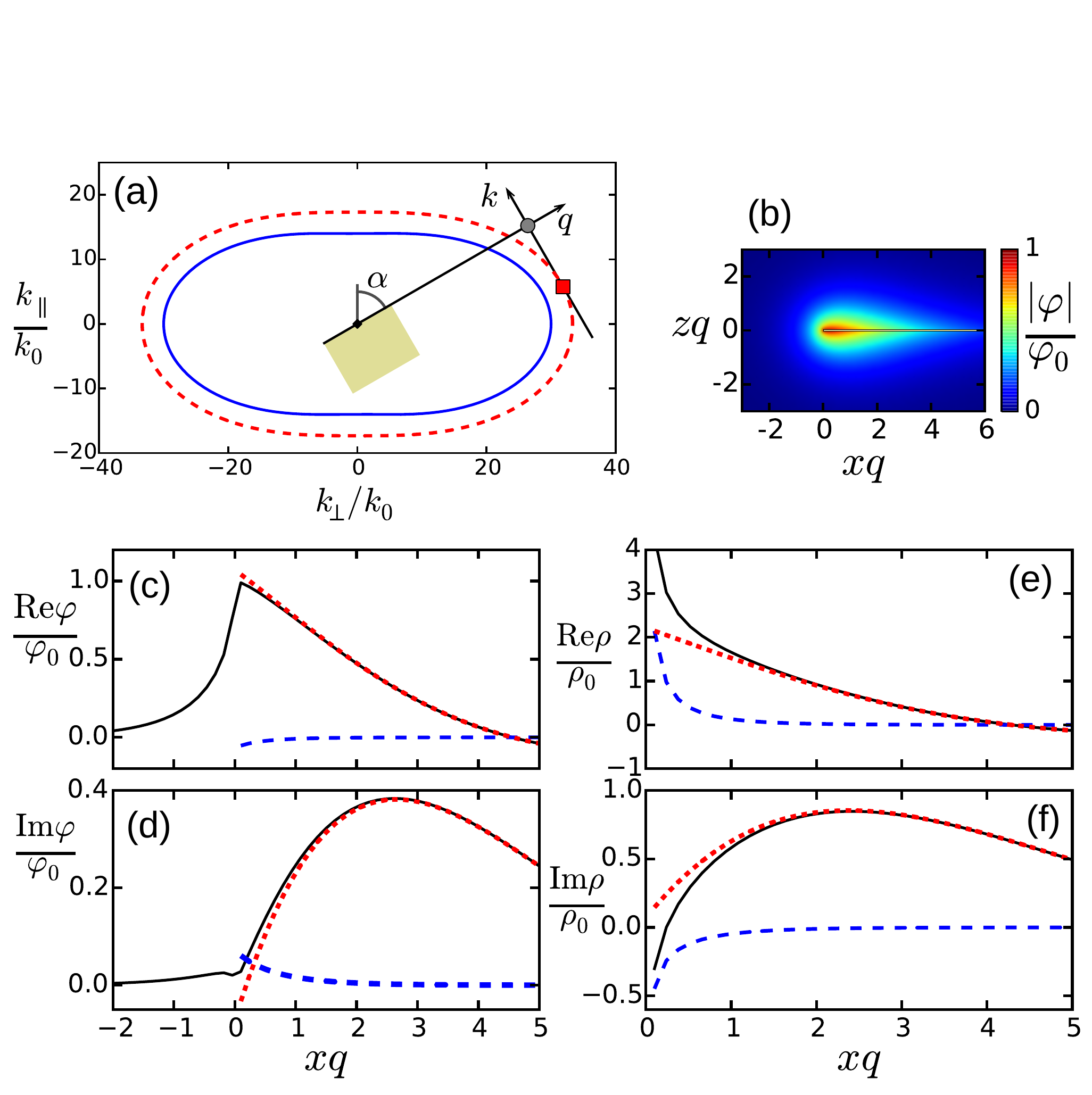}
\end{center}
\caption{\label{Fig6}The same as Fig.~\ref{Fig5} but at $\alpha=\pi/3$, 
$q=30.5k_0$, $\omega_\mathrm{e}=4\Omega_\perp$. At these parameters, only one evanescent 2D mode exists, while field confinement near the edge is moderate.}
\end{figure}

In Fig.~\ref{Fig5} we show the potential and density distributions at small nonzero $\alpha$ when Eq.~(\ref{evanescent}) has four solutions and two 
evanescent 2D waves are present. Fig.~\ref{Fig5}(a) demonstrates formation of 
the evanescent waves in the wave vector plane. According to the dispersions 
picture (see Fig.~\ref{Fig4}), all elliptic 2D modes have lower $y$-axis 
wave vectors $q$ than the edge mode with the same frequency, so in Fig.~\ref{Fig5}(a) the 2D mode isofrequency contour $\omega^\mathrm{2D}_+=\omega_\mathrm{e}$ cannot cross the line $k_y=q$ at any real $k=-k_x$. This $q$-wave vector mismatch along the $y$ axis can be surmounted when the $x$-axis wave vector $k_x=-k_i$ becomes complex. Its 
real part $\mathrm{Re}\,k_i$ can be considered as the point in the $\mathbf{k}$ 
plane, which is pictorially the ``closest'' to the isofrequency contour $\omega^\mathrm{2D}_+=\omega_\mathrm{e}$. The anisotropic and sometimes non-convex shape of this contour allows the existence of two ``closest'' points thus leading to the existence of two evanescent 2D modes at once [square and star in Fig.~\ref{Fig5}(a)].

On the other hand, we can consider not wave vector, but frequency mismatch between the edge mode and all elliptic 2D modes with the same $y$-axis wave vector $q$, demonstrated by Fig.~\ref{Fig4}. The nonzero imaginary part of an evanescent wave vector $k_i$ allows us to surmount this mismatch, which is shown in Fig.~\ref{Fig5}(a) as differences between solid, dotted, and dash-dotted isofrequency lines.

\begin{figure}
\begin{center}
\includegraphics[width=\columnwidth]{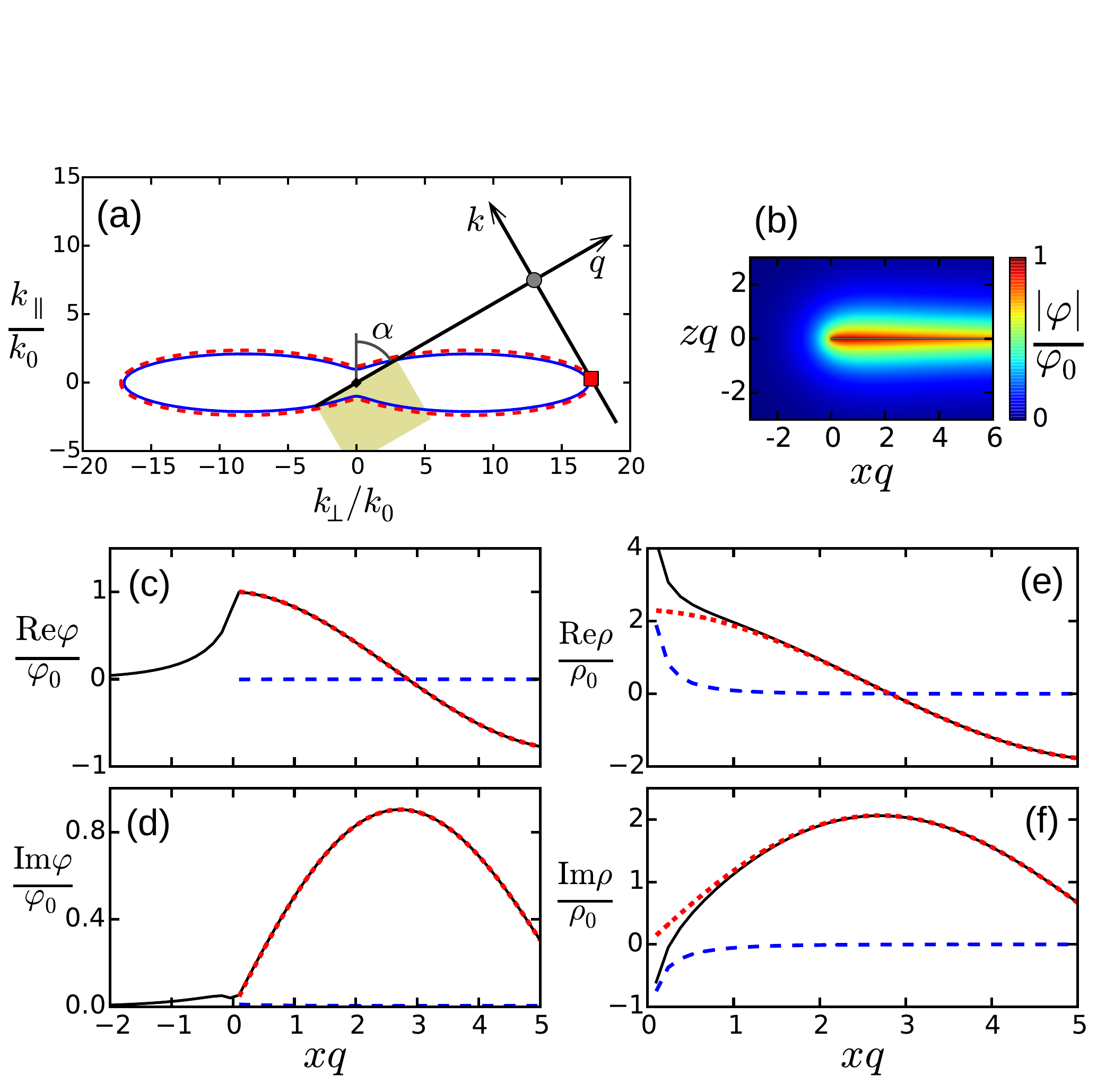}
\end{center}
\caption{\label{Fig7}The same as Fig.~\ref{Fig6} but at $\alpha=\pi/3$, 
$q=15k_0$, $\omega_\mathrm{e}=3.08\Omega_\perp$. The wave vector and frequency mismatches between the edge and 2D modes are very small, so the field confinement near the edge is very weak.}
\end{figure}

The spatial potential distribution in Fig.~\ref{Fig5}(b) shows that the edge 
mode field is well confined near the surface edge since the wave vector (or frequency) mismatch is relatively large. Decomposition of the field and charge density into the cut and pole (evanescent 2D wave) contributions in Fig.~\ref{Fig5}(c)--(f) shows that the cut contribution is indeed very rapidly decreasing.

It is remarkable that $\varphi(x)$ is linear at $x\rightarrow+0$ but behaves as $\sqrt{-x}$ at $x\rightarrow-0$, i.e. the electric field strength diverges as $E\propto (-x)^{-1/2}$ in the free-space direction near the sharp edge of the surface [see the asymptotic expressions (\ref{phi+a})--(\ref{phi-a})]. Charge density $\rho(x)$ behaves as $x^{-1/2}$ at $x\rightarrow+0$ [see Eq.~(\ref{rhoa})]. Similar features were noted in the case of edge magnetoplasmons \cite{Volkov}. Note that the Fetter approximation (\ref{Fetter_phi})--(\ref{Fetter_rho}) provides qualitatively different asymptotics of the field and density near the edge: the former turns out to be linear in $x$ from both sides of $x=0$, and the latter has $\delta$-functional and linear parts.

\begin{figure}
\begin{center}
\includegraphics[width=\columnwidth]{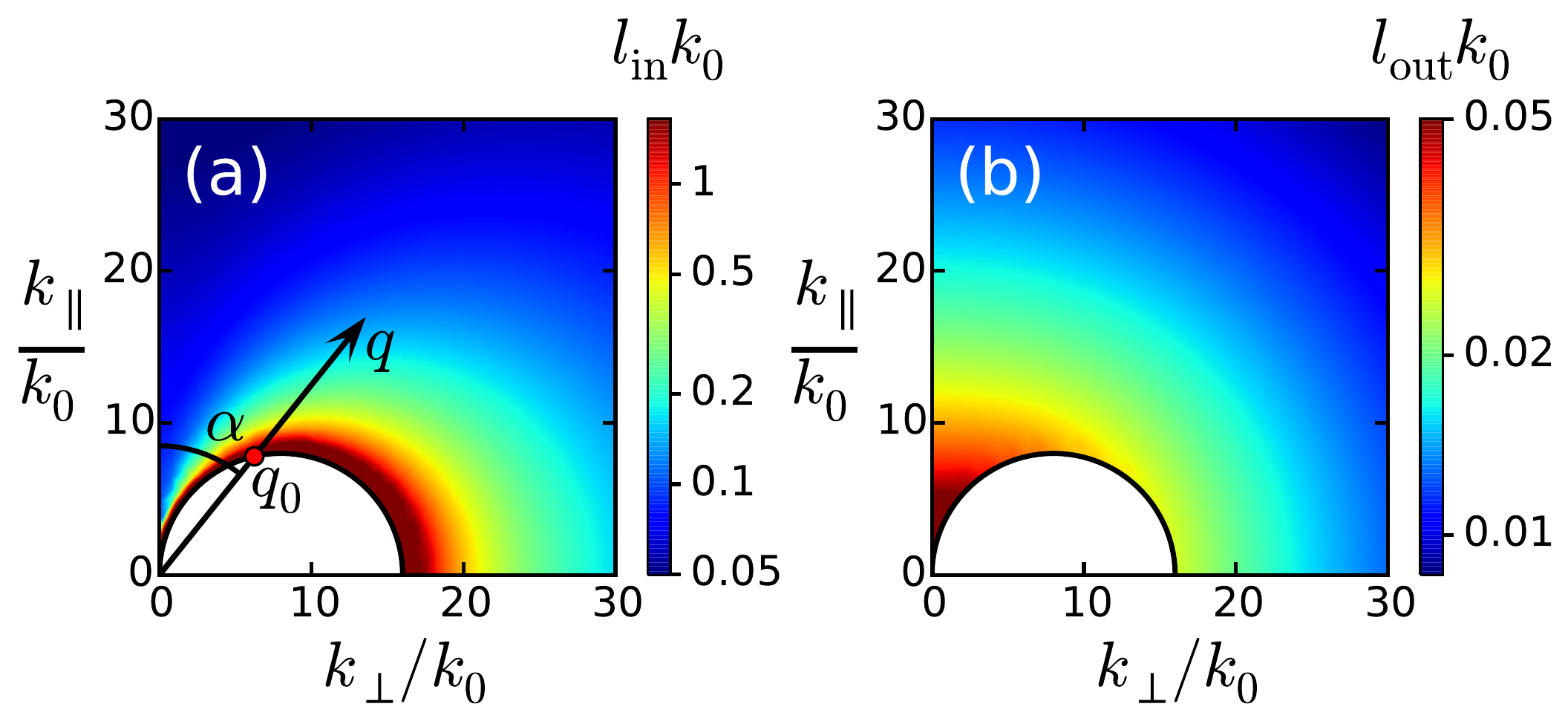}
\end{center}
\caption{\label{Fig8} Edge mode field localization lengths (a) $l_\mathrm{in}$ and (b) $l_\mathrm{out}$ defined as the distances from the edge measured, respectively, inside ($x>0$) and outside ($x<0$) the 2D layer, where $|\varphi|$ decays $e$ times: $|\varphi(l_\mathrm{in})|=|\varphi(-l_\mathrm{out})|=\varphi_0/e$. The $(k_\perp,k_\parallel)$ plane spans different edge mode wave vectors $q=[k_\perp^2+k_\parallel^2]^{1/2}$ exceeding the threshold $q_0$, and edge directions $\alpha=\arctan(k_\perp/k_\parallel)$.}
\end{figure}

Another case, where only one evanescent 2D mode exists, is shown in Fig.~\ref{Fig6}. The edge mode confinement in this case is moderate. Fig.~\ref{Fig7} shows what happens at the same $\alpha$ but at lower wave vector $q$, where (see the panel $\alpha=\pi/3$ of Fig.~\ref{Fig4}) the continuum of 2D modes comes much closer to the edge mode dispersion. Since the wave vector (or frequency) mismatch significantly reduces, as seen in Fig.~\ref{Fig7}(a), the edge mode confinement drops dramatically, and spatial decay of field and density along the surface becomes very slow, as seen in Fig.~\ref{Fig7}(b)--(f).

Although the field of the edge mode does not take a simple exponentially decaying form upon moving away from the edge, we can still characterize the degree of the field confinement near the edge by the localization lengths $l_\mathrm{in}$ (on top of the 2D layer) and $l_\mathrm{out}$ (in the outer space). They are defined, respectively, as the distances measured in the positive (negative) $x$ direction from the edge where amplitude of the potential drops $e$ times with respect to the peak value at the edge: $|\varphi(l_\mathrm{in})|=|\varphi(-l_\mathrm{out})|=\varphi_0/e$. Localization lengths in both directions are shown in Fig.~\ref{Fig8} for different edge mode wave vectors $q$ and edge directions $\alpha$. Field localization in empty space [Fig.~\ref{Fig8}(b)] is by about one order of magnitude stronger than along the 2D layer [Fig.~\ref{Fig8}(a)]. In the vicinity of the threshold wave vector $q_0$, the field delocalizes at $x>0$ because the edge mode becomes almost indistinguishable from the freely propagating elliptic 2D mode. 

\begin{figure}
\begin{center}
\includegraphics[width=\columnwidth]{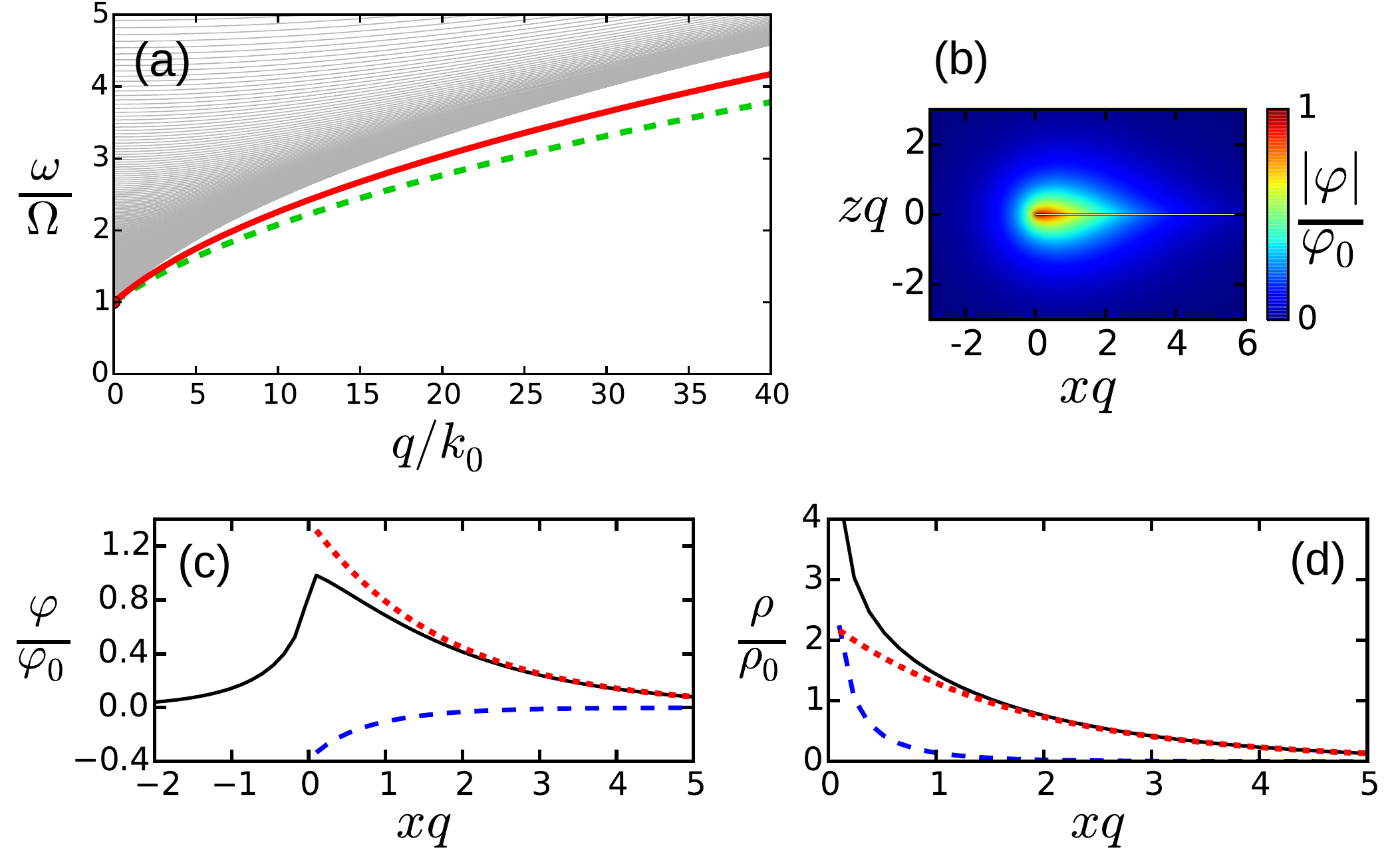}
\end{center}
\caption{\label{Fig9}Dispersions and field distribution in the isotropic case, 
when $\Omega_\parallel=\Omega_\perp\equiv\Omega$, at $q=30k_0$, 
$\omega_\mathrm{e}=3.65\Omega$, $k_0=\varepsilon_\mathrm{b}\Omega^2/cA$. (a) 
Dispersion of the edge mode (solid line) and its approximation in the Fetter 
approach (dashed line); thin lines show the continuum of 2D modes with the same wave vector $q$ along the edge and different perpendicular wave vectors $k$. (b) Spatial distribution of the amplitude $|\varphi|$ of the edge mode field in the $(x,z)$ plane. (c) The real potential $\varphi(x)$ (solid line) and its 
decomposition into the rapidly decaying (dashed line) and oscillating (dotted line) parts; the latter one corresponds to purely decaying evanescent 2D wave. (d) The same as (c) but for the charge density $\rho(x)$ in the units of 
$\rho_0=\varphi_0q\varepsilon_\mathrm{b}/4\pi$.}
\end{figure}

Note that near the threshold $l_\mathrm{in}$ and $l_\mathrm{out}$ depend on $\alpha$ differently: the former one is larger near $\alpha=\pi/2$, while the latter one is larger near $\alpha=0$. It can be attributed to the difference of the conductivities $\sigma_{\perp,\parallel}$: while $\sigma_\parallel\rightarrow\infty$ at $\omega\rightarrow\Omega_\parallel$ near the threshold, $\sigma_\perp$ remains finite. Consequently, at $\alpha\approx0$ the currents perpendicular to the edge, which are responsible for charge density oscillations and thus for edge mode formation, flow presumably along the $\perp$ axis, where the conductivity is smaller, so the charge density oscillations are better confined to the edge. At $\alpha\approx\pi/2$ these currents flow presumably along the $\parallel$ axis, where the conductivity is high, so the charge density is broadly distributed along the 2D layer.

Finally, we compare our results for anisotropic 2D layer to those for isotropic surface, considered in Sec.~\ref{sec_isotropic}. Dispersion curves, shown in Fig.~\ref{Fig9}(a), start at $q=0$ without the threshold. They are qualitatively similar to those in Fig.~\ref{Fig4} at $\alpha=0$. The field profile in Fig.~\ref{Fig9}(b) is qualitatively similar to that in the anisotropic case, while $\varphi(x,z)$ and $\rho(x)$ are everywhere in phase [Fig.~\ref{Fig9}(c)--(d)]. It means that the evanescent 2D wave field is purely decaying without oscillations in the $x$ direction and the edge mode wave fronts in the $(x,y)$ plane are always perpendicular to the edge in the isotropic case.

\section{Discussion and conclusion}\label{sec_discussion}
We analyzed the modes of plasmonic type (i.e. accompanied by charge density oscillations) propagating along the edge, which is arbitrary inclined with respect to the optical axes of a 2D layer with time-reversal symmetric anisotropic conductivity. We assumed the Lorentz-type conductivities with different resonance frequencies $\Omega_\parallel>\Omega_\perp$, so the hyperbolic behaviour occurs at $\Omega_\perp<\omega<\Omega_\parallel$. Our results can be easily generalized to the case of Drude conductivity along one of the axes $\Omega_\perp=0$ (Drude-Lorentz type of hyperbolic material in terms of Ref.~\cite{sun2014}).

The integro-differential equations describing coupled dynamics of electric field and charge density in the half-plane geometry were solved both exactly using the Wiener-Hopf method and approximately using the Fetter approach. We neglected EM field retardation, that is expected to be justified far enough from the light cone, when $q\gg\omega/c$. Edge mode dispersions as well as spatial profiles of electric potential and charge density were calculated numerically and analyzed at different angles $\alpha$ between the edge directions and the optical axis $\parallel$.

We show that the edge modes exist only in the inductive elliptic frequency region, where both conductivities $\sigma_{\perp,\parallel}$ have positive imaginary parts (in contrast to edge magnetoplasmons, which exist below the cyclotron frequency). Edge mode frequencies lie between the boundary $\omega=\Omega_\parallel$ of the elliptic region and the continuum of 2D waves, which freely propagate along the surface and have the same projection of wave vector on the edge. The edge mode dispersions monotonously increase and cross $\omega=\Omega_\parallel$ at the threshold wave vector $q_0\propto(\Omega_\parallel^2-\Omega_\perp^2)|\sin\alpha|$, which coincides with the point of hybridization of elliptic and hyperbolic 2D modes. Thus the anisotropy of 2D layer makes the edge mode existing only at wave vectors $q$ exceeding this threshold, which is absent in the case of isotropic surface.

Spatial profiles of electric field and charge density oscillations corresponding to the edge mode can be naturally divided into two parts (the similar field decomposition was noted in Ref.~\cite{svintsov2020}). The first part is always strongly confined to the edge, and the second part corresponds to evanescent 2D waves with complex wave vector projections perpendicular to the edge. Oscillations of these evanescent waves in a combination with their propagation along the edge lead to inclination of the edge mode wave fronts with respect to the edge. Near the threshold wave vector, these evanescent waves become weakly localized and almost indistinguishable from the freely propagating 2D waves, because the edge mode dispersion approaches the continuum. Depending on $q$ and $\alpha$, there exist one or two evanescent 2D waves (the latter case is specific for anisotropic surface), which can be explained by considering wave vector or frequency mismatch between edge and 2D modes.

Degree of confinement of edge mode field near the edge was analyzed. The field is shown to be always highly confined towards the empty space. Confinement along the 2D layer weakens at small wave vectors when the edge mode dispersion comes close to the continuum of elliptic 2D modes, because in this case the evanescent 2D modes contributing to the edge mode field become weakly decaying on departing from the edge. Remarkably, due to the anisotropy of conductivities, at $\alpha\approx0$ the field confinement can remain strong even at small wave vectors.

The approximated Fetter approach provides a qualitatively correct picture for the edge mode dispersion with slightly red-shifted frequencies. The frequency region of existence of edge modes, $\mathrm{Im}\,\sigma_\perp>0$, $\mathrm{Im}\,\sigma_\parallel>0$, is also correctly predicted in this approach. However the angular dependence of the threshold wave vector $q_0^\mathrm{F}\propto(\Omega_\parallel^2-\Omega_\perp^2)(3-\cos2\alpha)$ is different from the exact one. Charge density distribution, obtained in this approach, has a $\delta$-functional singularity on the edge, which is replaced by a softer $x^{-1/2}$ singularity in exact solution. Besides, the Fetter approach reproduces a qualitative picture of evanescent 2D waves and accompanying inclined wave fronts of edge modes, as discussed in detail in Ref.~\cite{stauber2019}, although with quantitatively incorrect wave vectors.

Our results for the Lorentz-Lorentz type of anisotropic conductivity tensor can be applied to a wide class of hyperbolic layers (e.g., metal nanodisk metasurfaces or natural polar materials such as $\rm MoO_3$) and to various types of polaritons (plasmon-, phonon-, exciton-polaritons, etc.) at large enough wave vectors. Moreover, the predicted existence of the threshold wave vector should take place in a more general case where conductivity in one direction has the resonant Lorentz-type form and conductivity in the perpendicular direction is arbitrary and smoothly varying near the resonance. This scenario is realized, e.g., in metal strip metasurfaces or in such natural 2D materials as phosphorene. In contrast, where both conductivities vary smoothly in the elliptic region of spectrum (e.g., hyperbolic layer of Drude-Drude type \cite{sun2014}), the edge mode dispersion should be dependent on edge direction $\alpha$ and always starting at $q=0$ without the threshold. It is also of interest to extend our analysis on the case of 2D layers with arbitrary conductivity tensor, not necessarily respecting time reversal symmetry, or to the case of 1D interface of two surfaces with different conductivities. General mathematical approach to such analysis was presented in the recent paper \cite{margetis2020}. Notice that the considered modes are of plasmonic type and have TM polarization in the 2D layer plane. The extension of our approach accounting for EM field retardation would allow us to study TE or hybrid TM-TE edge modes. 

The superior field confinement in all directions perpendicular to the edge makes the considered modes quite promising for various applications. Large field gradients near the edges can be applied for optical manipulation and trapping of nanoparticles. In natural 2D materials edge modes can be also of interest due to strong dependence of edge properties on its atomic-scale details. Namely zigzag-terminated edges of $\rm MoS_2$ monolayers have metallic and ferromagnetic nature, while armchair edges display semiconducting and nonmagnetic behavior \cite{bollinger2001,rossi2017}. Metallic edges are chemically active, so they exhibit high electro- and photo-catalytic activity \cite{zhou2013}, as well as high gas sensitivity \cite{donarelli2018}. Besides, due to symmetry breaking at the edges, they are also promising for nonlinear optical applications, such as enhanced second harmonic generation \cite{yin2014}.

Special attention should be paid to robust directional propagation of the edge modes in a wide variety of both magnetic and nonmagnetic materials. The former are materials where time reversal symmetry can be broken by external magnetic field, giving regular Hall effect, or by a non-trivial Berry curvature of the electronic band structure, resulting in anomalous Hall effect. Quantum Hall systems host gapless unidirectional (non-reciprocal) edge magnetoplasmons \cite{Fetter,Volkov,wassermeier1990,ashoori1992,talyanskii1992,kukushkin2008,wang2012,yan2012,lin2013,kumada2014,jin2016,cohen2018,sokolik2019}. Anomalous Hall systems with the spin-induced \cite{zhang2018,mahoney2017} or valley-induced \cite{kumar2016,song2016} Berry curvature support similar chiral edge plasmons. In ferromagnetic materials the Berry curvature is generated by strong spin-orbit interaction and has opposite signs for carriers with spins parallel or antiparallel to the magnetization. By analogy in gapped graphene or transition metal dichalcogenides, the Berry curvature induced by circularly polarized light has opposite signs in different electron valleys.  

Nonmagnetic materials with preserved time reversal symmetry can also support unidirectional edge EM modes owing to the photonic spin-orbit coupling (photonic counterpart of quantum spin Hall effect), where the role of (pseudo)spin is played by the orbital angular momentum of light in systems with nontrivial topology of photonic band structure \cite{hafezi2013,yang2018,parappurath2020}. Edge EM modes in such systems have the advantage of being topologically protected against defects that do not couple modes with opposite pseudo-spins \cite{lu2014,ota2020}. Additionally, the effect of spin-momentum coupling can be used for unidirectional excitation of edge modes in an arbitrary 2D system by means of a circularly polarized dipole \cite{stauber2019}. The ordinary reciprocal edge modes, such as those studied in this article, can be excited and detected with different methods \cite{Talebi_book} used for any evanescent plasmonic waves, e.g., scanning near-field optical microscopy \cite{yao2020} or electron energy-loss spectroscopy \cite{aken2016,lu2018}. Numerical simulations of edge mode excitation by dipole emitters located near the edge would allow us to estimate and optimize its efficiency.

\section*{Acknowledgments}
The work was supported by the Russian Science Foundation (Grant 17-12-01393). 

\appendix
\section{Wiener-Hopf solution for edge modes}\label{Appendix_A}
The method \cite{Volkov,Carrier} starts from dividing the potential $\varphi(x,z)=\varphi_+(x,z)+\varphi_-(x,z)$ into the components $\varphi_\pm$ which are nonzero at, respectively, $x\geqslant0$ and $x\leqslant0$. After the Fourier transform
\begin{equation}
\Phi_\pm(\xi)=\int dx\:e^{-i\xi qx}\varphi_\pm(x)\label{Fourier1}
\end{equation}
we obtain the functions $\Phi_\pm(\xi)$, which are analytical, respectively, in the upper and lower half-planes of the complex $\xi$. The variable $\xi$ has the meaning of $x$-projection of the field wave vector in the units of $q$ with the minus sign. Similar transform $Q_+(\xi)=\int_0^\infty dx\:e^{-i\xi qx}\rho(x)$ for the charge density allows us to rewrite Eq.~(\ref{cont}) after integration by parts as
\begin{align}
i\omega Q_+(\xi)=&q^2\{\xi^2\sigma_{xx}-\xi(\sigma_{xy}+\sigma_{yx})+\sigma_{yy}\}
\Phi_+(\xi)\nonumber
\\[0.2em]
&-iq(\xi\sigma_{xx}-\sigma_{yx})\varphi_0,\label{WH1}
\end{align}
where $\varphi_0\equiv\varphi(x=0)$.

The Poisson equation (\ref{Poisson}) with taking into account (\ref{kernel}) after the Fourier transform becomes
\begin{equation}
\Phi_+(\xi)+\Phi_-(\xi)=\frac{2\pi}{\varepsilon_\mathrm{b}q}
\frac{Q_+(\xi)}{\sqrt{\xi^2+1}}.\label{WH2}
\end{equation}
Substituting (\ref{WH1}) into (\ref{WH2}), we obtain
\begin{equation}
\varepsilon(\xi)\Phi_+(\xi)+\Phi_-(\xi)=
-\frac{i(\xi\eta_{xx}-\eta_{yx})}{2q\sqrt{\xi^2+1}}\varphi_0,\label{WH3}
\end{equation}
where the 2D dielectric function
\begin{equation}
\varepsilon(\xi)=1-\frac{\xi^2\eta_{xx}-\xi(\eta_{xy}+\eta_{yx})+\eta_{yy}}
{2\sqrt{\xi^2+1}}\label{epsilon1}
\end{equation}
and dimensionless conductivities
\begin{equation}
\eta_{\alpha\beta}=\frac{4\pi q\sigma_{\alpha\beta}}{i\varepsilon_\mathrm{b}\omega}
\label{etas}
\end{equation}
are introduced.

At this point we assume that the conductivity tensor in (\ref{etas}) has anisotropic time-reversal symmetric form (\ref{sigma_tensor}). The general case was considered in the recent paper Ref.~\cite{margetis2020}. Let us define the roots $\xi_{1,2}$ of ${\xi^2\eta_{xx}-\xi(\eta_{xy}+\eta_{yx})+\eta_{yy}=0}$, equal to
\begin{equation}
\xi_{1,2}=\frac{(\eta_\perp-\eta_\parallel)\sin\alpha\cos\alpha
\pm i\sqrt{\eta_\parallel\eta_\perp}}{\eta},\label{xi12}
\end{equation}
where 
\begin{equation}
\eta\equiv\eta_{xx}=\eta_\perp\cos^2\alpha+\eta_\parallel\sin^2\alpha,
\end{equation}
and
\begin{equation}
\eta_{\perp,\parallel}=\frac{4\pi q\sigma_{\perp,\parallel}}{i\varepsilon_\mathrm{b}\omega}=
\frac{cq}{\varepsilon_\mathrm{b}}
\frac{A_{\perp,\parallel}}{\omega^2-\Omega_{\perp,\parallel}^2},\label{eta_perp_para}
\end{equation}
[we take the conductivities (\ref{sigma_perp_para}) at $\gamma=0$, as explained in the beginning of Sec.~\ref{sec_edge_modes}]. Thus (\ref{epsilon1}) takes the form
\begin{equation}
\varepsilon(\xi)=1-\frac{\eta(\xi-\xi_1)(\xi-\xi_2)}
{2\sqrt{\xi^2+1}}.\label{epsilon2}
\end{equation}

\begin{figure}[!t]
\begin{center}
\includegraphics[width=0.7\columnwidth]{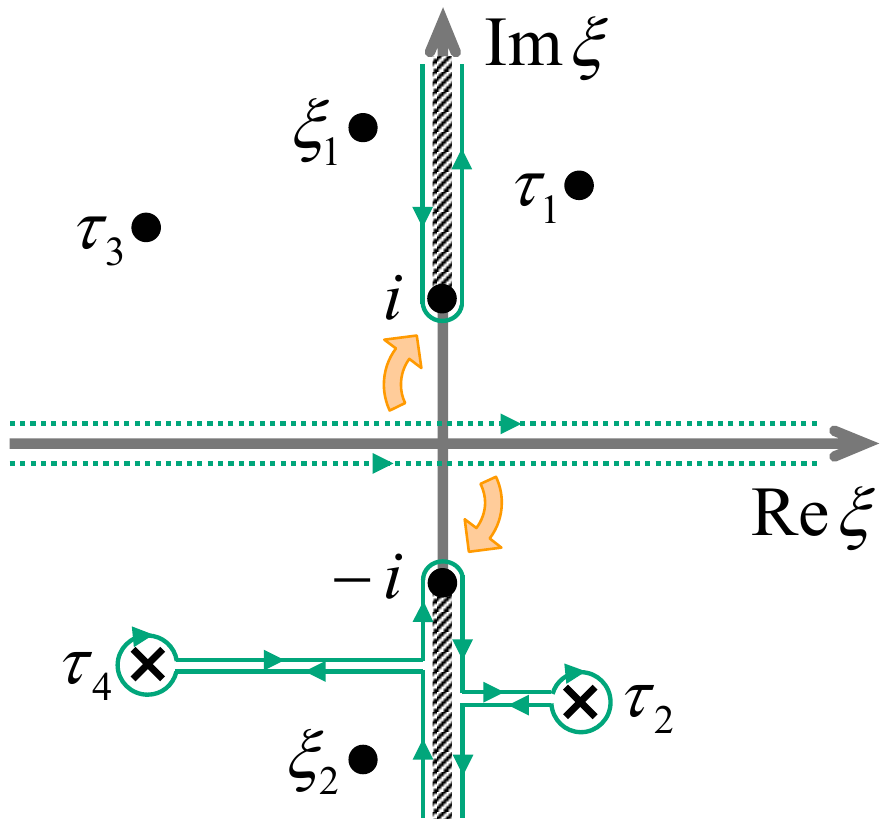}
\end{center}
\caption{\label{Fig10}Branch cuts (thick hatched lines), poles of $\Phi_+(\xi)$ (crosses), and other relevant points on a complex plane of $\xi$ in the four-root case. Initial (dotted lines) and deformed (solid lines) integration contours used to calculate $\varphi_\pm(x)$ are shown in, respectively, lower and upper half-planes. In the two-root case the points $\tau_{3,4}$ are absent.}
\end{figure}

The key step in the Wiener-Hopf method consists of kernel decomposition of the kind
\cite{Carrier}
\begin{equation}
\varepsilon(\xi)=F_+(\xi)/F_-(\xi),\label{WH_decomp1}
\end{equation}
where the functions $F_\pm(\xi)$ are analytical, respectively, in the upper and lower half-planes. This process is facilitated by finding the complex roots $\tau_i$ of $\varepsilon(\xi)=0$. Depending on the parameters of the problem, we can have either two ($\tau_{1,2}$) or four ($\tau_{1\ldots4}$) roots, coming in complex conjugated pairs $\{\tau_1,\tau_2\}$ and $\{\tau_3,\tau_4\}$ (if present). We denote by $\tau_{1,3}$ ($\tau_{2,4}$) the roots with positive (negative) imaginary parts, see Fig.~\ref{Fig10}. It is convenient to define
\begin{align}
F_+(\xi)&=\sqrt{\frac\eta2}\frac{\xi-\tau_2}{\sqrt{1-i\xi}}G_+(\xi),
\label{Fp1}\\[0.2em]
F_-(\xi)&=-\sqrt{\frac2\eta}\frac{\sqrt{1+i\xi}}{\xi-\tau_1}G_-(\xi)\label{Fm1}
\end{align}
in the two-root case and
\begin{align}
F_+(\xi)&=\sqrt{\frac\eta2}\frac{(\xi-\tau_2)(\xi-\tau_4)}
{\sqrt{1-i\xi}(\xi+i|\xi_1|)}G_+(\xi),\label{Fp2}\\[0.2em]
F_-(\xi)&=-\sqrt{\frac2\eta}\frac{\sqrt{1+i\xi}(\xi-i|\xi_1|)}
{(\xi-\tau_1)(\xi-\tau_3)}G_-(\xi)\label{Fm2}
\end{align}
in the four-root case. The prefactors here are responsible for zeros and poles of $F_\pm(\xi)$, while the remaining functions $G_\pm(\xi)$ have only branch cuts along the imaginary axis, respectively, $(-i,-i\infty)$ and $(i,i\infty)$, and quickly tend to 1 at $|\xi|\rightarrow\infty$. This allows us to deform the integration contours for these functions, which are initially defined at real arguments as \cite{Carrier}
\begin{equation}
G_\pm(\xi)=\exp\left\{\frac1{2\pi i}\int_{-\infty}^{+\infty}
\frac{d\xi'}{\xi'-\xi\mp i\delta}\ln G(\xi')\right\},\label{WH4}
\end{equation}
from the real axis to these cuts (similarly to what is shown in Fig.~\ref{Fig10} but without going around the poles $\tau_{2,4}$). As a result, we obtain the analytical continuation of (\ref{WH4}) to complex arguments $\xi$, valid in both two- and four-root cases:
\begin{equation}
G_\pm(\xi)=\exp\left\{\frac1\pi\int_1^\infty\!\!\!\!\frac{du}{i\xi\mp u}\arctan\frac{2\sqrt{u^2-1}}{\eta(u\mp i\xi_1)(u\mp i\xi_2)}\right\}.\label{Gpm}
\end{equation}
The functions (\ref{Gpm}) can be quickly calculated numerically. Substituting (\ref{WH_decomp1}) into (\ref{WH3}) and using (\ref{epsilon2}), we obtain
\begin{align}
&F_+(\xi)\Phi_+(\xi)+F_-(\xi)\Phi_-(\xi)\nonumber
\\[0.2em]
&=\frac{i\varphi_0}q\left(\xi-\frac{\eta_{yx}}\eta\right)
\frac{F_+(\xi)-F_-(\xi)}{(\xi-\xi_1)(\xi-\xi_2)}.\label{WH5}
\end{align}

The next step is to decompose the right-hand side of (\ref{WH5}) into a sum $P_+(\xi)+P_-(\xi)$ of functions $P_\pm$, which are analytical in, respectively, upper and lower half-planes. For real $\xi$ this decomposition can be performed analogously to (\ref{WH4}). The integrals over $\xi'$ can be reduced to residues in the poles $\xi\pm i\delta$ and $\xi_{1,2}$, if we take into account that $\xi_1$ ($\xi_2$) has a positive (negative) imaginary part and close the integration contour by infinitely large half-circle in the upper (lower) half-plane for the terms of the integrand containing $F_+$ ($F_-$):
\begin{equation}
P_\pm(\xi)=\pm\frac{i\varphi_0}{2q}
\left\{\frac{F_\pm(\xi)-F_+(\xi_1)}{\xi-\xi_1}+\frac{F_\pm(\xi)-F_-(\xi_2)}{\xi-\xi_2}\right\}.\label{Ppm}
\end{equation}
According to these formulas, both $P_+(\xi)$ and $P_-(\xi)$ tend to zero at $|\xi|\rightarrow0$, so the Liouville theorem \cite{Carrier} states that the ``$+$'' and ``$-$'' parts of Eq.~(\ref{WH5}), which can be rewritten as $F_+(\xi)\Phi_+(\xi)+F_-(\xi)\Phi_-(\xi)=P_+(\xi)+P_-(\xi)$, should both be equal to zero at any $\xi$. This allows us to find $\Phi_\pm(\xi)=P_\pm(\xi)/F_\pm(\xi)$ as
\begin{align}
\Phi_\pm(\xi)=\pm\frac{i\varphi_0}{2q}&\left\{\frac1{\xi-\xi_1}
\left[1-\frac{F_+(\xi_1)}{F_\pm(\xi)}\right]\right.\nonumber
\\[0.2em]
&\left.+\frac1{\xi-\xi_2}\left[1-\frac{F_-(\xi_2)}{F_\pm(\xi)}\right]\right\}.
\label{Phipm}
\end{align}

The Fourier transform inverse to Eq.~(\ref{Fourier1}) allows us to obtain the asymptotics of $\varphi(x)$ at $x\rightarrow\pm0$. From the asymptotics of (\ref{Phipm}) at $|\xi|\rightarrow\infty$ with taking into account (\ref{Fp1})--(\ref{Fm2}), and (\ref{Gpm}),
\begin{align}
\Phi_+(\xi)&=\frac{i\varphi_0}{q\xi}
+\frac{i\varphi_0C}{\sqrt{2\eta}q(-i\xi)^{3/2}}
+\mathcal{O}(\xi^{-2}),\label{Phip_decomp}
\\[0.2em]
\Phi_-(\xi)&=-\frac{i\varphi_0}{q\xi}
-\frac{i\varphi_0\sqrt\eta C}{2\sqrt2q\sqrt{i\xi}}
+\mathcal{O}(\xi^{-3/2}),\label{Phim_decomp}
\end{align}
where $C=F_+(\xi_1)+F_-(\xi_2)$, we can obtain asymptotics of $\varphi_\pm$:
\begin{align}
\varphi_+(x)&=\varphi_0\left\{1+i\sqrt{\frac{2qx}{\pi\eta}}C
+\mathcal{O}(x)\right\},\label{phipa1}
\\[0.2em]
\varphi_-(x)&=\varphi_0\left\{1-\frac{i}2\sqrt{\frac\eta{2\pi qx}}C
+\mathcal{O}(\sqrt{x})\right\}.\label{phima1}
\end{align}
When $C=0$, both potentials (\ref{phipa1})--(\ref{phima1}) correctly tend to $\varphi_0$ at ${x\rightarrow0}$. This matching condition provides the edge mode dispersion equation:
\begin{equation}
F_+(\xi_1)+F_-(\xi_2)=0.\label{disp2}
\end{equation}
With (\ref{WH_decomp1})--(\ref{WH4}), it can be rewritten as
\begin{equation}
\cosh\left\{\frac{\xi_1-\xi_2}{4\pi i}\int\frac{d\xi}{(\xi-\xi_1)(\xi-\xi_2)}
\ln\varepsilon(\xi)\right\}=0.
\end{equation}
Taking into account that $\cosh z=0$ at $z=\pi i(n+1/2)$ and that $-\pi<\mathrm{Im}\,\ln\varepsilon(k)<\pi$, we get
\begin{equation}
\int\frac{d\xi}{(\xi-\xi_1)(\xi-\xi_2)}
\ln\left\{-\varepsilon(\xi)\right\}=0.\label{disp1}
\end{equation}
Using (\ref{epsilon1}) and (\ref{etas}), we obtain the final dispersion equations (\ref{disp})--(\ref{epsilon}).

A numerical solution for the edge mode dispersion $\omega_\mathrm{e}(q)$ of (\ref{disp1}) exists only in the inductive elliptic range, where $\eta_{\perp,\parallel}>0$ and $\xi_{1,2}$ in (\ref{xi12}) are complex conjugated. Given the monotonous increase of $\omega_\mathrm{e}(q)$, there exists the threshold wave vector $q_0$ where $\omega_\mathrm{e}\rightarrow\Omega_\parallel$. In this limit, according to (\ref{xi12})--(\ref{eta_perp_para}), $\eta_\parallel\rightarrow\infty$, $\eta\approx\eta_\parallel\sin^2\alpha$, $\xi_{1,2}\approx-\cot\alpha\pm i\sqrt{\eta_\perp/\eta_\parallel}/\sin^2\alpha$, and the integral (\ref{disp1}) is dominated by close vicinity of $\xi=\mathrm{Re}\,\xi_{1,2}$, so after the change of variable $\xi=\mathrm{Re}\,\xi_1+u\,\mathrm{Im}\,\xi_1$ it takes the asymptotic form
\begin{equation}
\int\frac{du}{u^2+1}\ln\left\{a(u^2+1)-1\right\},
\end{equation}
where $a=\eta_\perp(\Omega_\parallel)/2|\sin\alpha|=cA_\perp q/2\varepsilon_\mathrm{b}(\Omega_\parallel^2-\Omega_\perp^2)|\sin\alpha|$. This integral vanishes at $a=1$, and the threshold wave vector (\ref{q0gen})--(\ref{q0}) is obtained from this condition.
\vspace{-0.5em}
\section{Edge modes in the Fetter approximation}\label{Appendix_B}
\vspace{-0.5em}
The simplified method to solve equations (\ref{cont})--(\ref{Poisson}), which was proposed by Fetter in Ref. \cite{Fetter}, is frequently applied for the problems of edge modes \cite{wang2011,cohen2018,stauber2019,zabolotnykh2016}. This method consists of approximating the nonlocal integral equation (\ref{Poisson}) by the local differential one:
\begin{equation}
(\partial_x^2-2q^2)\varphi(x)=
-\frac{4\pi q}{\varepsilon_\mathrm{b}}\rho(x)\Theta(x).\label{Fetter1}
\end{equation}
The solution of Eqs. (\ref{cont}) and (\ref{Fetter1}) is:
\begin{align}
&\varphi(x)=\varphi_0\left\{
e^{-k_\mathrm{F}x}\Theta(x)+e^{\sqrt{2}qx}\Theta(-x)\right\},
\label{Fetter_phi}
\\[0.2em]
&\rho(x)=\frac{\varphi_0\varepsilon_\mathrm{b}(k_\mathrm{F}+\sqrt2q)}{4\pi q}
\left\{\delta(x)-(k_\mathrm{F}-\sqrt2q)e^{-k_\mathrm{F}x}\right\},
\label{Fetter_rho}
\end{align}
where
\begin{align}
k_\mathrm{F}=q\frac{\sqrt2+i\eta_{xy}}{\eta_{xx}-1}\label{k_F}
\end{align}
is the inverse edge mode localization length [we use notations (\ref{etas})]. These functions satisfy Eq.~(\ref{cont}) when the condition
\begin{align}
(\eta_{xx}-1)(\eta_{yy}-2)-(\eta_{xy}-i\sqrt2)(\eta_{yx}+i\sqrt2)=0\label{Fetter2}
\\[0.1em] \nonumber
\end{align}
is met. It gives the dispersion equation (\ref{Fetter_disp_equation}). In order for the formal solution (\ref{disp_Fetter}) of the dispersion equation to be physical, $k_\mathrm{F}$ should have a positive real part. Using (\ref{disp_Fetter}), it can be shown that $\mathrm{Re}\,k_\mathrm{F}>0$ only at sufficiently large wave vectors when $\omega>\Omega_\parallel$ and $q>q_0^\mathrm{F}$, where the threshold wave vector $q_0^\mathrm{F}$ is given by Eq.~(\ref{q0F}).

\section{Calculation of field and density distributions}\label{Appendix_C}
According to the Maxwell equations in the non-retarded limit $q\gg\omega/c$, each harmonic $\varphi(x)=e^{i(-kx+qy)}$ of the potential on the 2D layer plane corresponds to evanescent field $\varphi(x,z)=e^{i(-kx+qy)-\sqrt{k^2+q^2}|z|}$ in space. Therefore to obtain the spatial potential distributions $\varphi_\pm(x,z)$ at, respectively, $x\geqslant0$ and $x\leqslant0$ we need to carry out the inverse Fourier transform
\begin{equation}
\varphi_\pm(x,z)=\int\frac{q\,d\xi}{2\pi}e^{-i\xi qx-\sqrt{\xi^2+1}q|z|}\Phi_\pm(\xi)
\end{equation}
with $\Phi_\pm(\xi)$ given by (\ref{Phipm}). It is convenient to deform, as shown in Fig.~\ref{Fig10}, the integration contours for $\varphi_\pm$ from the real axis to the lower (upper) complex half-planes down to the cuts $(\mp i,\mp\infty)$ and use (\ref{WH_decomp1}), (\ref{disp2}). In the ``+'' case we need to take into account the presence of the poles $\tau_j$, where $j=2$ and $j=2,4$ in the two- and four-root cases (see Appendix~\ref{Appendix_A}) of the function $\Phi_+(\xi)$ in the lower half-plane originating from the zeros of (\ref{Fp1}) or (\ref{Fp2}). Residues in these poles contribute to $\varphi_+(x,z)$ as the waves $e^{-i\tau_jqx-\sqrt{\tau_j^2+1}q|z|}$, which exponentially decay at $x\rightarrow+\infty$. These terms correspond to evanescent 2D waves with the complex wave vectors  $k_i=q\tau_i$ considered in Sec.~\ref{sec_field_dens}. The remaining contribution to $\varphi_+(x,z)$ from the cut integration over $0\leqslant u<\infty$ decays faster than the exponent $e^{-qx}$:	
\onecolumngrid
\begin{align}
\varphi_+(x,z)=\varphi_0F_+(\xi_1)\,\mathrm{Im}\,\xi_1&\left\{\sum_j\frac{-ie^{-i\tau_jqx-\sqrt{\tau_j^2+1}q|z|}}
{\varepsilon'(\tau_j)F_-(\tau_j)(\tau_j-\xi_1)(\tau_j-\xi_2)}+\frac\eta{2\pi}\int\limits_1^\infty\frac{du\:e^{-uqx}}{F_-(-iu)\left\{u^2-1+\frac14\eta^2(iu+\xi_1)^2(iu+\xi_2)^2\right\}}\right.\nonumber
\\[0.2em]
&\left.\times\left[\sqrt{u^2-1}\cos(\sqrt{u^2-1}q|z|)-\frac\eta2(iu+\xi_1)(iu+\xi_2)\sin(\sqrt{u^2-1}q|z|)\right]\vphantom{\int\limits_1^\infty}\right\}.\label{phi+}
\end{align}
Here $j=2$ and $j=2,4$ for, respectively, the two- and four-root cases.
In calculating $\varphi_-(x,z)$ the poles do not appear, since (\ref{Fm1}) and 
(\ref{Fm2}) have poles at $\tau_1$ or $\tau_{1,3}$ in the upper 
half-plane, so $\Phi_-(\xi)$ in (\ref{Phipm}) has no poles at these points. Physically this corresponds to the absence of evanescent 2D waves in the empty space. The only remaining cut integration gives:
\begin{equation}
\varphi_-(x,z)=\varphi_0F_+(\xi_1)\,\mathrm{Im}\,\xi_1\frac\eta{2\pi}\int\limits_1^\infty\frac{du\:e^{uqx}\cos(\sqrt{u^2-1}q|z|)}{F_+(iu)\sqrt{u^2-1}}.
\end{equation}
The density distribution $\rho(x)$ can be obtained by the inverse Fourier 
transform $\rho(x)=\int(q\,d\xi/2\pi)e^{-iqx\xi}Q_+(\xi)$, where $Q_+(\xi)$ is 
found from (\ref{WH1}) or (\ref{WH2}) by taking into account the dispersion 
equation (\ref{disp2}):
\begin{equation}
Q_+(\xi)=\frac{\varphi_0\varepsilon_\mathrm{b}\eta\,
F_+(\xi_1)\,\mathrm{Im}\,\xi_1}{4\pi F_+(\xi)}.\label{dens_Q}
\end{equation}
By deforming the integration contour into the lower complex half-plane (Fig.~\ref{Fig10}), taking into account the residues at poles $\tau_j$, and using (\ref{WH_decomp1}), we obtain
\begin{equation}
\rho(x)=\frac{\varphi_0q\varepsilon_\mathrm{b}\eta 
F_+(\xi_1)\,\mathrm{Im}\,\xi_1}
{4\pi}\left\{\sum_j\frac{-ie^{-iqx\tau_j}}
{\varepsilon'(\tau_j)F_-(\tau_j)}+\frac\eta{2\pi}\int\limits_1^\infty
\frac{du\:e^{-uqx}\sqrt{u^2-1}(iu+\xi_1)(iu+\xi_2)}{F_-(-iu)
\left\{u^2-1+\frac14\eta^2(iu+\xi_1)^2(iu+\xi_2)^2\right\}}\right\},
\label{rho}
\end{equation}
where $j=2$ and $j=2,4$ for, respectively, the two- and four-root cases. 
Similarly to (\ref{phi+}), we obtain the density distribution as a sum of the 
decaying oscillating terms $e^{-iqx\tau_j}$ originating from the poles 
and the rapidly decaying term coming from the cut integration.

From decompositions of (\ref{Phipm}) at $|\xi|\rightarrow\infty$, which are more accurate versions of (\ref{Phip_decomp})--(\ref{Phim_decomp}), we can obtain the following asymptotics of $\varphi_\pm(x)$ at $x\rightarrow0$:
\begin{align}
\varphi_+(x)=\varphi_0&\left\{1-ix\,\mathrm{Re}\,\xi_1
\vphantom{\frac{8iF_+(\xi_1)\,\mathrm{Im}\,\xi_1}{3\sqrt{2\pi\eta}}}
+\frac{8iF_+(\xi_1)\,\mathrm{Im}\,\xi_1}{3\sqrt{2\pi\eta}}x^{3/2}+
\mathcal{O}(x^2)\right\},\label{phi+a}
\\[0.2em]
\varphi_-(x)=\varphi_0&\left\{1-i\sqrt{\frac{2\eta}\pi}
F_-(\xi_2)\,\mathrm{Im}\,\xi_1\sqrt{-x}+\mathcal{O}(x)\right\}.\label{phi-a}
\end{align}
Similar expansion for charge density (\ref{dens_Q})--(\ref{rho}) yields
\begin{equation}
\rho(x)=-i\sqrt{\frac{\eta q}x}\frac{\varphi_0\varepsilon_\mathrm{b}
F_+(\xi_1)\,\mathrm{Im}\,\xi_1}{(2\pi)^{3/2}}+\mathcal{O}(1).\label{rhoa}
\end{equation}
\twocolumngrid

\begin{thebibliography}{98}%
	\makeatletter
	\providecommand \@ifxundefined [1]{%
		\@ifx{#1\undefined}
	}%
	\providecommand \@ifnum [1]{%
		\ifnum #1\expandafter \@firstoftwo
		\else \expandafter \@secondoftwo
		\fi
	}%
	\providecommand \@ifx [1]{%
		\ifx #1\expandafter \@firstoftwo
		\else \expandafter \@secondoftwo
		\fi
	}%
	\providecommand \natexlab [1]{#1}%
	\providecommand \enquote  [1]{``#1''}%
	\providecommand \bibnamefont  [1]{#1}%
	\providecommand \bibfnamefont [1]{#1}%
	\providecommand \citenamefont [1]{#1}%
	\providecommand \href@noop [0]{\@secondoftwo}%
	\providecommand \href [0]{\begingroup \@sanitize@url \@href}%
	\providecommand \@href[1]{\@@startlink{#1}\@@href}%
	\providecommand \@@href[1]{\endgroup#1\@@endlink}%
	\providecommand \@sanitize@url [0]{\catcode `\\12\catcode `\$12\catcode
		`\&12\catcode `\#12\catcode `\^12\catcode `\_12\catcode `\%12\relax}%
	\providecommand \@@startlink[1]{}%
	\providecommand \@@endlink[0]{}%
	\providecommand \url  [0]{\begingroup\@sanitize@url \@url }%
	\providecommand \@url [1]{\endgroup\@href {#1}{\urlprefix }}%
	\providecommand \urlprefix  [0]{URL }%
	\providecommand \Eprint [0]{\href }%
	\providecommand \doibase [0]{http://dx.doi.org/}%
	\providecommand \selectlanguage [0]{\@gobble}%
	\providecommand \bibinfo  [0]{\@secondoftwo}%
	\providecommand \bibfield  [0]{\@secondoftwo}%
	\providecommand \translation [1]{[#1]}%
	\providecommand \BibitemOpen [0]{}%
	\providecommand \bibitemStop [0]{}%
	\providecommand \bibitemNoStop [0]{.\EOS\space}%
	\providecommand \EOS [0]{\spacefactor3000\relax}%
	\providecommand \BibitemShut  [1]{\csname bibitem#1\endcsname}%
	\let\auto@bib@innerbib\@empty
	\bibitem [{\citenamefont {Schuller}\ \emph {et~al.}(2010)\citenamefont
		{Schuller}, \citenamefont {Barnard}, \citenamefont {Cai}, \citenamefont
		{Jun}, \citenamefont {White},\ and\ \citenamefont
		{Brongersma}}]{schuller2010}%
	\BibitemOpen
	\bibfield  {author} {\bibinfo {author} {\bibfnamefont {J.~A.}\ \bibnamefont
			{Schuller}}, \bibinfo {author} {\bibfnamefont {E.~S.}\ \bibnamefont
			{Barnard}}, \bibinfo {author} {\bibfnamefont {W.}~\bibnamefont {Cai}},
		\bibinfo {author} {\bibfnamefont {Y.~C.}\ \bibnamefont {Jun}}, \bibinfo
		{author} {\bibfnamefont {J.~S.}\ \bibnamefont {White}}, \ and\ \bibinfo
		{author} {\bibfnamefont {M.~L.}\ \bibnamefont {Brongersma}},\ }\href
	{\doibase 10.1038/nmat2630} {\bibfield  {journal} {\bibinfo  {journal} {Nat.
				Mater.}\ }\textbf {\bibinfo {volume} {9}},\ \bibinfo {pages} {193} (\bibinfo
		{year} {2010})}\BibitemShut {NoStop}%
	\bibitem [{\citenamefont {Basov}\ \emph {et~al.}(2016)\citenamefont {Basov},
		\citenamefont {Fogler},\ and\ \citenamefont {de~Abajo}}]{basov2016}%
	\BibitemOpen
	\bibfield  {author} {\bibinfo {author} {\bibfnamefont {D.~N.}\ \bibnamefont
			{Basov}}, \bibinfo {author} {\bibfnamefont {M.~M.}\ \bibnamefont {Fogler}}, \
		and\ \bibinfo {author} {\bibfnamefont {F.~J.~G.}\ \bibnamefont {de~Abajo}},\
	}\href {\doibase 10.1126/science.aag1992} {\bibfield  {journal} {\bibinfo
			{journal} {Science}\ }\textbf {\bibinfo {volume} {354}},\ \bibinfo {pages}
		{aag1992} (\bibinfo {year} {2016})}\BibitemShut {NoStop}%
	\bibitem [{\citenamefont {Low}\ \emph {et~al.}(2016)\citenamefont {Low},
		\citenamefont {Chaves}, \citenamefont {Caldwell}, \citenamefont {Kumar},
		\citenamefont {Fang}, \citenamefont {Avouris}, \citenamefont {Heinz},
		\citenamefont {Guinea}, \citenamefont {Martin-Moreno},\ and\ \citenamefont
		{Koppens}}]{low2016}%
	\BibitemOpen
	\bibfield  {author} {\bibinfo {author} {\bibfnamefont {T.}~\bibnamefont
			{Low}}, \bibinfo {author} {\bibfnamefont {A.}~\bibnamefont {Chaves}},
		\bibinfo {author} {\bibfnamefont {J.~D.}\ \bibnamefont {Caldwell}}, \bibinfo
		{author} {\bibfnamefont {A.}~\bibnamefont {Kumar}}, \bibinfo {author}
		{\bibfnamefont {N.~X.}\ \bibnamefont {Fang}}, \bibinfo {author}
		{\bibfnamefont {P.}~\bibnamefont {Avouris}}, \bibinfo {author} {\bibfnamefont
			{T.~F.}\ \bibnamefont {Heinz}}, \bibinfo {author} {\bibfnamefont
			{F.}~\bibnamefont {Guinea}}, \bibinfo {author} {\bibfnamefont
			{L.}~\bibnamefont {Martin-Moreno}}, \ and\ \bibinfo {author} {\bibfnamefont
			{F.}~\bibnamefont {Koppens}},\ }\href {\doibase 10.1038/nmat4792} {\bibfield
		{journal} {\bibinfo  {journal} {Nat. Mater.}\ }\textbf {\bibinfo {volume}
			{16}},\ \bibinfo {pages} {182} (\bibinfo {year} {2016})}\BibitemShut
	{NoStop}%
	\bibitem [{\citenamefont {Maier}(2007)}]{Maier}%
	\BibitemOpen
	\bibfield  {author} {\bibinfo {author} {\bibfnamefont {S.~A.}\ \bibnamefont
			{Maier}},\ }\href@noop {} {\emph {\bibinfo {title} {Plasmonics: Fundamentals
				and Applications}}}\ (\bibinfo  {publisher} {Springer},\ \bibinfo {address}
	{New York},\ \bibinfo {year} {2007})\BibitemShut {NoStop}%
	\bibitem [{\citenamefont {Bozhevolnyi}(2008)}]{Bozhevolnyi}%
	\BibitemOpen
	\bibfield  {author} {\bibinfo {author} {\bibfnamefont {S.~I.}\ \bibnamefont
			{Bozhevolnyi}},\ }\href@noop {} {\emph {\bibinfo {title} {Plasmonic
				Nanoguides and Circuits}}}\ (\bibinfo  {publisher} {Pan Stanford},\ \bibinfo
	{address} {Singapore},\ \bibinfo {year} {2008})\BibitemShut {NoStop}%
	\bibitem [{\citenamefont {Ferrari}\ \emph {et~al.}(2015)\citenamefont
		{Ferrari}, \citenamefont {Bonaccorso}, \citenamefont {Fal'ko}, \citenamefont
		{Novoselov},\ and\ \citenamefont {et~al.}}]{Roadmap2D}%
	\BibitemOpen
	\bibfield  {author} {\bibinfo {author} {\bibfnamefont {A.~C.}\ \bibnamefont
			{Ferrari}}, \bibinfo {author} {\bibfnamefont {F.}~\bibnamefont {Bonaccorso}},
		\bibinfo {author} {\bibfnamefont {V.}~\bibnamefont {Fal'ko}}, \bibinfo
		{author} {\bibfnamefont {K.~S.}\ \bibnamefont {Novoselov}}, \ and\ \bibinfo
		{author} {\bibnamefont {et~al.}},\ }\href {\doibase 10.1039/c4nr01600a}
	{\bibfield  {journal} {\bibinfo  {journal} {Nanoscale}\ }\textbf {\bibinfo
			{volume} {7}},\ \bibinfo {pages} {4598} (\bibinfo {year} {2015})}\BibitemShut
	{NoStop}%
	\bibitem [{\citenamefont {Geim}\ and\ \citenamefont
		{Grigorieva}(2013)}]{Geim2013}%
	\BibitemOpen
	\bibfield  {author} {\bibinfo {author} {\bibfnamefont {A.~K.}\ \bibnamefont
			{Geim}}\ and\ \bibinfo {author} {\bibfnamefont {I.~V.}\ \bibnamefont
			{Grigorieva}},\ }\href {\doibase 10.1038/nature12385} {\bibfield  {journal}
		{\bibinfo  {journal} {Nature}\ }\textbf {\bibinfo {volume} {499}},\ \bibinfo
		{pages} {419} (\bibinfo {year} {2013})}\BibitemShut {NoStop}%
	\bibitem [{\citenamefont {Novoselov}\ \emph {et~al.}(2016)\citenamefont
		{Novoselov}, \citenamefont {Mishchenko}, \citenamefont {Carvalho},\ and\
		\citenamefont {Neto}}]{novoselov2016}%
	\BibitemOpen
	\bibfield  {author} {\bibinfo {author} {\bibfnamefont {K.~S.}\ \bibnamefont
			{Novoselov}}, \bibinfo {author} {\bibfnamefont {A.}~\bibnamefont
			{Mishchenko}}, \bibinfo {author} {\bibfnamefont {A.}~\bibnamefont
			{Carvalho}}, \ and\ \bibinfo {author} {\bibfnamefont {A.~H.~C.}\ \bibnamefont
			{Neto}},\ }\href {\doibase 10.1126/science.aac9439} {\bibfield  {journal}
		{\bibinfo  {journal} {Science}\ }\textbf {\bibinfo {volume} {353}},\ \bibinfo
		{pages} {aac9439} (\bibinfo {year} {2016})}\BibitemShut {NoStop}%
	\bibitem [{\citenamefont {Caldwell}\ \emph {et~al.}(2015)\citenamefont
		{Caldwell}, \citenamefont {Lindsay}, \citenamefont {Giannini}, \citenamefont
		{Vurgaftman}, \citenamefont {Reinecke}, \citenamefont {Maier},\ and\
		\citenamefont {Glembocki}}]{Rev2015_PhP}%
	\BibitemOpen
	\bibfield  {author} {\bibinfo {author} {\bibfnamefont {J.~D.}\ \bibnamefont
			{Caldwell}}, \bibinfo {author} {\bibfnamefont {L.}~\bibnamefont {Lindsay}},
		\bibinfo {author} {\bibfnamefont {V.}~\bibnamefont {Giannini}}, \bibinfo
		{author} {\bibfnamefont {I.}~\bibnamefont {Vurgaftman}}, \bibinfo {author}
		{\bibfnamefont {T.~L.}\ \bibnamefont {Reinecke}}, \bibinfo {author}
		{\bibfnamefont {S.~A.}\ \bibnamefont {Maier}}, \ and\ \bibinfo {author}
		{\bibfnamefont {O.~J.}\ \bibnamefont {Glembocki}},\ }\href {\doibase
		10.1515/nanoph-2014-0003} {\bibfield  {journal} {\bibinfo  {journal}
			{Nanophotonics}\ }\textbf {\bibinfo {volume} {4}},\ \bibinfo {pages} {44}
		(\bibinfo {year} {2015})}\BibitemShut {NoStop}%
	\bibitem [{\citenamefont {Foteinopoulou}\ \emph {et~al.}(2019)\citenamefont
		{Foteinopoulou}, \citenamefont {Devarapu}, \citenamefont {Subramania},
		\citenamefont {Krishna},\ and\ \citenamefont {Wasserman}}]{Rev2019_PhP}%
	\BibitemOpen
	\bibfield  {author} {\bibinfo {author} {\bibfnamefont {S.}~\bibnamefont
			{Foteinopoulou}}, \bibinfo {author} {\bibfnamefont {G.~C.~R.}\ \bibnamefont
			{Devarapu}}, \bibinfo {author} {\bibfnamefont {G.~S.}\ \bibnamefont
			{Subramania}}, \bibinfo {author} {\bibfnamefont {S.}~\bibnamefont {Krishna}},
		\ and\ \bibinfo {author} {\bibfnamefont {D.}~\bibnamefont {Wasserman}},\
	}\href {\doibase 10.1515/nanoph-2019-0232} {\bibfield  {journal} {\bibinfo
			{journal} {Nanophotonics}\ }\textbf {\bibinfo {volume} {8}},\ \bibinfo
		{pages} {2129} (\bibinfo {year} {2019})}\BibitemShut {NoStop}%
	\bibitem [{\citenamefont {Dai}\ \emph {et~al.}(2014)\citenamefont {Dai},
		\citenamefont {Fei}, \citenamefont {Ma}, \citenamefont {Rodin}, \citenamefont
		{Wagner}, \citenamefont {McLeod}, \citenamefont {Liu}, \citenamefont
		{Gannett}, \citenamefont {Regan}, \citenamefont {Watanabe}, \citenamefont
		{Taniguchi}, \citenamefont {Thiemens}, \citenamefont {Dominguez},
		\citenamefont {Neto}, \citenamefont {Zettl}, \citenamefont {Keilmann},
		\citenamefont {Jarillo-Herrero}, \citenamefont {Fogler},\ and\ \citenamefont
		{Basov}}]{dai2014}%
	\BibitemOpen
	\bibfield  {author} {\bibinfo {author} {\bibfnamefont {S.}~\bibnamefont
			{Dai}}, \bibinfo {author} {\bibfnamefont {Z.}~\bibnamefont {Fei}}, \bibinfo
		{author} {\bibfnamefont {Q.}~\bibnamefont {Ma}}, \bibinfo {author}
		{\bibfnamefont {A.~S.}\ \bibnamefont {Rodin}}, \bibinfo {author}
		{\bibfnamefont {M.}~\bibnamefont {Wagner}}, \bibinfo {author} {\bibfnamefont
			{A.~S.}\ \bibnamefont {McLeod}}, \bibinfo {author} {\bibfnamefont {M.~K.}\
			\bibnamefont {Liu}}, \bibinfo {author} {\bibfnamefont {W.}~\bibnamefont
			{Gannett}}, \bibinfo {author} {\bibfnamefont {W.}~\bibnamefont {Regan}},
		\bibinfo {author} {\bibfnamefont {K.}~\bibnamefont {Watanabe}}, \bibinfo
		{author} {\bibfnamefont {T.}~\bibnamefont {Taniguchi}}, \bibinfo {author}
		{\bibfnamefont {M.}~\bibnamefont {Thiemens}}, \bibinfo {author}
		{\bibfnamefont {G.}~\bibnamefont {Dominguez}}, \bibinfo {author}
		{\bibfnamefont {A.~H.~C.}\ \bibnamefont {Neto}}, \bibinfo {author}
		{\bibfnamefont {A.}~\bibnamefont {Zettl}}, \bibinfo {author} {\bibfnamefont
			{F.}~\bibnamefont {Keilmann}}, \bibinfo {author} {\bibfnamefont
			{P.}~\bibnamefont {Jarillo-Herrero}}, \bibinfo {author} {\bibfnamefont
			{M.~M.}\ \bibnamefont {Fogler}}, \ and\ \bibinfo {author} {\bibfnamefont
			{D.~N.}\ \bibnamefont {Basov}},\ }\href {\doibase 10.1126/science.1246833}
	{\bibfield  {journal} {\bibinfo  {journal} {Science}\ }\textbf {\bibinfo
			{volume} {343}},\ \bibinfo {pages} {1125} (\bibinfo {year}
		{2014})}\BibitemShut {NoStop}%
	\bibitem [{\citenamefont {Li}\ \emph {et~al.}(2015)\citenamefont {Li},
		\citenamefont {Lewin}, \citenamefont {Kretinin}, \citenamefont {Caldwell},
		\citenamefont {Novoselov}, \citenamefont {Taniguchi}, \citenamefont
		{Watanabe}, \citenamefont {Gaussmann},\ and\ \citenamefont
		{Taubner}}]{li2015}%
	\BibitemOpen
	\bibfield  {author} {\bibinfo {author} {\bibfnamefont {P.}~\bibnamefont
			{Li}}, \bibinfo {author} {\bibfnamefont {M.}~\bibnamefont {Lewin}}, \bibinfo
		{author} {\bibfnamefont {A.~V.}\ \bibnamefont {Kretinin}}, \bibinfo {author}
		{\bibfnamefont {J.~D.}\ \bibnamefont {Caldwell}}, \bibinfo {author}
		{\bibfnamefont {K.~S.}\ \bibnamefont {Novoselov}}, \bibinfo {author}
		{\bibfnamefont {T.}~\bibnamefont {Taniguchi}}, \bibinfo {author}
		{\bibfnamefont {K.}~\bibnamefont {Watanabe}}, \bibinfo {author}
		{\bibfnamefont {F.}~\bibnamefont {Gaussmann}}, \ and\ \bibinfo {author}
		{\bibfnamefont {T.}~\bibnamefont {Taubner}},\ }\href {\doibase
		10.1038/ncomms8507} {\bibfield  {journal} {\bibinfo  {journal} {Nat.
				Commun.}\ }\textbf {\bibinfo {volume} {6}},\ \bibinfo {pages} {7507}
		(\bibinfo {year} {2015})}\BibitemShut {NoStop}%
	\bibitem [{\citenamefont {Caldwell}\ \emph {et~al.}(2019)\citenamefont
		{Caldwell}, \citenamefont {Aharonovich}, \citenamefont {Cassabois},
		\citenamefont {Edgar}, \citenamefont {Gil},\ and\ \citenamefont
		{Basov}}]{Rev2019_hBN}%
	\BibitemOpen
	\bibfield  {author} {\bibinfo {author} {\bibfnamefont {J.~D.}\ \bibnamefont
			{Caldwell}}, \bibinfo {author} {\bibfnamefont {I.}~\bibnamefont
			{Aharonovich}}, \bibinfo {author} {\bibfnamefont {G.}~\bibnamefont
			{Cassabois}}, \bibinfo {author} {\bibfnamefont {J.~H.}\ \bibnamefont
			{Edgar}}, \bibinfo {author} {\bibfnamefont {B.}~\bibnamefont {Gil}}, \ and\
		\bibinfo {author} {\bibfnamefont {D.~N.}\ \bibnamefont {Basov}},\ }\href
	{\doibase 10.1038/s41578-019-0124-1} {\bibfield  {journal} {\bibinfo
			{journal} {Nat. Rev. Mater.}\ }\textbf {\bibinfo {volume} {4}},\ \bibinfo
		{pages} {552} (\bibinfo {year} {2019})}\BibitemShut {NoStop}%
	\bibitem [{\citenamefont {Li}\ \emph {et~al.}(2018)\citenamefont {Li},
		\citenamefont {Dolado}, \citenamefont {Alfaro-Mozaz}, \citenamefont
		{Casanova}, \citenamefont {Hueso}, \citenamefont {Liu}, \citenamefont
		{Edgar}, \citenamefont {Nikitin}, \citenamefont {V{\'{e}}lez},\ and\
		\citenamefont {Hillenbrand}}]{nikitin2018}%
	\BibitemOpen
	\bibfield  {author} {\bibinfo {author} {\bibfnamefont {P.}~\bibnamefont
			{Li}}, \bibinfo {author} {\bibfnamefont {I.}~\bibnamefont {Dolado}}, \bibinfo
		{author} {\bibfnamefont {F.~J.}\ \bibnamefont {Alfaro-Mozaz}}, \bibinfo
		{author} {\bibfnamefont {F.}~\bibnamefont {Casanova}}, \bibinfo {author}
		{\bibfnamefont {L.~E.}\ \bibnamefont {Hueso}}, \bibinfo {author}
		{\bibfnamefont {S.}~\bibnamefont {Liu}}, \bibinfo {author} {\bibfnamefont
			{J.~H.}\ \bibnamefont {Edgar}}, \bibinfo {author} {\bibfnamefont {A.~Y.}\
			\bibnamefont {Nikitin}}, \bibinfo {author} {\bibfnamefont {S.}~\bibnamefont
			{V{\'{e}}lez}}, \ and\ \bibinfo {author} {\bibfnamefont {R.}~\bibnamefont
			{Hillenbrand}},\ }\href {\doibase 10.1126/science.aaq1704} {\bibfield
		{journal} {\bibinfo  {journal} {Science}\ }\textbf {\bibinfo {volume}
			{359}},\ \bibinfo {pages} {892} (\bibinfo {year} {2018})}\BibitemShut
	{NoStop}%
	\bibitem [{\citenamefont {Ma}\ \emph {et~al.}(2018)\citenamefont {Ma},
		\citenamefont {Alonso-Gonz{\'{a}}lez}, \citenamefont {Li}, \citenamefont
		{Nikitin}, \citenamefont {Yuan}, \citenamefont {Mart{\'{\i}}n-S{\'{a}}nchez},
		\citenamefont {Taboada-Guti{\'{e}}rrez}, \citenamefont {Amenabar},
		\citenamefont {Li}, \citenamefont {V{\'{e}}lez}, \citenamefont {Tollan},
		\citenamefont {Dai}, \citenamefont {Zhang}, \citenamefont {Sriram},
		\citenamefont {Kalantar-Zadeh}, \citenamefont {Lee}, \citenamefont
		{Hillenbrand},\ and\ \citenamefont {Bao}}]{ma2018}%
	\BibitemOpen
	\bibfield  {author} {\bibinfo {author} {\bibfnamefont {W.}~\bibnamefont
			{Ma}}, \bibinfo {author} {\bibfnamefont {P.}~\bibnamefont
			{Alonso-Gonz{\'{a}}lez}}, \bibinfo {author} {\bibfnamefont {S.}~\bibnamefont
			{Li}}, \bibinfo {author} {\bibfnamefont {A.~Y.}\ \bibnamefont {Nikitin}},
		\bibinfo {author} {\bibfnamefont {J.}~\bibnamefont {Yuan}}, \bibinfo {author}
		{\bibfnamefont {J.}~\bibnamefont {Mart{\'{\i}}n-S{\'{a}}nchez}}, \bibinfo
		{author} {\bibfnamefont {J.}~\bibnamefont {Taboada-Guti{\'{e}}rrez}},
		\bibinfo {author} {\bibfnamefont {I.}~\bibnamefont {Amenabar}}, \bibinfo
		{author} {\bibfnamefont {P.}~\bibnamefont {Li}}, \bibinfo {author}
		{\bibfnamefont {S.}~\bibnamefont {V{\'{e}}lez}}, \bibinfo {author}
		{\bibfnamefont {C.}~\bibnamefont {Tollan}}, \bibinfo {author} {\bibfnamefont
			{Z.}~\bibnamefont {Dai}}, \bibinfo {author} {\bibfnamefont {Y.}~\bibnamefont
			{Zhang}}, \bibinfo {author} {\bibfnamefont {S.}~\bibnamefont {Sriram}},
		\bibinfo {author} {\bibfnamefont {K.}~\bibnamefont {Kalantar-Zadeh}},
		\bibinfo {author} {\bibfnamefont {S.-T.}\ \bibnamefont {Lee}}, \bibinfo
		{author} {\bibfnamefont {R.}~\bibnamefont {Hillenbrand}}, \ and\ \bibinfo
		{author} {\bibfnamefont {Q.}~\bibnamefont {Bao}},\ }\href {\doibase
		10.1038/s41586-018-0618-9} {\bibfield  {journal} {\bibinfo  {journal}
			{Nature}\ }\textbf {\bibinfo {volume} {562}},\ \bibinfo {pages} {557}
		(\bibinfo {year} {2018})}\BibitemShut {NoStop}%
	\bibitem [{\citenamefont {Zheng}\ \emph {et~al.}(2019)\citenamefont {Zheng},
		\citenamefont {Xu}, \citenamefont {Oscurato}, \citenamefont {Tamagnone},
		\citenamefont {Sun}, \citenamefont {Jiang}, \citenamefont {Ke}, \citenamefont
		{Chen}, \citenamefont {Huang}, \citenamefont {Wilson}, \citenamefont
		{Ambrosio}, \citenamefont {Deng},\ and\ \citenamefont {Chen}}]{zheng2019}%
	\BibitemOpen
	\bibfield  {author} {\bibinfo {author} {\bibfnamefont {Z.}~\bibnamefont
			{Zheng}}, \bibinfo {author} {\bibfnamefont {N.}~\bibnamefont {Xu}}, \bibinfo
		{author} {\bibfnamefont {S.~L.}\ \bibnamefont {Oscurato}}, \bibinfo {author}
		{\bibfnamefont {M.}~\bibnamefont {Tamagnone}}, \bibinfo {author}
		{\bibfnamefont {F.}~\bibnamefont {Sun}}, \bibinfo {author} {\bibfnamefont
			{Y.}~\bibnamefont {Jiang}}, \bibinfo {author} {\bibfnamefont
			{Y.}~\bibnamefont {Ke}}, \bibinfo {author} {\bibfnamefont {J.}~\bibnamefont
			{Chen}}, \bibinfo {author} {\bibfnamefont {W.}~\bibnamefont {Huang}},
		\bibinfo {author} {\bibfnamefont {W.~L.}\ \bibnamefont {Wilson}}, \bibinfo
		{author} {\bibfnamefont {A.}~\bibnamefont {Ambrosio}}, \bibinfo {author}
		{\bibfnamefont {S.}~\bibnamefont {Deng}}, \ and\ \bibinfo {author}
		{\bibfnamefont {H.}~\bibnamefont {Chen}},\ }\href {\doibase
		10.1126/sciadv.aav8690} {\bibfield  {journal} {\bibinfo  {journal} {Sci.
				Adv.}\ }\textbf {\bibinfo {volume} {5}},\ \bibinfo {pages} {eaav8690}
		(\bibinfo {year} {2019})}\BibitemShut {NoStop}%
	\bibitem [{\citenamefont {Duan}\ \emph {et~al.}(2020)\citenamefont {Duan},
		\citenamefont {Capote-Robayna}, \citenamefont {Taboada-Guti{\'e}rrez},
		\citenamefont {{\'A}lvarez-P{\'e}rez}, \citenamefont {Prieto}, \citenamefont
		{Mart{\'i}n-S{\'a}nchez}, \citenamefont {Nikitin},\ and\ \citenamefont
		{Alonso-Gonz{\'a}lez}}]{nikitin2020}%
	\BibitemOpen
	\bibfield  {author} {\bibinfo {author} {\bibfnamefont {J.}~\bibnamefont
			{Duan}}, \bibinfo {author} {\bibfnamefont {N.}~\bibnamefont
			{Capote-Robayna}}, \bibinfo {author} {\bibfnamefont {J.}~\bibnamefont
			{Taboada-Guti{\'e}rrez}}, \bibinfo {author} {\bibfnamefont {G.}~\bibnamefont
			{{\'A}lvarez-P{\'e}rez}}, \bibinfo {author} {\bibfnamefont {I.}~\bibnamefont
			{Prieto}}, \bibinfo {author} {\bibfnamefont {J.}~\bibnamefont
			{Mart{\'i}n-S{\'a}nchez}}, \bibinfo {author} {\bibfnamefont {A.~Y.}\
			\bibnamefont {Nikitin}}, \ and\ \bibinfo {author} {\bibfnamefont
			{P.}~\bibnamefont {Alonso-Gonz{\'a}lez}},\ }\href {\doibase
		10.1021/acs.nanolett.0c01673} {\bibfield  {journal} {\bibinfo  {journal}
			{Nano Lett.}\ }\textbf {\bibinfo {volume} {20}},\ \bibinfo {pages} {5323}
		(\bibinfo {year} {2020})}\BibitemShut {NoStop}%
	\bibitem [{\citenamefont {Dubrovkin}\ \emph {et~al.}(2018)\citenamefont
		{Dubrovkin}, \citenamefont {Qiang}, \citenamefont {Krishnamoorthy},
		\citenamefont {Zheludev},\ and\ \citenamefont {Wang}}]{dubrovkin2018}%
	\BibitemOpen
	\bibfield  {author} {\bibinfo {author} {\bibfnamefont {A.~M.}\ \bibnamefont
			{Dubrovkin}}, \bibinfo {author} {\bibfnamefont {B.}~\bibnamefont {Qiang}},
		\bibinfo {author} {\bibfnamefont {H.~N.~S.}\ \bibnamefont {Krishnamoorthy}},
		\bibinfo {author} {\bibfnamefont {N.~I.}\ \bibnamefont {Zheludev}}, \ and\
		\bibinfo {author} {\bibfnamefont {Q.~J.}\ \bibnamefont {Wang}},\ }\href
	{\doibase 10.1038/s41467-018-04168-x} {\bibfield  {journal} {\bibinfo
			{journal} {Nat. Commun.}\ }\textbf {\bibinfo {volume} {9}},\ \bibinfo {pages}
		{1762} (\bibinfo {year} {2018})}\BibitemShut {NoStop}%
	\bibitem [{\citenamefont {Lee}\ \emph {et~al.}(2020)\citenamefont {Lee},
		\citenamefont {He}, \citenamefont {Zhang}, \citenamefont {Luo}, \citenamefont
		{Liu}, \citenamefont {Edgar}, \citenamefont {Wang}, \citenamefont {Avouris},
		\citenamefont {Low}, \citenamefont {Caldwell},\ and\ \citenamefont
		{Oh}}]{lee2020}%
	\BibitemOpen
	\bibfield  {author} {\bibinfo {author} {\bibfnamefont {I.-H.}\ \bibnamefont
			{Lee}}, \bibinfo {author} {\bibfnamefont {M.}~\bibnamefont {He}}, \bibinfo
		{author} {\bibfnamefont {X.}~\bibnamefont {Zhang}}, \bibinfo {author}
		{\bibfnamefont {Y.}~\bibnamefont {Luo}}, \bibinfo {author} {\bibfnamefont
			{S.}~\bibnamefont {Liu}}, \bibinfo {author} {\bibfnamefont {J.~H.}\
			\bibnamefont {Edgar}}, \bibinfo {author} {\bibfnamefont {K.}~\bibnamefont
			{Wang}}, \bibinfo {author} {\bibfnamefont {P.}~\bibnamefont {Avouris}},
		\bibinfo {author} {\bibfnamefont {T.}~\bibnamefont {Low}}, \bibinfo {author}
		{\bibfnamefont {J.~D.}\ \bibnamefont {Caldwell}}, \ and\ \bibinfo {author}
		{\bibfnamefont {S.-H.}\ \bibnamefont {Oh}},\ }\href {\doibase
		10.1038/s41467-020-17424-w} {\bibfield  {journal} {\bibinfo  {journal} {Nat.
				Commun.}\ }\textbf {\bibinfo {volume} {11}},\ \bibinfo {pages} {3649}
		(\bibinfo {year} {2020})}\BibitemShut {NoStop}%
	\bibitem [{\citenamefont {Giles}\ \emph {et~al.}(2017)\citenamefont {Giles},
		\citenamefont {Dai}, \citenamefont {Vurgaftman}, \citenamefont {Hoffman},
		\citenamefont {Liu}, \citenamefont {Lindsay}, \citenamefont {Ellis},
		\citenamefont {Assefa}, \citenamefont {Chatzakis}, \citenamefont {Reinecke},
		\citenamefont {Tischler}, \citenamefont {Fogler}, \citenamefont {Edgar},
		\citenamefont {Basov},\ and\ \citenamefont {Caldwell}}]{giles2017}%
	\BibitemOpen
	\bibfield  {author} {\bibinfo {author} {\bibfnamefont {A.~J.}\ \bibnamefont
			{Giles}}, \bibinfo {author} {\bibfnamefont {S.}~\bibnamefont {Dai}}, \bibinfo
		{author} {\bibfnamefont {I.}~\bibnamefont {Vurgaftman}}, \bibinfo {author}
		{\bibfnamefont {T.}~\bibnamefont {Hoffman}}, \bibinfo {author} {\bibfnamefont
			{S.}~\bibnamefont {Liu}}, \bibinfo {author} {\bibfnamefont {L.}~\bibnamefont
			{Lindsay}}, \bibinfo {author} {\bibfnamefont {C.~T.}\ \bibnamefont {Ellis}},
		\bibinfo {author} {\bibfnamefont {N.}~\bibnamefont {Assefa}}, \bibinfo
		{author} {\bibfnamefont {I.}~\bibnamefont {Chatzakis}}, \bibinfo {author}
		{\bibfnamefont {T.~L.}\ \bibnamefont {Reinecke}}, \bibinfo {author}
		{\bibfnamefont {J.~G.}\ \bibnamefont {Tischler}}, \bibinfo {author}
		{\bibfnamefont {M.~M.}\ \bibnamefont {Fogler}}, \bibinfo {author}
		{\bibfnamefont {J.~H.}\ \bibnamefont {Edgar}}, \bibinfo {author}
		{\bibfnamefont {D.~N.}\ \bibnamefont {Basov}}, \ and\ \bibinfo {author}
		{\bibfnamefont {J.~D.}\ \bibnamefont {Caldwell}},\ }\href {\doibase
		10.1038/nmat5047} {\bibfield  {journal} {\bibinfo  {journal} {Nat. Mater.}\
		}\textbf {\bibinfo {volume} {17}},\ \bibinfo {pages} {134} (\bibinfo {year}
		{2017})}\BibitemShut {NoStop}%
	\bibitem [{\citenamefont {Hu}\ \emph {et~al.}(2017)\citenamefont {Hu},
		\citenamefont {Luan}, \citenamefont {Scott}, \citenamefont {Yan},
		\citenamefont {Mandrus}, \citenamefont {Xu},\ and\ \citenamefont
		{Fei}}]{hu2017}%
	\BibitemOpen
	\bibfield  {author} {\bibinfo {author} {\bibfnamefont {F.}~\bibnamefont
			{Hu}}, \bibinfo {author} {\bibfnamefont {Y.}~\bibnamefont {Luan}}, \bibinfo
		{author} {\bibfnamefont {M.~E.}\ \bibnamefont {Scott}}, \bibinfo {author}
		{\bibfnamefont {J.}~\bibnamefont {Yan}}, \bibinfo {author} {\bibfnamefont
			{D.~G.}\ \bibnamefont {Mandrus}}, \bibinfo {author} {\bibfnamefont
			{X.}~\bibnamefont {Xu}}, \ and\ \bibinfo {author} {\bibfnamefont
			{Z.}~\bibnamefont {Fei}},\ }\href {\doibase 10.1038/nphoton.2017.65}
	{\bibfield  {journal} {\bibinfo  {journal} {Nat. Photonics}\ }\textbf
		{\bibinfo {volume} {11}},\ \bibinfo {pages} {356} (\bibinfo {year}
		{2017})}\BibitemShut {NoStop}%
	\bibitem [{\citenamefont {Smith}\ and\ \citenamefont
		{Schurig}(2003)}]{Smith_2003}%
	\BibitemOpen
	\bibfield  {author} {\bibinfo {author} {\bibfnamefont {D.~R.}\ \bibnamefont
			{Smith}}\ and\ \bibinfo {author} {\bibfnamefont {D.}~\bibnamefont
			{Schurig}},\ }\href {\doibase 10.1103/PhysRevLett.90.077405} {\bibfield
		{journal} {\bibinfo  {journal} {Phys. Rev. Lett.}\ }\textbf {\bibinfo
			{volume} {90}},\ \bibinfo {pages} {077405} (\bibinfo {year}
		{2003})}\BibitemShut {NoStop}%
	\bibitem [{\citenamefont {Krishnamoorthy}\ \emph {et~al.}(2012)\citenamefont
		{Krishnamoorthy}, \citenamefont {Jacob}, \citenamefont {Narimanov},
		\citenamefont {Kretzschmar},\ and\ \citenamefont {Menon}}]{Jacob_2012}%
	\BibitemOpen
	\bibfield  {author} {\bibinfo {author} {\bibfnamefont {H.~N.~S.}\
			\bibnamefont {Krishnamoorthy}}, \bibinfo {author} {\bibfnamefont
			{Z.}~\bibnamefont {Jacob}}, \bibinfo {author} {\bibfnamefont
			{E.}~\bibnamefont {Narimanov}}, \bibinfo {author} {\bibfnamefont
			{I.}~\bibnamefont {Kretzschmar}}, \ and\ \bibinfo {author} {\bibfnamefont
			{V.~M.}\ \bibnamefont {Menon}},\ }\href {\doibase 10.1126/science.1219171}
	{\bibfield  {journal} {\bibinfo  {journal} {Science}\ }\textbf {\bibinfo
			{volume} {336}},\ \bibinfo {pages} {205} (\bibinfo {year}
		{2012})}\BibitemShut {NoStop}%
	\bibitem [{\citenamefont {Hoffman}\ \emph {et~al.}(2007)\citenamefont
		{Hoffman}, \citenamefont {Alekseyev}, \citenamefont {Howard}, \citenamefont
		{Franz}, \citenamefont {Wasserman}, \citenamefont {Podolskiy}, \citenamefont
		{Narimanov}, \citenamefont {Sivco},\ and\ \citenamefont
		{Gmachl}}]{hoffman2007}%
	\BibitemOpen
	\bibfield  {author} {\bibinfo {author} {\bibfnamefont {A.~J.}\ \bibnamefont
			{Hoffman}}, \bibinfo {author} {\bibfnamefont {L.}~\bibnamefont {Alekseyev}},
		\bibinfo {author} {\bibfnamefont {S.~S.}\ \bibnamefont {Howard}}, \bibinfo
		{author} {\bibfnamefont {K.~J.}\ \bibnamefont {Franz}}, \bibinfo {author}
		{\bibfnamefont {D.}~\bibnamefont {Wasserman}}, \bibinfo {author}
		{\bibfnamefont {V.~A.}\ \bibnamefont {Podolskiy}}, \bibinfo {author}
		{\bibfnamefont {E.~E.}\ \bibnamefont {Narimanov}}, \bibinfo {author}
		{\bibfnamefont {D.~L.}\ \bibnamefont {Sivco}}, \ and\ \bibinfo {author}
		{\bibfnamefont {C.}~\bibnamefont {Gmachl}},\ }\href {\doibase
		10.1038/nmat2033} {\bibfield  {journal} {\bibinfo  {journal} {Nat. Mater.}\
		}\textbf {\bibinfo {volume} {6}},\ \bibinfo {pages} {946} (\bibinfo {year}
		{2007})}\BibitemShut {NoStop}%
	\bibitem [{\citenamefont {Fang}\ \emph {et~al.}(2009)\citenamefont {Fang},
		\citenamefont {Koschny},\ and\ \citenamefont {Soukoulis}}]{fang2009}%
	\BibitemOpen
	\bibfield  {author} {\bibinfo {author} {\bibfnamefont {A.}~\bibnamefont
			{Fang}}, \bibinfo {author} {\bibfnamefont {T.}~\bibnamefont {Koschny}}, \
		and\ \bibinfo {author} {\bibfnamefont {C.~M.}\ \bibnamefont {Soukoulis}},\
	}\href {\doibase 10.1103/PhysRevB.79.245127} {\bibfield  {journal} {\bibinfo
			{journal} {Phys. Rev. B}\ }\textbf {\bibinfo {volume} {79}},\ \bibinfo
		{pages} {245127} (\bibinfo {year} {2009})}\BibitemShut {NoStop}%
	\bibitem [{\citenamefont {Lin}\ \emph {et~al.}(2017)\citenamefont {Lin},
		\citenamefont {Yang}, \citenamefont {Rivera}, \citenamefont {L{\'{o}}pez},
		\citenamefont {Shen}, \citenamefont {Kaminer}, \citenamefont {Chen},
		\citenamefont {Zhang}, \citenamefont {Joannopoulos},\ and\ \citenamefont
		{Solja{\v{c}}i{\'{c}}}}]{lin2017}%
	\BibitemOpen
	\bibfield  {author} {\bibinfo {author} {\bibfnamefont {X.}~\bibnamefont
			{Lin}}, \bibinfo {author} {\bibfnamefont {Y.}~\bibnamefont {Yang}}, \bibinfo
		{author} {\bibfnamefont {N.}~\bibnamefont {Rivera}}, \bibinfo {author}
		{\bibfnamefont {J.~J.}\ \bibnamefont {L{\'{o}}pez}}, \bibinfo {author}
		{\bibfnamefont {Y.}~\bibnamefont {Shen}}, \bibinfo {author} {\bibfnamefont
			{I.}~\bibnamefont {Kaminer}}, \bibinfo {author} {\bibfnamefont
			{H.}~\bibnamefont {Chen}}, \bibinfo {author} {\bibfnamefont {B.}~\bibnamefont
			{Zhang}}, \bibinfo {author} {\bibfnamefont {J.~D.}\ \bibnamefont
			{Joannopoulos}}, \ and\ \bibinfo {author} {\bibfnamefont {M.}~\bibnamefont
			{Solja{\v{c}}i{\'{c}}}},\ }\href {\doibase 10.1073/pnas.1701830114}
	{\bibfield  {journal} {\bibinfo  {journal} {Proc. Natl. Acad. Sci.}\ }\textbf
		{\bibinfo {volume} {114}},\ \bibinfo {pages} {6717} (\bibinfo {year}
		{2017})}\BibitemShut {NoStop}%
	\bibitem [{\citenamefont {Noginov}\ \emph {et~al.}(2010)\citenamefont
		{Noginov}, \citenamefont {Li}, \citenamefont {Barnakov}, \citenamefont
		{Dryden}, \citenamefont {Nataraj}, \citenamefont {Zhu}, \citenamefont
		{Bonner}, \citenamefont {Mayy}, \citenamefont {Jacob},\ and\ \citenamefont
		{Narimanov}}]{noginov2010}%
	\BibitemOpen
	\bibfield  {author} {\bibinfo {author} {\bibfnamefont {M.~A.}\ \bibnamefont
			{Noginov}}, \bibinfo {author} {\bibfnamefont {H.}~\bibnamefont {Li}},
		\bibinfo {author} {\bibfnamefont {Y.~A.}\ \bibnamefont {Barnakov}}, \bibinfo
		{author} {\bibfnamefont {D.}~\bibnamefont {Dryden}}, \bibinfo {author}
		{\bibfnamefont {G.}~\bibnamefont {Nataraj}}, \bibinfo {author} {\bibfnamefont
			{G.}~\bibnamefont {Zhu}}, \bibinfo {author} {\bibfnamefont {C.~E.}\
			\bibnamefont {Bonner}}, \bibinfo {author} {\bibfnamefont {M.}~\bibnamefont
			{Mayy}}, \bibinfo {author} {\bibfnamefont {Z.}~\bibnamefont {Jacob}}, \ and\
		\bibinfo {author} {\bibfnamefont {E.~E.}\ \bibnamefont {Narimanov}},\ }\href
	{\doibase 10.1364/OL.35.001863} {\bibfield  {journal} {\bibinfo  {journal}
			{Opt. Lett.}\ }\textbf {\bibinfo {volume} {35}},\ \bibinfo {pages} {1863}
		(\bibinfo {year} {2010})}\BibitemShut {NoStop}%
	\bibitem [{\citenamefont {Jacob}\ \emph {et~al.}(2012)\citenamefont {Jacob},
		\citenamefont {Smolyaninov},\ and\ \citenamefont {Narimanov}}]{jacob2012}%
	\BibitemOpen
	\bibfield  {author} {\bibinfo {author} {\bibfnamefont {Z.}~\bibnamefont
			{Jacob}}, \bibinfo {author} {\bibfnamefont {I.~I.}\ \bibnamefont
			{Smolyaninov}}, \ and\ \bibinfo {author} {\bibfnamefont {E.~E.}\ \bibnamefont
			{Narimanov}},\ }\href {\doibase 10.1063/1.4710548} {\bibfield  {journal}
		{\bibinfo  {journal} {Appl. Phys. Lett.}\ }\textbf {\bibinfo {volume}
			{100}},\ \bibinfo {pages} {181105} (\bibinfo {year} {2012})}\BibitemShut
	{NoStop}%
	\bibitem [{\citenamefont {Guo}\ \emph {et~al.}(2012)\citenamefont {Guo},
		\citenamefont {Cortes}, \citenamefont {Molesky},\ and\ \citenamefont
		{Jacob}}]{guo2012}%
	\BibitemOpen
	\bibfield  {author} {\bibinfo {author} {\bibfnamefont {Y.}~\bibnamefont
			{Guo}}, \bibinfo {author} {\bibfnamefont {C.~L.}\ \bibnamefont {Cortes}},
		\bibinfo {author} {\bibfnamefont {S.}~\bibnamefont {Molesky}}, \ and\
		\bibinfo {author} {\bibfnamefont {Z.}~\bibnamefont {Jacob}},\ }\href
	{\doibase 10.1063/1.4754616} {\bibfield  {journal} {\bibinfo  {journal}
			{Appl. Phys. Lett.}\ }\textbf {\bibinfo {volume} {101}},\ \bibinfo {pages}
		{131106} (\bibinfo {year} {2012})}\BibitemShut {NoStop}%
	\bibitem [{\citenamefont {Smith}\ \emph {et~al.}(2004)\citenamefont {Smith},
		\citenamefont {Schurig}, \citenamefont {Mock}, \citenamefont {Kolinko},\ and\
		\citenamefont {Rye}}]{smith2004}%
	\BibitemOpen
	\bibfield  {author} {\bibinfo {author} {\bibfnamefont {D.~R.}\ \bibnamefont
			{Smith}}, \bibinfo {author} {\bibfnamefont {D.}~\bibnamefont {Schurig}},
		\bibinfo {author} {\bibfnamefont {J.~J.}\ \bibnamefont {Mock}}, \bibinfo
		{author} {\bibfnamefont {P.}~\bibnamefont {Kolinko}}, \ and\ \bibinfo
		{author} {\bibfnamefont {P.}~\bibnamefont {Rye}},\ }\href {\doibase
		10.1063/1.1690471} {\bibfield  {journal} {\bibinfo  {journal} {Appl. Phys.
				Lett.}\ }\textbf {\bibinfo {volume} {84}},\ \bibinfo {pages} {2244} (\bibinfo
		{year} {2004})}\BibitemShut {NoStop}%
	\bibitem [{\citenamefont {Liu}\ \emph {et~al.}(2007)\citenamefont {Liu},
		\citenamefont {Lee}, \citenamefont {Xiong}, \citenamefont {Sun},\ and\
		\citenamefont {Zhang}}]{liu2007}%
	\BibitemOpen
	\bibfield  {author} {\bibinfo {author} {\bibfnamefont {Z.}~\bibnamefont
			{Liu}}, \bibinfo {author} {\bibfnamefont {H.}~\bibnamefont {Lee}}, \bibinfo
		{author} {\bibfnamefont {Y.}~\bibnamefont {Xiong}}, \bibinfo {author}
		{\bibfnamefont {C.}~\bibnamefont {Sun}}, \ and\ \bibinfo {author}
		{\bibfnamefont {X.}~\bibnamefont {Zhang}},\ }\href {\doibase
		10.1126/science.1137368} {\bibfield  {journal} {\bibinfo  {journal}
			{Science}\ }\textbf {\bibinfo {volume} {315}},\ \bibinfo {pages} {1686}
		(\bibinfo {year} {2007})}\BibitemShut {NoStop}%
	\bibitem [{\citenamefont {Dai}\ \emph {et~al.}(2015)\citenamefont {Dai},
		\citenamefont {Ma}, \citenamefont {Andersen}, \citenamefont {Mcleod},
		\citenamefont {Fei}, \citenamefont {Liu}, \citenamefont {Wagner},
		\citenamefont {Watanabe}, \citenamefont {Taniguchi}, \citenamefont
		{Thiemens}, \citenamefont {Keilmann}, \citenamefont {Jarillo-Herrero},
		\citenamefont {Fogler},\ and\ \citenamefont {Basov}}]{dai2015}%
	\BibitemOpen
	\bibfield  {author} {\bibinfo {author} {\bibfnamefont {S.}~\bibnamefont
			{Dai}}, \bibinfo {author} {\bibfnamefont {Q.}~\bibnamefont {Ma}}, \bibinfo
		{author} {\bibfnamefont {T.}~\bibnamefont {Andersen}}, \bibinfo {author}
		{\bibfnamefont {A.~S.}\ \bibnamefont {Mcleod}}, \bibinfo {author}
		{\bibfnamefont {Z.}~\bibnamefont {Fei}}, \bibinfo {author} {\bibfnamefont
			{M.~K.}\ \bibnamefont {Liu}}, \bibinfo {author} {\bibfnamefont
			{M.}~\bibnamefont {Wagner}}, \bibinfo {author} {\bibfnamefont
			{K.}~\bibnamefont {Watanabe}}, \bibinfo {author} {\bibfnamefont
			{T.}~\bibnamefont {Taniguchi}}, \bibinfo {author} {\bibfnamefont
			{M.}~\bibnamefont {Thiemens}}, \bibinfo {author} {\bibfnamefont
			{F.}~\bibnamefont {Keilmann}}, \bibinfo {author} {\bibfnamefont
			{P.}~\bibnamefont {Jarillo-Herrero}}, \bibinfo {author} {\bibfnamefont
			{M.~M.}\ \bibnamefont {Fogler}}, \ and\ \bibinfo {author} {\bibfnamefont
			{D.~N.}\ \bibnamefont {Basov}},\ }\href {\doibase 10.1038/ncomms7963}
	{\bibfield  {journal} {\bibinfo  {journal} {Nat. Commun.}\ }\textbf {\bibinfo
			{volume} {6}},\ \bibinfo {pages} {6963} (\bibinfo {year} {2015})}\BibitemShut
	{NoStop}%
	\bibitem [{\citenamefont {Stein}\ \emph {et~al.}(2012)\citenamefont {Stein},
		\citenamefont {Devaux}, \citenamefont {Genet},\ and\ \citenamefont
		{Ebbesen}}]{stein2012}%
	\BibitemOpen
	\bibfield  {author} {\bibinfo {author} {\bibfnamefont {B.}~\bibnamefont
			{Stein}}, \bibinfo {author} {\bibfnamefont {E.}~\bibnamefont {Devaux}},
		\bibinfo {author} {\bibfnamefont {C.}~\bibnamefont {Genet}}, \ and\ \bibinfo
		{author} {\bibfnamefont {T.~W.}\ \bibnamefont {Ebbesen}},\ }\href {\doibase
		10.1364/ol.37.001916} {\bibfield  {journal} {\bibinfo  {journal} {Opt.
				Lett.}\ }\textbf {\bibinfo {volume} {37}},\ \bibinfo {pages} {1916} (\bibinfo
		{year} {2012})}\BibitemShut {NoStop}%
	\bibitem [{\citenamefont {Forati}\ \emph {et~al.}(2014)\citenamefont {Forati},
		\citenamefont {Hanson}, \citenamefont {Yakovlev},\ and\ \citenamefont
		{Al\`u}}]{forati2014}%
	\BibitemOpen
	\bibfield  {author} {\bibinfo {author} {\bibfnamefont {E.}~\bibnamefont
			{Forati}}, \bibinfo {author} {\bibfnamefont {G.~W.}\ \bibnamefont {Hanson}},
		\bibinfo {author} {\bibfnamefont {A.~B.}\ \bibnamefont {Yakovlev}}, \ and\
		\bibinfo {author} {\bibfnamefont {A.}~\bibnamefont {Al\`u}},\ }\href
	{\doibase 10.1103/PhysRevB.89.081410} {\bibfield  {journal} {\bibinfo
			{journal} {Phys. Rev. B}\ }\textbf {\bibinfo {volume} {89}},\ \bibinfo
		{pages} {081410} (\bibinfo {year} {2014})}\BibitemShut {NoStop}%
	\bibitem [{\citenamefont {Correas-Serrano}\ \emph {et~al.}(2017)\citenamefont
		{Correas-Serrano}, \citenamefont {Al\`u},\ and\ \citenamefont
		{Gomez-Diaz}}]{alu2017}%
	\BibitemOpen
	\bibfield  {author} {\bibinfo {author} {\bibfnamefont {D.}~\bibnamefont
			{Correas-Serrano}}, \bibinfo {author} {\bibfnamefont {A.}~\bibnamefont
			{Al\`u}}, \ and\ \bibinfo {author} {\bibfnamefont {J.~S.}\ \bibnamefont
			{Gomez-Diaz}},\ }\href {\doibase 10.1103/PhysRevB.96.075436} {\bibfield
		{journal} {\bibinfo  {journal} {Phys. Rev. B}\ }\textbf {\bibinfo {volume}
			{96}},\ \bibinfo {pages} {075436} (\bibinfo {year} {2017})}\BibitemShut
	{NoStop}%
	\bibitem [{\citenamefont {Yermakov}\ \emph {et~al.}(2016)\citenamefont
		{Yermakov}, \citenamefont {Ovcharenko}, \citenamefont {Bogdanov},
		\citenamefont {Iorsh}, \citenamefont {Bliokh},\ and\ \citenamefont
		{Kivshar}}]{ITMO_spin}%
	\BibitemOpen
	\bibfield  {author} {\bibinfo {author} {\bibfnamefont {O.~Y.}\ \bibnamefont
			{Yermakov}}, \bibinfo {author} {\bibfnamefont {A.~I.}\ \bibnamefont
			{Ovcharenko}}, \bibinfo {author} {\bibfnamefont {A.~A.}\ \bibnamefont
			{Bogdanov}}, \bibinfo {author} {\bibfnamefont {I.~V.}\ \bibnamefont {Iorsh}},
		\bibinfo {author} {\bibfnamefont {K.~Y.}\ \bibnamefont {Bliokh}}, \ and\
		\bibinfo {author} {\bibfnamefont {Y.~S.}\ \bibnamefont {Kivshar}},\ }\href
	{\doibase 10.1103/PhysRevB.94.075446} {\bibfield  {journal} {\bibinfo
			{journal} {Phys. Rev. B}\ }\textbf {\bibinfo {volume} {94}},\ \bibinfo
		{pages} {075446} (\bibinfo {year} {2016})}\BibitemShut {NoStop}%
	\bibitem [{\citenamefont {Narimanov}\ and\ \citenamefont
		{Kildishev}(2015)}]{narimanov2015}%
	\BibitemOpen
	\bibfield  {author} {\bibinfo {author} {\bibfnamefont {E.~E.}\ \bibnamefont
			{Narimanov}}\ and\ \bibinfo {author} {\bibfnamefont {A.~V.}\ \bibnamefont
			{Kildishev}},\ }\href {\doibase 10.1038/nphoton.2015.56} {\bibfield
		{journal} {\bibinfo  {journal} {Nat. Photonics}\ }\textbf {\bibinfo {volume}
			{9}},\ \bibinfo {pages} {214} (\bibinfo {year} {2015})}\BibitemShut {NoStop}%
	\bibitem [{\citenamefont {Korzeb}\ \emph {et~al.}(2015)\citenamefont {Korzeb},
		\citenamefont {Gajc},\ and\ \citenamefont {Pawlak}}]{korzeb2015}%
	\BibitemOpen
	\bibfield  {author} {\bibinfo {author} {\bibfnamefont {K.}~\bibnamefont
			{Korzeb}}, \bibinfo {author} {\bibfnamefont {M.}~\bibnamefont {Gajc}}, \ and\
		\bibinfo {author} {\bibfnamefont {D.~A.}\ \bibnamefont {Pawlak}},\ }\href
	{\doibase 10.1364/oe.23.025406} {\bibfield  {journal} {\bibinfo  {journal}
			{Opt. Express}\ }\textbf {\bibinfo {volume} {23}},\ \bibinfo {pages} {25406}
		(\bibinfo {year} {2015})}\BibitemShut {NoStop}%
	\bibitem [{\citenamefont {Gjerding}\ \emph {et~al.}(2017)\citenamefont
		{Gjerding}, \citenamefont {Petersen}, \citenamefont {Pedersen}, \citenamefont
		{Mortensen},\ and\ \citenamefont {Thygesen}}]{gjerding2017}%
	\BibitemOpen
	\bibfield  {author} {\bibinfo {author} {\bibfnamefont {M.~N.}\ \bibnamefont
			{Gjerding}}, \bibinfo {author} {\bibfnamefont {R.}~\bibnamefont {Petersen}},
		\bibinfo {author} {\bibfnamefont {T.~G.}\ \bibnamefont {Pedersen}}, \bibinfo
		{author} {\bibfnamefont {N.~A.}\ \bibnamefont {Mortensen}}, \ and\ \bibinfo
		{author} {\bibfnamefont {K.~S.}\ \bibnamefont {Thygesen}},\ }\href {\doibase
		10.1038/s41467-017-00412-y} {\bibfield  {journal} {\bibinfo  {journal} {Nat.
				Commun.}\ }\textbf {\bibinfo {volume} {8}},\ \bibinfo {pages} {320} (\bibinfo
		{year} {2017})}\BibitemShut {NoStop}%
	\bibitem [{\citenamefont {Sun}\ \emph {et~al.}(2014)\citenamefont {Sun},
		\citenamefont {Litchinitser},\ and\ \citenamefont {Zhou}}]{sun2014}%
	\BibitemOpen
	\bibfield  {author} {\bibinfo {author} {\bibfnamefont {J.}~\bibnamefont
			{Sun}}, \bibinfo {author} {\bibfnamefont {N.~M.}\ \bibnamefont
			{Litchinitser}}, \ and\ \bibinfo {author} {\bibfnamefont {J.}~\bibnamefont
			{Zhou}},\ }\href {\doibase 10.1021/ph4000983} {\bibfield  {journal} {\bibinfo
			{journal} {{ACS} Photonics}\ }\textbf {\bibinfo {volume} {1}},\ \bibinfo
		{pages} {293} (\bibinfo {year} {2014})}\BibitemShut {NoStop}%
	\bibitem [{\citenamefont {Correas-Serrano}\ \emph {et~al.}(2016)\citenamefont
		{Correas-Serrano}, \citenamefont {Gomez-Diaz}, \citenamefont {Melcon},\ and\
		\citenamefont {Al\`u}}]{Alu_BP}%
	\BibitemOpen
	\bibfield  {author} {\bibinfo {author} {\bibfnamefont {D.}~\bibnamefont
			{Correas-Serrano}}, \bibinfo {author} {\bibfnamefont {J.~S.}\ \bibnamefont
			{Gomez-Diaz}}, \bibinfo {author} {\bibfnamefont {A.~A.}\ \bibnamefont
			{Melcon}}, \ and\ \bibinfo {author} {\bibfnamefont {A.}~\bibnamefont
			{Al\`u}},\ }\href {\doibase 10.1088/2040-8978/18/10/104006} {\bibfield
		{journal} {\bibinfo  {journal} {J. Opt.}\ }\textbf {\bibinfo {volume} {18}},\
		\bibinfo {pages} {104006} (\bibinfo {year} {2016})}\BibitemShut {NoStop}%
	\bibitem [{\citenamefont {van Veen}\ \emph {et~al.}(2019)\citenamefont {van
			Veen}, \citenamefont {Nemilentsau}, \citenamefont {Kumar}, \citenamefont
		{Rold\'an}, \citenamefont {Katsnelson}, \citenamefont {Low},\ and\
		\citenamefont {Yuan}}]{Katsnelson_BP}%
	\BibitemOpen
	\bibfield  {author} {\bibinfo {author} {\bibfnamefont {E.}~\bibnamefont {van
				Veen}}, \bibinfo {author} {\bibfnamefont {A.}~\bibnamefont {Nemilentsau}},
		\bibinfo {author} {\bibfnamefont {A.}~\bibnamefont {Kumar}}, \bibinfo
		{author} {\bibfnamefont {R.}~\bibnamefont {Rold\'an}}, \bibinfo {author}
		{\bibfnamefont {M.~I.}\ \bibnamefont {Katsnelson}}, \bibinfo {author}
		{\bibfnamefont {T.}~\bibnamefont {Low}}, \ and\ \bibinfo {author}
		{\bibfnamefont {S.}~\bibnamefont {Yuan}},\ }\href {\doibase
		10.1103/PhysRevApplied.12.014011} {\bibfield  {journal} {\bibinfo  {journal}
			{Phys. Rev. Applied}\ }\textbf {\bibinfo {volume} {12}},\ \bibinfo {pages}
		{014011} (\bibinfo {year} {2019})}\BibitemShut {NoStop}%
	\bibitem [{\citenamefont {Esslinger}\ \emph {et~al.}(2014)\citenamefont
		{Esslinger}, \citenamefont {Vogelgesang}, \citenamefont {Talebi},
		\citenamefont {Khunsin}, \citenamefont {Gehring}, \citenamefont {de~Zuani},
		\citenamefont {Gompf},\ and\ \citenamefont {Kern}}]{esslinger2014}%
	\BibitemOpen
	\bibfield  {author} {\bibinfo {author} {\bibfnamefont {M.}~\bibnamefont
			{Esslinger}}, \bibinfo {author} {\bibfnamefont {R.}~\bibnamefont
			{Vogelgesang}}, \bibinfo {author} {\bibfnamefont {N.}~\bibnamefont {Talebi}},
		\bibinfo {author} {\bibfnamefont {W.}~\bibnamefont {Khunsin}}, \bibinfo
		{author} {\bibfnamefont {P.}~\bibnamefont {Gehring}}, \bibinfo {author}
		{\bibfnamefont {S.}~\bibnamefont {de~Zuani}}, \bibinfo {author}
		{\bibfnamefont {B.}~\bibnamefont {Gompf}}, \ and\ \bibinfo {author}
		{\bibfnamefont {K.}~\bibnamefont {Kern}},\ }\href {\doibase
		10.1021/ph500296e} {\bibfield  {journal} {\bibinfo  {journal} {{ACS}
				Photonics}\ }\textbf {\bibinfo {volume} {1}},\ \bibinfo {pages} {1285}
		(\bibinfo {year} {2014})}\BibitemShut {NoStop}%
	\bibitem [{\citenamefont {Holloway}\ \emph {et~al.}(2012)\citenamefont
		{Holloway}, \citenamefont {Kuester}, \citenamefont {Gordon}, \citenamefont
		{O'Hara}, \citenamefont {Booth},\ and\ \citenamefont
		{Smith}}]{RevMeta_Holloway}%
	\BibitemOpen
	\bibfield  {author} {\bibinfo {author} {\bibfnamefont {C.~L.}\ \bibnamefont
			{Holloway}}, \bibinfo {author} {\bibfnamefont {E.~F.}\ \bibnamefont
			{Kuester}}, \bibinfo {author} {\bibfnamefont {J.~A.}\ \bibnamefont {Gordon}},
		\bibinfo {author} {\bibfnamefont {J.}~\bibnamefont {O'Hara}}, \bibinfo
		{author} {\bibfnamefont {J.}~\bibnamefont {Booth}}, \ and\ \bibinfo {author}
		{\bibfnamefont {D.~R.}\ \bibnamefont {Smith}},\ }\href {\doibase
		10.1109/MAP.2012.6230714} {\bibfield  {journal} {\bibinfo  {journal} {IEEE
				Trans. Antennas Propag.}\ }\textbf {\bibinfo {volume} {54}},\ \bibinfo
		{pages} {10} (\bibinfo {year} {2012})}\BibitemShut {NoStop}%
	\bibitem [{\citenamefont {Glybovski}\ \emph {et~al.}(2016)\citenamefont
		{Glybovski}, \citenamefont {Tretyakov}, \citenamefont {Belov}, \citenamefont
		{Kivshar},\ and\ \citenamefont {Simovski}}]{RevMeta_ITMO}%
	\BibitemOpen
	\bibfield  {author} {\bibinfo {author} {\bibfnamefont {S.~B.}\ \bibnamefont
			{Glybovski}}, \bibinfo {author} {\bibfnamefont {S.~A.}\ \bibnamefont
			{Tretyakov}}, \bibinfo {author} {\bibfnamefont {P.~A.}\ \bibnamefont
			{Belov}}, \bibinfo {author} {\bibfnamefont {Y.~S.}\ \bibnamefont {Kivshar}},
		\ and\ \bibinfo {author} {\bibfnamefont {C.~R.}\ \bibnamefont {Simovski}},\
	}\href {\doibase http://dx.doi.org/10.1016/j.physrep.2016.04.004} {\bibfield
		{journal} {\bibinfo  {journal} {Phys. Rep.}\ }\textbf {\bibinfo {volume}
			{634}},\ \bibinfo {pages} {1 } (\bibinfo {year} {2016})}\BibitemShut
	{NoStop}%
	\bibitem [{\citenamefont {Gomez-Diaz}\ and\ \citenamefont
		{Al\`u}(2016)}]{Alu_Rev}%
	\BibitemOpen
	\bibfield  {author} {\bibinfo {author} {\bibfnamefont {J.~S.}\ \bibnamefont
			{Gomez-Diaz}}\ and\ \bibinfo {author} {\bibfnamefont {A.}~\bibnamefont
			{Al\`u}},\ }\href {\doibase 10.1021/acsphotonics.6b00645} {\bibfield
		{journal} {\bibinfo  {journal} {ACS Photonics}\ }\textbf {\bibinfo {volume}
			{3}},\ \bibinfo {pages} {2211} (\bibinfo {year} {2016})}\BibitemShut
	{NoStop}%
	\bibitem [{\citenamefont {Kotov}\ and\ \citenamefont
		{Lozovik}(2017)}]{Kotov2017}%
	\BibitemOpen
	\bibfield  {author} {\bibinfo {author} {\bibfnamefont {O.~V.}\ \bibnamefont
			{Kotov}}\ and\ \bibinfo {author} {\bibfnamefont {Y.~E.}\ \bibnamefont
			{Lozovik}},\ }\href {\doibase 10.1103/PhysRevB.96.235403} {\bibfield
		{journal} {\bibinfo  {journal} {Phys. Rev. B}\ }\textbf {\bibinfo {volume}
			{96}},\ \bibinfo {pages} {235403} (\bibinfo {year} {2017})}\BibitemShut
	{NoStop}%
	\bibitem [{\citenamefont {Huo}\ \emph {et~al.}(2019)\citenamefont {Huo},
		\citenamefont {Zhang}, \citenamefont {Liang}, \citenamefont {Lu},\ and\
		\citenamefont {Xu}}]{RevHMS_2019}%
	\BibitemOpen
	\bibfield  {author} {\bibinfo {author} {\bibfnamefont {P.}~\bibnamefont
			{Huo}}, \bibinfo {author} {\bibfnamefont {S.}~\bibnamefont {Zhang}}, \bibinfo
		{author} {\bibfnamefont {Y.}~\bibnamefont {Liang}}, \bibinfo {author}
		{\bibfnamefont {Y.}~\bibnamefont {Lu}}, \ and\ \bibinfo {author}
		{\bibfnamefont {T.}~\bibnamefont {Xu}},\ }\href {\doibase
		10.1002/adom.201801616} {\bibfield  {journal} {\bibinfo  {journal} {Adv. Opt.
				Mater.}\ }\textbf {\bibinfo {volume} {7}},\ \bibinfo {pages} {1801616}
		(\bibinfo {year} {2019})}\BibitemShut {NoStop}%
	\bibitem [{\citenamefont {Kotov}\ and\ \citenamefont
		{Lozovik}(2019)}]{Kotov2019}%
	\BibitemOpen
	\bibfield  {author} {\bibinfo {author} {\bibfnamefont {O.~V.}\ \bibnamefont
			{Kotov}}\ and\ \bibinfo {author} {\bibfnamefont {Y.~E.}\ \bibnamefont
			{Lozovik}},\ }\href {\doibase 10.1103/PhysRevB.100.165424} {\bibfield
		{journal} {\bibinfo  {journal} {Phys. Rev. B}\ }\textbf {\bibinfo {volume}
			{100}},\ \bibinfo {pages} {165424} (\bibinfo {year} {2019})}\BibitemShut
	{NoStop}%
	\bibitem [{\citenamefont {Liu}\ and\ \citenamefont
		{Zhang}(2013)}]{HMS_Liu2013}%
	\BibitemOpen
	\bibfield  {author} {\bibinfo {author} {\bibfnamefont {Y.}~\bibnamefont
			{Liu}}\ and\ \bibinfo {author} {\bibfnamefont {X.}~\bibnamefont {Zhang}},\
	}\href {\doibase 10.1063/1.4821444} {\bibfield  {journal} {\bibinfo
			{journal} {Appl. Phys. Lett.}\ }\textbf {\bibinfo {volume} {103}},\ \bibinfo
		{eid} {141101} (\bibinfo {year} {2013})}\BibitemShut {NoStop}%
	\bibitem [{\citenamefont {High}\ \emph {et~al.}(2015)\citenamefont {High},
		\citenamefont {Devlin}, \citenamefont {Dibos}, \citenamefont {Polking},
		\citenamefont {Wild}, \citenamefont {Perczel}, \citenamefont {de~Leon},
		\citenamefont {Lukin},\ and\ \citenamefont {Park}}]{Visible_HMS}%
	\BibitemOpen
	\bibfield  {author} {\bibinfo {author} {\bibfnamefont {A.~A.}\ \bibnamefont
			{High}}, \bibinfo {author} {\bibfnamefont {R.~C.}\ \bibnamefont {Devlin}},
		\bibinfo {author} {\bibfnamefont {A.}~\bibnamefont {Dibos}}, \bibinfo
		{author} {\bibfnamefont {M.}~\bibnamefont {Polking}}, \bibinfo {author}
		{\bibfnamefont {D.~S.}\ \bibnamefont {Wild}}, \bibinfo {author}
		{\bibfnamefont {J.}~\bibnamefont {Perczel}}, \bibinfo {author} {\bibfnamefont
			{N.~P.}\ \bibnamefont {de~Leon}}, \bibinfo {author} {\bibfnamefont {M.~D.}\
			\bibnamefont {Lukin}}, \ and\ \bibinfo {author} {\bibfnamefont
			{H.}~\bibnamefont {Park}},\ }\href {http://dx.doi.org/10.1038/nature14477}
	{\bibfield  {journal} {\bibinfo  {journal} {Nature}\ }\textbf {\bibinfo
			{volume} {522}},\ \bibinfo {pages} {192} (\bibinfo {year}
		{2015})}\BibitemShut {NoStop}%
	\bibitem [{\citenamefont {Yermakov}\ \emph {et~al.}(2015)\citenamefont
		{Yermakov}, \citenamefont {Ovcharenko}, \citenamefont {Song}, \citenamefont
		{Bogdanov}, \citenamefont {Iorsh},\ and\ \citenamefont {Kivshar}}]{ITMO_PRB}%
	\BibitemOpen
	\bibfield  {author} {\bibinfo {author} {\bibfnamefont {O.~Y.}\ \bibnamefont
			{Yermakov}}, \bibinfo {author} {\bibfnamefont {A.~I.}\ \bibnamefont
			{Ovcharenko}}, \bibinfo {author} {\bibfnamefont {M.}~\bibnamefont {Song}},
		\bibinfo {author} {\bibfnamefont {A.~A.}\ \bibnamefont {Bogdanov}}, \bibinfo
		{author} {\bibfnamefont {I.~V.}\ \bibnamefont {Iorsh}}, \ and\ \bibinfo
		{author} {\bibfnamefont {Y.~S.}\ \bibnamefont {Kivshar}},\ }\href {\doibase
		10.1103/PhysRevB.91.235423} {\bibfield  {journal} {\bibinfo  {journal} {Phys.
				Rev. B}\ }\textbf {\bibinfo {volume} {91}},\ \bibinfo {pages} {235423}
		(\bibinfo {year} {2015})}\BibitemShut {NoStop}%
	\bibitem [{\citenamefont {Samusev}\ \emph {et~al.}(2017)\citenamefont
		{Samusev}, \citenamefont {Mukhin}, \citenamefont {Malureanu}, \citenamefont
		{Takayama}, \citenamefont {Permyakov}, \citenamefont {Sinev}, \citenamefont
		{Baranov}, \citenamefont {Yermakov}, \citenamefont {Iorsh}, \citenamefont
		{Bogdanov},\ and\ \citenamefont {Lavrinenko}}]{ITMO_expOpt}%
	\BibitemOpen
	\bibfield  {author} {\bibinfo {author} {\bibfnamefont {A.}~\bibnamefont
			{Samusev}}, \bibinfo {author} {\bibfnamefont {I.}~\bibnamefont {Mukhin}},
		\bibinfo {author} {\bibfnamefont {R.}~\bibnamefont {Malureanu}}, \bibinfo
		{author} {\bibfnamefont {O.}~\bibnamefont {Takayama}}, \bibinfo {author}
		{\bibfnamefont {D.~V.}\ \bibnamefont {Permyakov}}, \bibinfo {author}
		{\bibfnamefont {I.~S.}\ \bibnamefont {Sinev}}, \bibinfo {author}
		{\bibfnamefont {D.}~\bibnamefont {Baranov}}, \bibinfo {author} {\bibfnamefont
			{O.}~\bibnamefont {Yermakov}}, \bibinfo {author} {\bibfnamefont {I.~V.}\
			\bibnamefont {Iorsh}}, \bibinfo {author} {\bibfnamefont {A.~A.}\ \bibnamefont
			{Bogdanov}}, \ and\ \bibinfo {author} {\bibfnamefont {A.~V.}\ \bibnamefont
			{Lavrinenko}},\ }\href {\doibase 10.1364/OE.25.032631} {\bibfield  {journal}
		{\bibinfo  {journal} {Opt. Express}\ }\textbf {\bibinfo {volume} {25}},\
		\bibinfo {pages} {32631} (\bibinfo {year} {2017})}\BibitemShut {NoStop}%
	\bibitem [{\citenamefont {Nikitin}\ \emph {et~al.}(2011)\citenamefont
		{Nikitin}, \citenamefont {Guinea}, \citenamefont {Garc\'{\i}a-Vidal},\ and\
		\citenamefont {Mart\'{\i}n-Moreno}}]{nikitin2011}%
	\BibitemOpen
	\bibfield  {author} {\bibinfo {author} {\bibfnamefont {A.~Y.}\ \bibnamefont
			{Nikitin}}, \bibinfo {author} {\bibfnamefont {F.}~\bibnamefont {Guinea}},
		\bibinfo {author} {\bibfnamefont {F.~J.}\ \bibnamefont {Garc\'{\i}a-Vidal}},
		\ and\ \bibinfo {author} {\bibfnamefont {L.}~\bibnamefont
			{Mart\'{\i}n-Moreno}},\ }\href {\doibase 10.1103/PhysRevB.84.161407}
	{\bibfield  {journal} {\bibinfo  {journal} {Phys. Rev. B}\ }\textbf {\bibinfo
			{volume} {84}},\ \bibinfo {pages} {161407} (\bibinfo {year}
		{2011})}\BibitemShut {NoStop}%
	\bibitem [{\citenamefont {Mason}\ \emph {et~al.}(2014)\citenamefont {Mason},
		\citenamefont {Menabde}, \citenamefont {Yu},\ and\ \citenamefont
		{Park}}]{mason2014}%
	\BibitemOpen
	\bibfield  {author} {\bibinfo {author} {\bibfnamefont {D.~R.}\ \bibnamefont
			{Mason}}, \bibinfo {author} {\bibfnamefont {S.~G.}\ \bibnamefont {Menabde}},
		\bibinfo {author} {\bibfnamefont {S.}~\bibnamefont {Yu}}, \ and\ \bibinfo
		{author} {\bibfnamefont {N.}~\bibnamefont {Park}},\ }\href {\doibase
		10.1038/srep04536} {\bibfield  {journal} {\bibinfo  {journal} {Sci. Rep.}\
		}\textbf {\bibinfo {volume} {4}},\ \bibinfo {pages} {4536} (\bibinfo {year}
		{2014})}\BibitemShut {NoStop}%
	\bibitem [{\citenamefont {Nikitin}\ \emph {et~al.}(2016)\citenamefont
		{Nikitin}, \citenamefont {Alonso-Gonz{\'{a}}lez}, \citenamefont
		{V{\'{e}}lez}, \citenamefont {Mastel}, \citenamefont {Centeno}, \citenamefont
		{Pesquera}, \citenamefont {Zurutuza}, \citenamefont {Casanova}, \citenamefont
		{Hueso}, \citenamefont {Koppens},\ and\ \citenamefont
		{Hillenbrand}}]{nikitin2016}%
	\BibitemOpen
	\bibfield  {author} {\bibinfo {author} {\bibfnamefont {A.~Y.}\ \bibnamefont
			{Nikitin}}, \bibinfo {author} {\bibfnamefont {P.}~\bibnamefont
			{Alonso-Gonz{\'{a}}lez}}, \bibinfo {author} {\bibfnamefont {S.}~\bibnamefont
			{V{\'{e}}lez}}, \bibinfo {author} {\bibfnamefont {S.}~\bibnamefont {Mastel}},
		\bibinfo {author} {\bibfnamefont {A.}~\bibnamefont {Centeno}}, \bibinfo
		{author} {\bibfnamefont {A.}~\bibnamefont {Pesquera}}, \bibinfo {author}
		{\bibfnamefont {A.}~\bibnamefont {Zurutuza}}, \bibinfo {author}
		{\bibfnamefont {F.}~\bibnamefont {Casanova}}, \bibinfo {author}
		{\bibfnamefont {L.~E.}\ \bibnamefont {Hueso}}, \bibinfo {author}
		{\bibfnamefont {F.~H.~L.}\ \bibnamefont {Koppens}}, \ and\ \bibinfo {author}
		{\bibfnamefont {R.}~\bibnamefont {Hillenbrand}},\ }\href {\doibase
		10.1038/nphoton.2016.44} {\bibfield  {journal} {\bibinfo  {journal} {Nat.
				Photonics}\ }\textbf {\bibinfo {volume} {10}},\ \bibinfo {pages} {239}
		(\bibinfo {year} {2016})}\BibitemShut {NoStop}%
	\bibitem [{\citenamefont {Angelis}\ \emph {et~al.}(2016)\citenamefont
		{Angelis}, \citenamefont {Locatelli}, \citenamefont {Mutti},\ and\
		\citenamefont {Aceves}}]{angelis2016}%
	\BibitemOpen
	\bibfield  {author} {\bibinfo {author} {\bibfnamefont {C.~D.}\ \bibnamefont
			{Angelis}}, \bibinfo {author} {\bibfnamefont {A.}~\bibnamefont {Locatelli}},
		\bibinfo {author} {\bibfnamefont {A.}~\bibnamefont {Mutti}}, \ and\ \bibinfo
		{author} {\bibfnamefont {A.}~\bibnamefont {Aceves}},\ }\href {\doibase
		10.1364/ol.41.000480} {\bibfield  {journal} {\bibinfo  {journal} {Opt.
				Lett.}\ }\textbf {\bibinfo {volume} {41}},\ \bibinfo {pages} {480} (\bibinfo
		{year} {2016})}\BibitemShut {NoStop}%
	\bibitem [{\citenamefont {Gon{\c{c}}alves}\ \emph {et~al.}(2017)\citenamefont
		{Gon{\c{c}}alves}, \citenamefont {Xiao}, \citenamefont {Peres},\ and\
		\citenamefont {Mortensen}}]{mortensen2017}%
	\BibitemOpen
	\bibfield  {author} {\bibinfo {author} {\bibfnamefont {P.~A.~D.}\
			\bibnamefont {Gon{\c{c}}alves}}, \bibinfo {author} {\bibfnamefont
			{S.}~\bibnamefont {Xiao}}, \bibinfo {author} {\bibfnamefont {N.~M.~R.}\
			\bibnamefont {Peres}}, \ and\ \bibinfo {author} {\bibfnamefont {N.~A.}\
			\bibnamefont {Mortensen}},\ }\href {\doibase 10.1021/acsphotonics.7b00558}
	{\bibfield  {journal} {\bibinfo  {journal} {{ACS} Photonics}\ }\textbf
		{\bibinfo {volume} {4}},\ \bibinfo {pages} {3045} (\bibinfo {year}
		{2017})}\BibitemShut {NoStop}%
	\bibitem [{\citenamefont {Talebi}\ \emph {et~al.}(2016)\citenamefont {Talebi},
		\citenamefont {Ozsoy-Keskinbora}, \citenamefont {Benia}, \citenamefont
		{Kern}, \citenamefont {Koch},\ and\ \citenamefont {van Aken}}]{aken2016}%
	\BibitemOpen
	\bibfield  {author} {\bibinfo {author} {\bibfnamefont {N.}~\bibnamefont
			{Talebi}}, \bibinfo {author} {\bibfnamefont {C.}~\bibnamefont
			{Ozsoy-Keskinbora}}, \bibinfo {author} {\bibfnamefont {H.~M.}\ \bibnamefont
			{Benia}}, \bibinfo {author} {\bibfnamefont {K.}~\bibnamefont {Kern}},
		\bibinfo {author} {\bibfnamefont {C.~T.}\ \bibnamefont {Koch}}, \ and\
		\bibinfo {author} {\bibfnamefont {P.~A.}\ \bibnamefont {van Aken}},\ }\href
	{\doibase 10.1021/acsnano.6b02968} {\bibfield  {journal} {\bibinfo  {journal}
			{{ACS} Nano}\ }\textbf {\bibinfo {volume} {10}},\ \bibinfo {pages} {6988}
		(\bibinfo {year} {2016})}\BibitemShut {NoStop}%
	\bibitem [{\citenamefont {Lu}\ \emph {et~al.}(2018)\citenamefont {Lu},
		\citenamefont {Hao}, \citenamefont {Cen}, \citenamefont {Zhang},
		\citenamefont {Sun}, \citenamefont {Mao}, \citenamefont {Cao}, \citenamefont
		{Zhou}, \citenamefont {Jiang}, \citenamefont {Yang},\ and\ \citenamefont
		{Bao}}]{lu2018}%
	\BibitemOpen
	\bibfield  {author} {\bibinfo {author} {\bibfnamefont {X.}~\bibnamefont
			{Lu}}, \bibinfo {author} {\bibfnamefont {Q.}~\bibnamefont {Hao}}, \bibinfo
		{author} {\bibfnamefont {M.}~\bibnamefont {Cen}}, \bibinfo {author}
		{\bibfnamefont {G.}~\bibnamefont {Zhang}}, \bibinfo {author} {\bibfnamefont
			{J.}~\bibnamefont {Sun}}, \bibinfo {author} {\bibfnamefont {L.}~\bibnamefont
			{Mao}}, \bibinfo {author} {\bibfnamefont {T.}~\bibnamefont {Cao}}, \bibinfo
		{author} {\bibfnamefont {C.}~\bibnamefont {Zhou}}, \bibinfo {author}
		{\bibfnamefont {P.}~\bibnamefont {Jiang}}, \bibinfo {author} {\bibfnamefont
			{X.}~\bibnamefont {Yang}}, \ and\ \bibinfo {author} {\bibfnamefont
			{X.}~\bibnamefont {Bao}},\ }\href {\doibase 10.1021/acs.nanolett.8b00023}
	{\bibfield  {journal} {\bibinfo  {journal} {Nano Lett.}\ }\textbf {\bibinfo
			{volume} {18}},\ \bibinfo {pages} {2879} (\bibinfo {year}
		{2018})}\BibitemShut {NoStop}%
	\bibitem [{\citenamefont {Lingst{\"a}dt}\ \emph {et~al.}(2021)\citenamefont
		{Lingst{\"a}dt}, \citenamefont {Talebi}, \citenamefont {Hentschel},
		\citenamefont {Mashhadi}, \citenamefont {Gompf}, \citenamefont {Burghard},
		\citenamefont {Giessen},\ and\ \citenamefont {van Aken}}]{aken2020}%
	\BibitemOpen
	\bibfield  {author} {\bibinfo {author} {\bibfnamefont {R.}~\bibnamefont
			{Lingst{\"a}dt}}, \bibinfo {author} {\bibfnamefont {N.}~\bibnamefont
			{Talebi}}, \bibinfo {author} {\bibfnamefont {M.}~\bibnamefont {Hentschel}},
		\bibinfo {author} {\bibfnamefont {S.}~\bibnamefont {Mashhadi}}, \bibinfo
		{author} {\bibfnamefont {B.}~\bibnamefont {Gompf}}, \bibinfo {author}
		{\bibfnamefont {M.}~\bibnamefont {Burghard}}, \bibinfo {author}
		{\bibfnamefont {H.}~\bibnamefont {Giessen}}, \ and\ \bibinfo {author}
		{\bibfnamefont {P.~A.}\ \bibnamefont {van Aken}},\ }\href {\doibase
		10.1038/s43246-020-00108-9} {\bibfield  {journal} {\bibinfo  {journal}
			{Communications Materials}\ }\textbf {\bibinfo {volume} {2}},\ \bibinfo
		{pages} {5} (\bibinfo {year} {2021})}\BibitemShut {NoStop}%
	\bibitem [{\citenamefont {Bisharat}\ and\ \citenamefont
		{Sievenpiper}(2017)}]{bisharat2017}%
	\BibitemOpen
	\bibfield  {author} {\bibinfo {author} {\bibfnamefont {D.~J.}\ \bibnamefont
			{Bisharat}}\ and\ \bibinfo {author} {\bibfnamefont {D.~F.}\ \bibnamefont
			{Sievenpiper}},\ }\href {\doibase 10.1103/PhysRevLett.119.106802} {\bibfield
		{journal} {\bibinfo  {journal} {Phys. Rev. Lett.}\ }\textbf {\bibinfo
			{volume} {119}},\ \bibinfo {pages} {106802} (\bibinfo {year}
		{2017})}\BibitemShut {NoStop}%
	\bibitem [{\citenamefont {Fetter}(1985)}]{Fetter}%
	\BibitemOpen
	\bibfield  {author} {\bibinfo {author} {\bibfnamefont {A.~L.}\ \bibnamefont
			{Fetter}},\ }\href {\doibase 10.1103/PhysRevB.32.7676} {\bibfield  {journal}
		{\bibinfo  {journal} {Phys. Rev. B}\ }\textbf {\bibinfo {volume} {32}},\
		\bibinfo {pages} {7676} (\bibinfo {year} {1985})}\BibitemShut {NoStop}%
	\bibitem [{\citenamefont {{Volkov}}\ and\ \citenamefont
		{{Mikhailov}}(1988)}]{Volkov}%
	\BibitemOpen
	\bibfield  {author} {\bibinfo {author} {\bibfnamefont {V.~A.}\ \bibnamefont
			{{Volkov}}}\ and\ \bibinfo {author} {\bibfnamefont {S.~A.}\ \bibnamefont
			{{Mikhailov}}},\ }\href@noop {} {\bibfield  {journal} {\bibinfo  {journal}
			{Sov. Phys. JETP}\ }\textbf {\bibinfo {volume} {94}},\ \bibinfo {pages} {217}
		(\bibinfo {year} {1988})}\BibitemShut {NoStop}%
	\bibitem [{\citenamefont {Wassermeier}\ \emph {et~al.}(1990)\citenamefont
		{Wassermeier}, \citenamefont {Oshinowo}, \citenamefont {Kotthaus},
		\citenamefont {MacDonald}, \citenamefont {Foxon},\ and\ \citenamefont
		{Harris}}]{wassermeier1990}%
	\BibitemOpen
	\bibfield  {author} {\bibinfo {author} {\bibfnamefont {M.}~\bibnamefont
			{Wassermeier}}, \bibinfo {author} {\bibfnamefont {J.}~\bibnamefont
			{Oshinowo}}, \bibinfo {author} {\bibfnamefont {J.~P.}\ \bibnamefont
			{Kotthaus}}, \bibinfo {author} {\bibfnamefont {A.~H.}\ \bibnamefont
			{MacDonald}}, \bibinfo {author} {\bibfnamefont {C.~T.}\ \bibnamefont
			{Foxon}}, \ and\ \bibinfo {author} {\bibfnamefont {J.~J.}\ \bibnamefont
			{Harris}},\ }\href {\doibase 10.1103/PhysRevB.41.10287} {\bibfield  {journal}
		{\bibinfo  {journal} {Phys. Rev. B}\ }\textbf {\bibinfo {volume} {41}},\
		\bibinfo {pages} {10287} (\bibinfo {year} {1990})}\BibitemShut {NoStop}%
	\bibitem [{\citenamefont {Ashoori}\ \emph {et~al.}(1992)\citenamefont
		{Ashoori}, \citenamefont {Stormer}, \citenamefont {Pfeiffer}, \citenamefont
		{Baldwin},\ and\ \citenamefont {West}}]{ashoori1992}%
	\BibitemOpen
	\bibfield  {author} {\bibinfo {author} {\bibfnamefont {R.~C.}\ \bibnamefont
			{Ashoori}}, \bibinfo {author} {\bibfnamefont {H.~L.}\ \bibnamefont
			{Stormer}}, \bibinfo {author} {\bibfnamefont {L.~N.}\ \bibnamefont
			{Pfeiffer}}, \bibinfo {author} {\bibfnamefont {K.~W.}\ \bibnamefont
			{Baldwin}}, \ and\ \bibinfo {author} {\bibfnamefont {K.}~\bibnamefont
			{West}},\ }\href {\doibase 10.1103/physrevb.45.3894} {\bibfield  {journal}
		{\bibinfo  {journal} {Phys. Rev. B}\ }\textbf {\bibinfo {volume} {45}},\
		\bibinfo {pages} {3894} (\bibinfo {year} {1992})}\BibitemShut {NoStop}%
	\bibitem [{\citenamefont {Talyanskii}\ \emph {et~al.}(1992)\citenamefont
		{Talyanskii}, \citenamefont {Polisski}, \citenamefont {Arnone}, \citenamefont
		{Pepper}, \citenamefont {Smith}, \citenamefont {Ritchie}, \citenamefont
		{Frost},\ and\ \citenamefont {Jones}}]{talyanskii1992}%
	\BibitemOpen
	\bibfield  {author} {\bibinfo {author} {\bibfnamefont {V.~I.}\ \bibnamefont
			{Talyanskii}}, \bibinfo {author} {\bibfnamefont {A.~V.}\ \bibnamefont
			{Polisski}}, \bibinfo {author} {\bibfnamefont {D.~D.}\ \bibnamefont
			{Arnone}}, \bibinfo {author} {\bibfnamefont {M.}~\bibnamefont {Pepper}},
		\bibinfo {author} {\bibfnamefont {C.~G.}\ \bibnamefont {Smith}}, \bibinfo
		{author} {\bibfnamefont {D.~A.}\ \bibnamefont {Ritchie}}, \bibinfo {author}
		{\bibfnamefont {J.~E.}\ \bibnamefont {Frost}}, \ and\ \bibinfo {author}
		{\bibfnamefont {G.~A.~C.}\ \bibnamefont {Jones}},\ }\href {\doibase
		10.1103/physrevb.46.12427} {\bibfield  {journal} {\bibinfo  {journal} {Phys.
				Rev. B}\ }\textbf {\bibinfo {volume} {46}},\ \bibinfo {pages} {12427}
		(\bibinfo {year} {1992})}\BibitemShut {NoStop}%
	\bibitem [{\citenamefont {Muravev}\ \emph {et~al.}(2008)\citenamefont
		{Muravev}, \citenamefont {Fortunatov}, \citenamefont {Kukushkin},
		\citenamefont {Smet}, \citenamefont {Dietsche},\ and\ \citenamefont {von
			Klitzing}}]{kukushkin2008}%
	\BibitemOpen
	\bibfield  {author} {\bibinfo {author} {\bibfnamefont {V.~M.}\ \bibnamefont
			{Muravev}}, \bibinfo {author} {\bibfnamefont {A.~A.}\ \bibnamefont
			{Fortunatov}}, \bibinfo {author} {\bibfnamefont {I.~V.}\ \bibnamefont
			{Kukushkin}}, \bibinfo {author} {\bibfnamefont {J.~H.}\ \bibnamefont {Smet}},
		\bibinfo {author} {\bibfnamefont {W.}~\bibnamefont {Dietsche}}, \ and\
		\bibinfo {author} {\bibfnamefont {K.}~\bibnamefont {von Klitzing}},\ }\href
	{\doibase 10.1103/PhysRevLett.101.216801} {\bibfield  {journal} {\bibinfo
			{journal} {Phys. Rev. Lett.}\ }\textbf {\bibinfo {volume} {101}},\ \bibinfo
		{pages} {216801} (\bibinfo {year} {2008})}\BibitemShut {NoStop}%
	\bibitem [{\citenamefont {Wang}\ \emph {et~al.}(2012)\citenamefont {Wang},
		\citenamefont {Kinaret},\ and\ \citenamefont {Apell}}]{wang2012}%
	\BibitemOpen
	\bibfield  {author} {\bibinfo {author} {\bibfnamefont {W.}~\bibnamefont
			{Wang}}, \bibinfo {author} {\bibfnamefont {J.~M.}\ \bibnamefont {Kinaret}}, \
		and\ \bibinfo {author} {\bibfnamefont {S.~P.}\ \bibnamefont {Apell}},\ }\href
	{\doibase 10.1103/PhysRevB.85.235444} {\bibfield  {journal} {\bibinfo
			{journal} {Phys. Rev. B}\ }\textbf {\bibinfo {volume} {85}},\ \bibinfo
		{pages} {235444} (\bibinfo {year} {2012})}\BibitemShut {NoStop}%
	\bibitem [{\citenamefont {Yan}\ \emph {et~al.}(2012)\citenamefont {Yan},
		\citenamefont {Li}, \citenamefont {Li}, \citenamefont {Zhu}, \citenamefont
		{Avouris},\ and\ \citenamefont {Xia}}]{yan2012}%
	\BibitemOpen
	\bibfield  {author} {\bibinfo {author} {\bibfnamefont {H.}~\bibnamefont
			{Yan}}, \bibinfo {author} {\bibfnamefont {Z.}~\bibnamefont {Li}}, \bibinfo
		{author} {\bibfnamefont {X.}~\bibnamefont {Li}}, \bibinfo {author}
		{\bibfnamefont {W.}~\bibnamefont {Zhu}}, \bibinfo {author} {\bibfnamefont
			{P.}~\bibnamefont {Avouris}}, \ and\ \bibinfo {author} {\bibfnamefont
			{F.}~\bibnamefont {Xia}},\ }\href {\doibase 10.1021/nl3016335} {\bibfield
		{journal} {\bibinfo  {journal} {Nano Lett.}\ }\textbf {\bibinfo {volume}
			{12}},\ \bibinfo {pages} {3766} (\bibinfo {year} {2012})}\BibitemShut
	{NoStop}%
	\bibitem [{\citenamefont {Lin}\ \emph {et~al.}(2013)\citenamefont {Lin},
		\citenamefont {Xu}, \citenamefont {Zhang}, \citenamefont {Hao}, \citenamefont
		{Chen},\ and\ \citenamefont {Li}}]{lin2013}%
	\BibitemOpen
	\bibfield  {author} {\bibinfo {author} {\bibfnamefont {X.}~\bibnamefont
			{Lin}}, \bibinfo {author} {\bibfnamefont {Y.}~\bibnamefont {Xu}}, \bibinfo
		{author} {\bibfnamefont {B.}~\bibnamefont {Zhang}}, \bibinfo {author}
		{\bibfnamefont {R.}~\bibnamefont {Hao}}, \bibinfo {author} {\bibfnamefont
			{H.}~\bibnamefont {Chen}}, \ and\ \bibinfo {author} {\bibfnamefont
			{E.}~\bibnamefont {Li}},\ }\href {\doibase 10.1088/1367-2630/15/11/113003}
	{\bibfield  {journal} {\bibinfo  {journal} {New J. Phys}\ }\textbf {\bibinfo
			{volume} {15}},\ \bibinfo {pages} {113003} (\bibinfo {year}
		{2013})}\BibitemShut {NoStop}%
	\bibitem [{\citenamefont {Kumada}\ \emph {et~al.}(2014)\citenamefont {Kumada},
		\citenamefont {Roulleau}, \citenamefont {Roche}, \citenamefont {Hashisaka},
		\citenamefont {Hibino}, \citenamefont {Petkovi\ifmmode~\acute{c}\else
			\'{c}\fi{}},\ and\ \citenamefont {Glattli}}]{kumada2014}%
	\BibitemOpen
	\bibfield  {author} {\bibinfo {author} {\bibfnamefont {N.}~\bibnamefont
			{Kumada}}, \bibinfo {author} {\bibfnamefont {P.}~\bibnamefont {Roulleau}},
		\bibinfo {author} {\bibfnamefont {B.}~\bibnamefont {Roche}}, \bibinfo
		{author} {\bibfnamefont {M.}~\bibnamefont {Hashisaka}}, \bibinfo {author}
		{\bibfnamefont {H.}~\bibnamefont {Hibino}}, \bibinfo {author} {\bibfnamefont
			{I.}~\bibnamefont {Petkovi\ifmmode~\acute{c}\else \'{c}\fi{}}}, \ and\
		\bibinfo {author} {\bibfnamefont {D.~C.}\ \bibnamefont {Glattli}},\ }\href
	{\doibase 10.1103/PhysRevLett.113.266601} {\bibfield  {journal} {\bibinfo
			{journal} {Phys. Rev. Lett.}\ }\textbf {\bibinfo {volume} {113}},\ \bibinfo
		{pages} {266601} (\bibinfo {year} {2014})}\BibitemShut {NoStop}%
	\bibitem [{\citenamefont {Jin}\ \emph {et~al.}(2016)\citenamefont {Jin},
		\citenamefont {Lu}, \citenamefont {Wang}, \citenamefont {Fang}, \citenamefont
		{Joannopoulos}, \citenamefont {Solja{\v{c}}i{\'{c}}}, \citenamefont {Fu},\
		and\ \citenamefont {Fang}}]{jin2016}%
	\BibitemOpen
	\bibfield  {author} {\bibinfo {author} {\bibfnamefont {D.}~\bibnamefont
			{Jin}}, \bibinfo {author} {\bibfnamefont {L.}~\bibnamefont {Lu}}, \bibinfo
		{author} {\bibfnamefont {Z.}~\bibnamefont {Wang}}, \bibinfo {author}
		{\bibfnamefont {C.}~\bibnamefont {Fang}}, \bibinfo {author} {\bibfnamefont
			{J.~D.}\ \bibnamefont {Joannopoulos}}, \bibinfo {author} {\bibfnamefont
			{M.}~\bibnamefont {Solja{\v{c}}i{\'{c}}}}, \bibinfo {author} {\bibfnamefont
			{L.}~\bibnamefont {Fu}}, \ and\ \bibinfo {author} {\bibfnamefont {N.~X.}\
			\bibnamefont {Fang}},\ }\href {\doibase 10.1038/ncomms13486} {\bibfield
		{journal} {\bibinfo  {journal} {Nat. Commun.}\ }\textbf {\bibinfo {volume}
			{7}},\ \bibinfo {pages} {13486} (\bibinfo {year} {2016})}\BibitemShut
	{NoStop}%
	\bibitem [{\citenamefont {Cohen}\ and\ \citenamefont
		{Goldstein}(2018)}]{cohen2018}%
	\BibitemOpen
	\bibfield  {author} {\bibinfo {author} {\bibfnamefont {R.}~\bibnamefont
			{Cohen}}\ and\ \bibinfo {author} {\bibfnamefont {M.}~\bibnamefont
			{Goldstein}},\ }\href {\doibase 10.1103/PhysRevB.98.235103} {\bibfield
		{journal} {\bibinfo  {journal} {Phys. Rev. B}\ }\textbf {\bibinfo {volume}
			{98}},\ \bibinfo {pages} {235103} (\bibinfo {year} {2018})}\BibitemShut
	{NoStop}%
	\bibitem [{\citenamefont {Sokolik}\ and\ \citenamefont
		{Lozovik}(2019)}]{sokolik2019}%
	\BibitemOpen
	\bibfield  {author} {\bibinfo {author} {\bibfnamefont {A.~A.}\ \bibnamefont
			{Sokolik}}\ and\ \bibinfo {author} {\bibfnamefont {Y.~E.}\ \bibnamefont
			{Lozovik}},\ }\href {\doibase 10.1103/PhysRevB.100.125409} {\bibfield
		{journal} {\bibinfo  {journal} {Phys. Rev. B}\ }\textbf {\bibinfo {volume}
			{100}},\ \bibinfo {pages} {125409} (\bibinfo {year} {2019})}\BibitemShut
	{NoStop}%
	\bibitem [{\citenamefont {Wang}\ \emph {et~al.}(2011)\citenamefont {Wang},
		\citenamefont {Apell},\ and\ \citenamefont {Kinaret}}]{wang2011}%
	\BibitemOpen
	\bibfield  {author} {\bibinfo {author} {\bibfnamefont {W.}~\bibnamefont
			{Wang}}, \bibinfo {author} {\bibfnamefont {P.}~\bibnamefont {Apell}}, \ and\
		\bibinfo {author} {\bibfnamefont {J.}~\bibnamefont {Kinaret}},\ }\href
	{\doibase 10.1103/PhysRevB.84.085423} {\bibfield  {journal} {\bibinfo
			{journal} {Phys. Rev. B}\ }\textbf {\bibinfo {volume} {84}},\ \bibinfo
		{pages} {085423} (\bibinfo {year} {2011})}\BibitemShut {NoStop}%
	\bibitem [{\citenamefont {Stauber}\ \emph {et~al.}(2019)\citenamefont
		{Stauber}, \citenamefont {Nemilentsau}, \citenamefont {Low},\ and\
		\citenamefont {G{\'{o}}mez-Santos}}]{stauber2019}%
	\BibitemOpen
	\bibfield  {author} {\bibinfo {author} {\bibfnamefont {T.}~\bibnamefont
			{Stauber}}, \bibinfo {author} {\bibfnamefont {A.}~\bibnamefont
			{Nemilentsau}}, \bibinfo {author} {\bibfnamefont {T.}~\bibnamefont {Low}}, \
		and\ \bibinfo {author} {\bibfnamefont {G.}~\bibnamefont
			{G{\'{o}}mez-Santos}},\ }\href {\doibase 10.1088/2053-1583/ab2f05} {\bibfield
		{journal} {\bibinfo  {journal} {2D Mater.}\ }\textbf {\bibinfo {volume}
			{6}},\ \bibinfo {pages} {045023} (\bibinfo {year} {2019})}\BibitemShut
	{NoStop}%
	\bibitem [{\citenamefont {Zabolotnykh}\ and\ \citenamefont
		{Volkov}(2016)}]{zabolotnykh2016}%
	\BibitemOpen
	\bibfield  {author} {\bibinfo {author} {\bibfnamefont {A.~A.}\ \bibnamefont
			{Zabolotnykh}}\ and\ \bibinfo {author} {\bibfnamefont {V.~A.}\ \bibnamefont
			{Volkov}},\ }\href {\doibase 10.1134/s0021364016180144} {\bibfield  {journal}
		{\bibinfo  {journal} {JETP Lett.}\ }\textbf {\bibinfo {volume} {104}},\
		\bibinfo {pages} {411} (\bibinfo {year} {2016})}\BibitemShut {NoStop}%
	\bibitem [{\citenamefont {Margetis}\ \emph {et~al.}(2020)\citenamefont
		{Margetis}, \citenamefont {Maier}, \citenamefont {Stauber}, \citenamefont
		{Low},\ and\ \citenamefont {Luskin}}]{stauber2020}%
	\BibitemOpen
	\bibfield  {author} {\bibinfo {author} {\bibfnamefont {D.}~\bibnamefont
			{Margetis}}, \bibinfo {author} {\bibfnamefont {M.}~\bibnamefont {Maier}},
		\bibinfo {author} {\bibfnamefont {T.}~\bibnamefont {Stauber}}, \bibinfo
		{author} {\bibfnamefont {T.}~\bibnamefont {Low}}, \ and\ \bibinfo {author}
		{\bibfnamefont {M.}~\bibnamefont {Luskin}},\ }\href {\doibase
		10.1088/1751-8121/ab5ff9} {\bibfield  {journal} {\bibinfo  {journal} {J.
				Phys. A Math. Theor.}\ }\textbf {\bibinfo {volume} {53}},\ \bibinfo {pages}
		{055201} (\bibinfo {year} {2020})}\BibitemShut {NoStop}%
	\bibitem [{\citenamefont {Margetis}(2020)}]{margetis2020}%
	\BibitemOpen
	\bibfield  {author} {\bibinfo {author} {\bibfnamefont {D.}~\bibnamefont
			{Margetis}},\ }\href {\doibase 10.1063/1.5128895} {\bibfield  {journal}
		{\bibinfo  {journal} {J. Math. Phys.}\ }\textbf {\bibinfo {volume} {61}},\
		\bibinfo {pages} {062901} (\bibinfo {year} {2020})}\BibitemShut {NoStop}%
	\bibitem [{\citenamefont {Nikulin}\ \emph {et~al.}(2021)\citenamefont
		{Nikulin}, \citenamefont {Mylnikov}, \citenamefont {Bandurin},\ and\
		\citenamefont {Svintsov}}]{svintsov2020}%
	\BibitemOpen
	\bibfield  {author} {\bibinfo {author} {\bibfnamefont {E.}~\bibnamefont
			{Nikulin}}, \bibinfo {author} {\bibfnamefont {D.}~\bibnamefont {Mylnikov}},
		\bibinfo {author} {\bibfnamefont {D.}~\bibnamefont {Bandurin}}, \ and\
		\bibinfo {author} {\bibfnamefont {D.}~\bibnamefont {Svintsov}},\ }\href
	{\doibase 10.1103/PhysRevB.103.085306} {\bibfield  {journal} {\bibinfo
			{journal} {Phys. Rev. B}\ }\textbf {\bibinfo {volume} {103}},\ \bibinfo
		{pages} {085306} (\bibinfo {year} {2021})}\BibitemShut {NoStop}%
	\bibitem [{\citenamefont {Bollinger}\ \emph {et~al.}(2001)\citenamefont
		{Bollinger}, \citenamefont {Lauritsen}, \citenamefont {Jacobsen},
		\citenamefont {N\o{}rskov}, \citenamefont {Helveg},\ and\ \citenamefont
		{Besenbacher}}]{bollinger2001}%
	\BibitemOpen
	\bibfield  {author} {\bibinfo {author} {\bibfnamefont {M.~V.}\ \bibnamefont
			{Bollinger}}, \bibinfo {author} {\bibfnamefont {J.~V.}\ \bibnamefont
			{Lauritsen}}, \bibinfo {author} {\bibfnamefont {K.~W.}\ \bibnamefont
			{Jacobsen}}, \bibinfo {author} {\bibfnamefont {J.~K.}\ \bibnamefont
			{N\o{}rskov}}, \bibinfo {author} {\bibfnamefont {S.}~\bibnamefont {Helveg}},
		\ and\ \bibinfo {author} {\bibfnamefont {F.}~\bibnamefont {Besenbacher}},\
	}\href {\doibase 10.1103/PhysRevLett.87.196803} {\bibfield  {journal}
		{\bibinfo  {journal} {Phys. Rev. Lett.}\ }\textbf {\bibinfo {volume} {87}},\
		\bibinfo {pages} {196803} (\bibinfo {year} {2001})}\BibitemShut {NoStop}%
	\bibitem [{\citenamefont {Rossi}\ \emph {et~al.}(2017)\citenamefont {Rossi},
		\citenamefont {Winther}, \citenamefont {Jacobsen}, \citenamefont {Nieminen},
		\citenamefont {Puska},\ and\ \citenamefont {Thygesen}}]{rossi2017}%
	\BibitemOpen
	\bibfield  {author} {\bibinfo {author} {\bibfnamefont {T.~P.}\ \bibnamefont
			{Rossi}}, \bibinfo {author} {\bibfnamefont {K.~T.}\ \bibnamefont {Winther}},
		\bibinfo {author} {\bibfnamefont {K.~W.}\ \bibnamefont {Jacobsen}}, \bibinfo
		{author} {\bibfnamefont {R.~M.}\ \bibnamefont {Nieminen}}, \bibinfo {author}
		{\bibfnamefont {M.~J.}\ \bibnamefont {Puska}}, \ and\ \bibinfo {author}
		{\bibfnamefont {K.~S.}\ \bibnamefont {Thygesen}},\ }\href {\doibase
		10.1103/PhysRevB.96.155407} {\bibfield  {journal} {\bibinfo  {journal} {Phys.
				Rev. B}\ }\textbf {\bibinfo {volume} {96}},\ \bibinfo {pages} {155407}
		(\bibinfo {year} {2017})}\BibitemShut {NoStop}%
	\bibitem [{\citenamefont {Zhou}\ \emph {et~al.}(2012)\citenamefont {Zhou},
		\citenamefont {Yin}, \citenamefont {Du}, \citenamefont {Huang}, \citenamefont
		{Zeng}, \citenamefont {Fan}, \citenamefont {Liu}, \citenamefont {Wang},\ and\
		\citenamefont {Zhang}}]{zhou2013}%
	\BibitemOpen
	\bibfield  {author} {\bibinfo {author} {\bibfnamefont {W.}~\bibnamefont
			{Zhou}}, \bibinfo {author} {\bibfnamefont {Z.}~\bibnamefont {Yin}}, \bibinfo
		{author} {\bibfnamefont {Y.}~\bibnamefont {Du}}, \bibinfo {author}
		{\bibfnamefont {X.}~\bibnamefont {Huang}}, \bibinfo {author} {\bibfnamefont
			{Z.}~\bibnamefont {Zeng}}, \bibinfo {author} {\bibfnamefont {Z.}~\bibnamefont
			{Fan}}, \bibinfo {author} {\bibfnamefont {H.}~\bibnamefont {Liu}}, \bibinfo
		{author} {\bibfnamefont {J.}~\bibnamefont {Wang}}, \ and\ \bibinfo {author}
		{\bibfnamefont {H.}~\bibnamefont {Zhang}},\ }\href {\doibase
		10.1002/smll.201201161} {\bibfield  {journal} {\bibinfo  {journal} {Small}\
		}\textbf {\bibinfo {volume} {9}},\ \bibinfo {pages} {140} (\bibinfo {year}
		{2012})}\BibitemShut {NoStop}%
	\bibitem [{\citenamefont {Donarelli}\ and\ \citenamefont
		{Ottaviano}(2018)}]{donarelli2018}%
	\BibitemOpen
	\bibfield  {author} {\bibinfo {author} {\bibfnamefont {M.}~\bibnamefont
			{Donarelli}}\ and\ \bibinfo {author} {\bibfnamefont {L.}~\bibnamefont
			{Ottaviano}},\ }\href {\doibase 10.3390/s18113638} {\bibfield  {journal}
		{\bibinfo  {journal} {Sensors}\ }\textbf {\bibinfo {volume} {18}},\ \bibinfo
		{pages} {3638} (\bibinfo {year} {2018})}\BibitemShut {NoStop}%
	\bibitem [{\citenamefont {Yin}\ \emph {et~al.}(2014)\citenamefont {Yin},
		\citenamefont {Ye}, \citenamefont {Chenet}, \citenamefont {Ye}, \citenamefont
		{O{\textquotesingle}Brien}, \citenamefont {Hone},\ and\ \citenamefont
		{Zhang}}]{yin2014}%
	\BibitemOpen
	\bibfield  {author} {\bibinfo {author} {\bibfnamefont {X.}~\bibnamefont
			{Yin}}, \bibinfo {author} {\bibfnamefont {Z.}~\bibnamefont {Ye}}, \bibinfo
		{author} {\bibfnamefont {D.~A.}\ \bibnamefont {Chenet}}, \bibinfo {author}
		{\bibfnamefont {Y.}~\bibnamefont {Ye}}, \bibinfo {author} {\bibfnamefont
			{K.}~\bibnamefont {O{\textquotesingle}Brien}}, \bibinfo {author}
		{\bibfnamefont {J.~C.}\ \bibnamefont {Hone}}, \ and\ \bibinfo {author}
		{\bibfnamefont {X.}~\bibnamefont {Zhang}},\ }\href {\doibase
		10.1126/science.1250564} {\bibfield  {journal} {\bibinfo  {journal}
			{Science}\ }\textbf {\bibinfo {volume} {344}},\ \bibinfo {pages} {488}
		(\bibinfo {year} {2014})}\BibitemShut {NoStop}%
	\bibitem [{\citenamefont {Zhang}\ and\ \citenamefont
		{Vignale}(2018)}]{zhang2018}%
	\BibitemOpen
	\bibfield  {author} {\bibinfo {author} {\bibfnamefont {S.~S.-L.}\
			\bibnamefont {Zhang}}\ and\ \bibinfo {author} {\bibfnamefont
			{G.}~\bibnamefont {Vignale}},\ }\href {\doibase 10.1103/PhysRevB.97.224408}
	{\bibfield  {journal} {\bibinfo  {journal} {Phys. Rev. B}\ }\textbf {\bibinfo
			{volume} {97}},\ \bibinfo {pages} {224408} (\bibinfo {year}
		{2018})}\BibitemShut {NoStop}%
	\bibitem [{\citenamefont {Mahoney}\ \emph {et~al.}(2017)\citenamefont
		{Mahoney}, \citenamefont {Colless}, \citenamefont {Peeters}, \citenamefont
		{Pauka}, \citenamefont {Fox}, \citenamefont {Kou}, \citenamefont {Pan},
		\citenamefont {Wang}, \citenamefont {Goldhaber-Gordon},\ and\ \citenamefont
		{Reilly}}]{mahoney2017}%
	\BibitemOpen
	\bibfield  {author} {\bibinfo {author} {\bibfnamefont {A.~C.}\ \bibnamefont
			{Mahoney}}, \bibinfo {author} {\bibfnamefont {J.~I.}\ \bibnamefont
			{Colless}}, \bibinfo {author} {\bibfnamefont {L.}~\bibnamefont {Peeters}},
		\bibinfo {author} {\bibfnamefont {S.~J.}\ \bibnamefont {Pauka}}, \bibinfo
		{author} {\bibfnamefont {E.~J.}\ \bibnamefont {Fox}}, \bibinfo {author}
		{\bibfnamefont {X.}~\bibnamefont {Kou}}, \bibinfo {author} {\bibfnamefont
			{L.}~\bibnamefont {Pan}}, \bibinfo {author} {\bibfnamefont {K.~L.}\
			\bibnamefont {Wang}}, \bibinfo {author} {\bibfnamefont {D.}~\bibnamefont
			{Goldhaber-Gordon}}, \ and\ \bibinfo {author} {\bibfnamefont {D.~J.}\
			\bibnamefont {Reilly}},\ }\href {\doibase 10.1038/s41467-017-01984-5}
	{\bibfield  {journal} {\bibinfo  {journal} {Nat. Commun.}\ }\textbf {\bibinfo
			{volume} {8}},\ \bibinfo {pages} {1836} (\bibinfo {year} {2017})}\BibitemShut
	{NoStop}%
	\bibitem [{\citenamefont {Kumar}\ \emph {et~al.}(2016)\citenamefont {Kumar},
		\citenamefont {Nemilentsau}, \citenamefont {Fung}, \citenamefont {Hanson},
		\citenamefont {Fang},\ and\ \citenamefont {Low}}]{kumar2016}%
	\BibitemOpen
	\bibfield  {author} {\bibinfo {author} {\bibfnamefont {A.}~\bibnamefont
			{Kumar}}, \bibinfo {author} {\bibfnamefont {A.}~\bibnamefont {Nemilentsau}},
		\bibinfo {author} {\bibfnamefont {K.~H.}\ \bibnamefont {Fung}}, \bibinfo
		{author} {\bibfnamefont {G.}~\bibnamefont {Hanson}}, \bibinfo {author}
		{\bibfnamefont {N.~X.}\ \bibnamefont {Fang}}, \ and\ \bibinfo {author}
		{\bibfnamefont {T.}~\bibnamefont {Low}},\ }\href {\doibase
		10.1103/PhysRevB.93.041413} {\bibfield  {journal} {\bibinfo  {journal} {Phys.
				Rev. B}\ }\textbf {\bibinfo {volume} {93}},\ \bibinfo {pages} {041413}
		(\bibinfo {year} {2016})}\BibitemShut {NoStop}%
	\bibitem [{\citenamefont {Song}\ and\ \citenamefont {Rudner}(2016)}]{song2016}%
	\BibitemOpen
	\bibfield  {author} {\bibinfo {author} {\bibfnamefont {J.~C.~W.}\
			\bibnamefont {Song}}\ and\ \bibinfo {author} {\bibfnamefont {M.~S.}\
			\bibnamefont {Rudner}},\ }\href {\doibase 10.1073/pnas.1519086113} {\bibfield
		{journal} {\bibinfo  {journal} {Proc. Natl. Acad. Sci.}\ }\textbf {\bibinfo
			{volume} {113}},\ \bibinfo {pages} {4658} (\bibinfo {year}
		{2016})}\BibitemShut {NoStop}%
	\bibitem [{\citenamefont {Hafezi}\ \emph {et~al.}(2013)\citenamefont {Hafezi},
		\citenamefont {Mittal}, \citenamefont {Fan}, \citenamefont {Migdall},\ and\
		\citenamefont {Taylor}}]{hafezi2013}%
	\BibitemOpen
	\bibfield  {author} {\bibinfo {author} {\bibfnamefont {M.}~\bibnamefont
			{Hafezi}}, \bibinfo {author} {\bibfnamefont {S.}~\bibnamefont {Mittal}},
		\bibinfo {author} {\bibfnamefont {J.}~\bibnamefont {Fan}}, \bibinfo {author}
		{\bibfnamefont {A.}~\bibnamefont {Migdall}}, \ and\ \bibinfo {author}
		{\bibfnamefont {J.~M.}\ \bibnamefont {Taylor}},\ }\href {\doibase
		10.1038/nphoton.2013.274} {\bibfield  {journal} {\bibinfo  {journal} {Nat.
				Photonics}\ }\textbf {\bibinfo {volume} {7}},\ \bibinfo {pages} {1001}
		(\bibinfo {year} {2013})}\BibitemShut {NoStop}%
	\bibitem [{\citenamefont {Yang}\ \emph {et~al.}(2018)\citenamefont {Yang},
		\citenamefont {Xu}, \citenamefont {Xu}, \citenamefont {Wang}, \citenamefont
		{Jiang}, \citenamefont {Hu},\ and\ \citenamefont {Hang}}]{yang2018}%
	\BibitemOpen
	\bibfield  {author} {\bibinfo {author} {\bibfnamefont {Y.}~\bibnamefont
			{Yang}}, \bibinfo {author} {\bibfnamefont {Y.~F.}\ \bibnamefont {Xu}},
		\bibinfo {author} {\bibfnamefont {T.}~\bibnamefont {Xu}}, \bibinfo {author}
		{\bibfnamefont {H.-X.}\ \bibnamefont {Wang}}, \bibinfo {author}
		{\bibfnamefont {J.-H.}\ \bibnamefont {Jiang}}, \bibinfo {author}
		{\bibfnamefont {X.}~\bibnamefont {Hu}}, \ and\ \bibinfo {author}
		{\bibfnamefont {Z.~H.}\ \bibnamefont {Hang}},\ }\href {\doibase
		10.1103/PhysRevLett.120.217401} {\bibfield  {journal} {\bibinfo  {journal}
			{Phys. Rev. Lett.}\ }\textbf {\bibinfo {volume} {120}},\ \bibinfo {pages}
		{217401} (\bibinfo {year} {2018})}\BibitemShut {NoStop}%
	\bibitem [{\citenamefont {Parappurath}\ \emph {et~al.}(2020)\citenamefont
		{Parappurath}, \citenamefont {Alpeggiani}, \citenamefont {Kuipers},\ and\
		\citenamefont {Verhagen}}]{parappurath2020}%
	\BibitemOpen
	\bibfield  {author} {\bibinfo {author} {\bibfnamefont {N.}~\bibnamefont
			{Parappurath}}, \bibinfo {author} {\bibfnamefont {F.}~\bibnamefont
			{Alpeggiani}}, \bibinfo {author} {\bibfnamefont {L.}~\bibnamefont {Kuipers}},
		\ and\ \bibinfo {author} {\bibfnamefont {E.}~\bibnamefont {Verhagen}},\
	}\href {\doibase 10.1126/sciadv.aaw4137} {\bibfield  {journal} {\bibinfo
			{journal} {Sci. Adv.}\ }\textbf {\bibinfo {volume} {6}},\ \bibinfo {pages}
		{eaaw4137} (\bibinfo {year} {2020})}\BibitemShut {NoStop}%
	\bibitem [{\citenamefont {Lu}\ \emph {et~al.}(2014)\citenamefont {Lu},
		\citenamefont {Joannopoulos},\ and\ \citenamefont
		{Solja{\v{c}}i{\'{c}}}}]{lu2014}%
	\BibitemOpen
	\bibfield  {author} {\bibinfo {author} {\bibfnamefont {L.}~\bibnamefont
			{Lu}}, \bibinfo {author} {\bibfnamefont {J.~D.}\ \bibnamefont
			{Joannopoulos}}, \ and\ \bibinfo {author} {\bibfnamefont {M.}~\bibnamefont
			{Solja{\v{c}}i{\'{c}}}},\ }\href {\doibase 10.1038/nphoton.2014.248}
	{\bibfield  {journal} {\bibinfo  {journal} {Nat. Photonics}\ }\textbf
		{\bibinfo {volume} {8}},\ \bibinfo {pages} {821} (\bibinfo {year}
		{2014})}\BibitemShut {NoStop}%
	\bibitem [{\citenamefont {Ota}\ \emph {et~al.}(2020)\citenamefont {Ota},
		\citenamefont {Takata}, \citenamefont {Ozawa}, \citenamefont {Amo},
		\citenamefont {Jia}, \citenamefont {Kante}, \citenamefont {Notomi},
		\citenamefont {Arakawa},\ and\ \citenamefont {Iwamoto}}]{ota2020}%
	\BibitemOpen
	\bibfield  {author} {\bibinfo {author} {\bibfnamefont {Y.}~\bibnamefont
			{Ota}}, \bibinfo {author} {\bibfnamefont {K.}~\bibnamefont {Takata}},
		\bibinfo {author} {\bibfnamefont {T.}~\bibnamefont {Ozawa}}, \bibinfo
		{author} {\bibfnamefont {A.}~\bibnamefont {Amo}}, \bibinfo {author}
		{\bibfnamefont {Z.}~\bibnamefont {Jia}}, \bibinfo {author} {\bibfnamefont
			{B.}~\bibnamefont {Kante}}, \bibinfo {author} {\bibfnamefont
			{M.}~\bibnamefont {Notomi}}, \bibinfo {author} {\bibfnamefont
			{Y.}~\bibnamefont {Arakawa}}, \ and\ \bibinfo {author} {\bibfnamefont
			{S.}~\bibnamefont {Iwamoto}},\ }\href {\doibase 10.1515/nanoph-2019-0376}
	{\bibfield  {journal} {\bibinfo  {journal} {Nanophotonics}\ }\textbf
		{\bibinfo {volume} {9}},\ \bibinfo {pages} {547} (\bibinfo {year}
		{2020})}\BibitemShut {NoStop}%
	\bibitem [{\citenamefont {Talebi}(2019)}]{Talebi_book}%
	\BibitemOpen
	\bibfield  {author} {\bibinfo {author} {\bibfnamefont {N.}~\bibnamefont
			{Talebi}},\ }\href@noop {} {\emph {\bibinfo {title} {Near-Field-Mediated Photon–Electron Interactions}}}\ (\bibinfo  {publisher} {Springer},\
	\bibinfo {address} {Cham},\ \bibinfo {year} {2019})\BibitemShut {NoStop}%
	\bibitem [{\citenamefont {Yao}\ \emph {et~al.}(2020)\citenamefont {Yao},
		\citenamefont {Xu}, \citenamefont {Hu}, \citenamefont {Chen}, \citenamefont
		{Dai},\ and\ \citenamefont {Liu}}]{yao2020}%
	\BibitemOpen
	\bibfield  {author} {\bibinfo {author} {\bibfnamefont {Z.}~\bibnamefont
			{Yao}}, \bibinfo {author} {\bibfnamefont {S.}~\bibnamefont {Xu}}, \bibinfo
		{author} {\bibfnamefont {D.}~\bibnamefont {Hu}}, \bibinfo {author}
		{\bibfnamefont {X.}~\bibnamefont {Chen}}, \bibinfo {author} {\bibfnamefont
			{Q.}~\bibnamefont {Dai}}, \ and\ \bibinfo {author} {\bibfnamefont
			{M.}~\bibnamefont {Liu}},\ }\href {\doibase 10.1002/adom.201901042}
	{\bibfield  {journal} {\bibinfo  {journal} {Adv. Opt. Mater.}\ }\textbf
		{\bibinfo {volume} {8}},\ \bibinfo {pages} {1901042} (\bibinfo {year}
		{2020})}\BibitemShut {NoStop}%
	\bibitem [{\citenamefont {Carrier}\ \emph {et~al.}(1966)\citenamefont
		{Carrier}, \citenamefont {Krook},\ and\ \citenamefont {Pearson}}]{Carrier}%
	\BibitemOpen
	\bibfield  {author} {\bibinfo {author} {\bibfnamefont {G.}~\bibnamefont
			{Carrier}}, \bibinfo {author} {\bibfnamefont {M.}~\bibnamefont {Krook}}, \
		and\ \bibinfo {author} {\bibfnamefont {C.}~\bibnamefont {Pearson}},\
	}\href@noop {} {\emph {\bibinfo {title} {Functions of a Complex Variable}}}\
	(\bibinfo  {publisher} {McGraw-Hill},\ \bibinfo {address} {New York},\
	\bibinfo {year} {1966})\BibitemShut {NoStop}%
\end{thebibliography}

%

\end{document}